\documentclass[a4paper,12pt]{article}
\usepackage{a4wide,latexsym,graphicx,epsfig,psfrag,float}
\usepackage{mathrsfs}
\usepackage{latexsym}
\usepackage{indentfirst}
\usepackage{amsmath}
\usepackage{amsxtra}
\usepackage{amsbsy}
\usepackage{amscd}
\usepackage{color}
\usepackage{bbm}
\usepackage{hyperref}
\usepackage{multirow}
\usepackage{graphics}
\usepackage{amssymb}
\usepackage{amsfonts}
\usepackage{subfigure}
\usepackage{color}
\usepackage{times,fancyhdr}
\usepackage{amsfonts}
\usepackage{longtable}

\allowdisplaybreaks

\newcommand{\be}{\begin{equation}}
\newcommand{\ee}{\end{equation}}

\def\lr #1{\left( #1 \right)}
\def\no{\nonumber}
\def\bea{\arraycolsep .1em \begin{eqnarray}}
\def\eea{\end{eqnarray}}

\begin{document}

\begin{titlepage}
\begin{flushright}
{
JLAB-THY-14-1842
}
\end{flushright}
\vspace{8mm}

\setcounter{footnote}{0}
\renewcommand{\thefootnote}{\fnsymbol{footnote}}

\begin{center}
{\LARGE \bf Comprehensive Amplitude Analysis of\\[1.8mm] $\mathbf{\gamma\gamma \rightarrow \pi^+\pi^-, \pi^0\pi^0}$ and $\mathbf{{\overline K}K}$ below 1.5~GeV}
\vspace{10mm} \\
{\sc  Ling-Yun Dai}\footnote{Email:~~lingyun@jlab.org} and {\sc M.R. Pennington}\footnote{Email:~michaelp@jlab.org}
\vspace{7mm} \\ Theory Center,
Thomas Jefferson National Accelerator Facility,\\
 Newport News, VA 23606, USA \\[1cm]
\end{center}

\setcounter{footnote}{0}
\renewcommand{\thefootnote}{\arabic{footnote}}
\vspace{0.5cm}
\baselineskip=5.7mm
\begin{abstract}
In this paper we perform an amplitude analysis of essentially all published pion and kaon pair production data from two photon collisions below 1.5~GeV. This includes all the high statistics results from Belle, as well as older data from Mark II at SLAC, CELLO at DESY, Crystal Ball at SLAC. The purpose of this analysis is to provide as close to a model-independent determination of the $\gamma\gamma$ to meson pair amplitudes as possible. Having data with limited angular coverage, typically $|\cos \theta| < 0.6-0.8$, and no polarization information for reactions in which spin is an essential complication, the determination of the underlying amplitudes might appear an intractable problem. However, imposing the basic constraints required by analyticity,  unitarity, and crossing-symmetry makes up for the experimentally missing information. Above 1.5~GeV multi-meson production channels become important and we have too little information to resolve the amplitudes. Nevertheless, below 1.5~GeV the two photon production of hadron pairs serves as a paradigm for the application of $S$-matrix techniques.

Final state interactions among the meson pairs are critical to this analysis. To fix these, we include the latest $\pi\pi\to\pi\pi$, ${\overline K}K$ scattering amplitudes given by dispersive analyses, supplemented in the ${\overline K}K$ threshold region by the recent precision Dalitz plot analysis from BaBar.
With these hadronic amplitudes built into unitarity, we can constrain the overall description of $\gamma\gamma\to\pi\pi$ and $\overline{K}K$ datasets, both integrated and differential cross-sections, including the high statistics charged and  neutral pion, as well as $K_sK_s$, data from Belle.
Since this analysis invokes coupled hadronic channels, having data on both $\pi\pi$ and $\overline{K}K$ reduces the solution space to essentially a single form in the region where these channels saturate unitarity. For the $\pi\pi$ channel the separation of isopin 0 and 2, and helicity 0 and 2 components is complete.
We present the partial wave amplitudes, show how well they fit all the available data, and give the two photon couplings
of scalar and tensor resonances that appear. These partial waves are important inputs into forthcoming dispersive calculations of hadronic light-by-light scattering.
\end{abstract}
\vspace{0.5cm}

{\indent PACS~: 11.55.Fv, 14.40.Be, 11.80.Et, 13.60.Le \\
\indent Keywords~: Dispersion relations, Light mesons, Partial-wave analysis, Meson production.}

\end{titlepage}

\parskip=2mm
\baselineskip=5.5mm

\section{Introduction}\label{sec:0}
There has long been interest  both theoretically and experimentally in photon-photon interactions as one of the cleanest ways of  probing hadron structure. The  differing compositions of resonant states, whether ${\overline q}q$, ${\overline {qq}}qq$, $gg$, or hadronic molecules, are, in principle, revealed by their two photon couplings~\cite{MRP07Rev}. Consequently, it is important to be able to extract these couplings reliably from experiment. With incomplete data, this is far from straightforward, but it is nevertheless possible. That is the purpose of the present study. Once that is done, we will compare our results with the many model predictions for different compositions of these states.
Just as importantly, our partial waves serve as key input into future dispersive calculations of the hadronic light-by-light contribution to the anomalous magnetic moment of the muon, providing further motivation for a new study of greater  certainty.

\noindent Experimental effort has focused with most precision on $\pi\pi$ production with a series of  measurements of the cross-sections for $\gamma\gamma\to\pi^+\pi^-$~\cite{MarkII,Cello1,Cello2} and $\gamma\gamma\to\pi^0\pi^0$~\cite{CB88,CB92} reactions.
It is sometimes advertized that the all neutral channel is  ideal for spotting resonances. The fact that there is no direct electromagnetic interaction of the photons with $\pi^0$'s makes this process background free. This is in contrast to the charged pion channel, where just above threshold the cross-section is dominated
by the one pion exchange Born term. Thus from the fact that there is no low mass enhancement in the $\pi^0\pi^0$ channel is inferred the $\sigma$ must have very small coupling to $\gamma\gamma$ and so be a \lq\lq glueball''. Unfortunately such arguments are far too simplistic. Even at energies just above $\pi^+\pi^-$ threshold, the Born term is rapidly modified by final state interactions that include the $\sigma$. Moreover, the neutral pion cross-section is naturally non-zero, since $\gamma\gamma\to\pi^+\pi^-$ through the Born amplitude  and then the $\pi^+\pi^-$ pair can  scatter to $\pi^0\pi^0$. Indeed, these final state interactions are so important that their effect dominates the coupling of the $\sigma\to\pi\pi$. Remarkably, the near threshold process is precisely calculable~\cite{Mennessier83,MRP87,MRP91} without knowing the exact composition of the $\sigma$. As the energy increases the photons couple not to the charge of the final state hadron, but rather to their internal charged constituents. Consequently, the charged and neutral pion cross-sections become more similar with the $f_2(1270)$ emerging as the dominant structure.

\noindent To determine the two photon amplitudes, and hence the couplings of any resonances, requires a model-independent strategy for extracting the relevant partial wave amplitudes from experiment. Implementing the basic constraints of analyticity, unitarity, and crossing are essential to this program.  This has been set out earlier in a series of papers~\cite{MRP88,MRP90,MRP98} that we follow here. This formalism automatically takes into account the key final state interactions and allows for the only robust determination of resonance couplings. Moreover, as we present here, recent data from Belle on $\gamma\gamma\to\pi^+\pi^-$~\cite{Belle-pm}, $\pi^0\pi^0$~\cite{Belle-nn} and $K_sK_s$~\cite{Belle-KsKs}  considerably sharpen the analysis by an increase in statistics and precision of at least a factor of ten.

\newpage
\noindent One of the low energy states of particular interest is the $f_0(980)$. The Belle charged pion measurement has statistics sufficient to allow  the $\pi^+\pi^-$ invariant mass to be binned in 5 MeV steps and in angular intervals of 0.05 in $\cos \theta$, where $\theta$ is the center-of-mass scattering angle (often called $\theta^*$ in experimental papers).
This reveals a clear peak in the $0.93-1.03$~GeV region. Fitting this with a simple smooth background, Belle~\cite{Belle-pm} quote a  two-photon width for the $f_0(980)$ of $205 ~^{+95}_{-83}\lr{\rm stat}~_{-117}^{+147}\lr{\rm syst}$ eV. However, these data, while having unprecedented statistics, seem to have large distortions from $\mu^+\mu^-$ contamination, as discussed in ~\cite{MRP-KLOE2} and to be seen later in this paper in the figures to come.

\noindent A previous  amplitude analysis, by one of the present authors (MRP) and Belle colleagues based on a similar marriage of dispersion relations with unitarity~\cite{MRP08} we use here, fitted all the available cross-sections and angular distributions. The partial wave analysis exposed
a range of solutions with a two photon width for the $f_0(980)$  between 100~eV and 540~eV (not so dissimilar to the range allowed by Belle's simple resonance plus background fit with no partial wave separation). However, at the time only the charged pion data from Belle were available.  Now with the publication of their neutral pion results, it is appropriate to revisit this analysis. This is the motivation for the present work.

\noindent To be able to perform an Amplitude Analysis, two unavoidable problems have to be solved. First experiments have only limited angular coverage and secondly the polarization of the initial states is not measured. It is here that the theoretical constraints from analyticity, unitarity, crossing symmetry and Low's low energy theorem of QED come in.
These allow all  the partial waves to be calculated within tolerable uncertainties below 600~MeV~\cite{MRP87,MRP91,MRP06,MRP08}. Above that energy imposing a framework that ensures the two photon amplitudes are correctly related to hadronic scattering processes through coupled channel unitarity is sufficient to determine the partial waves, provided, of course, one has sufficient information about the corresponding hadronic reactions. Here we take advantage of recent dispersive studies of meson-meson scattering amplitudes that combine classic inputs from experiments like that of the CERN-Munich group on $\pi\pi$ production~\cite{CERN-Munich}, and ANL, BNL on the ${\overline K}K$ final state~\cite{Cohen80,Etkin82} with the latest low energy data from NA48-2~\cite{NA48}.
Pel\'aez and his collaborators~\cite{KPY} provide $\pi\pi$ scattering amplitudes for the $I=0,\ 2$ $S$ and $D$-waves we need here. These in turn require  additional information on inelasticities that we discuss in detail in Sect.~2. When these hadronic inputs are included in a coupled-channel $K$-matrix representation, we then have constraints on channels like ${\overline K}K\to{\overline K}K$ that we also require. Armed with these $T$-matrix elements, we can then fit the $\gamma\gamma\to\pi\pi$ (and ${\overline K}K$) amplitudes and determine their partial waves up to almost 1500 MeV. Having good $\pi^+\pi^-$ and $\pi^0\pi^0$ angular distributions, even of limited range, now allows the relative proportions of helicity zero to helicity two isoscalar $D$-wave through the $f_2(1270)$ region to be narrowed down, without the need to appeal to simple quark model assumptions.

\noindent The Belle data limit the range of possible solutions significantly compared to previous Amplitude Analyses.  Being a coupled channel analysis, it also relates the $\pi\pi$ information to one of the main inelastic channels, namely ${\overline K}K$, in the same energy range. However, while $\gamma\gamma\to\pi\pi$  involves even isospins, the inclusion of the ${\overline K}K$ channel brings in isospin 1 too. Nevertheless, the older, rather sparse, experimental measurements~\cite{ARGUS89}-\cite{TASSO86} of the $K^+K^-$ and ${\overline K}^0K^0$ channels limit our amplitudes to a small patch of solutions. When combined with the new high statistics results from Belle~\cite{Belle-KsKs} on $\gamma\gamma\to K_sK_s$, this space narrows to essentially a single solution (we call Solution I). How this solution fits all the available data on integrated and differential cross-sections is presented in Sect.~3.  This amplitude contains poles in the complex energy plane for the $\sigma$/$f_0(500)$, $f_0(980)$, $f_0(1370)$ and $f_2(1270)$ resonances. The residues of these poles fix the two photon couplings of each of these states. These are tabulated in Sect.~4 and are a main result of our study.

\noindent We then discuss the interpretation of these results for the composition of these key hadrons in Sect.~5. It is here too that we discuss the relationship of the present study to the work of others~\cite{Achasov07}-\cite{Igor12}.

\noindent However, it is important to bear in mind the key distinction between the work reported here and that discussed in Sect.~5 is that this is an Amplitude Analysis. It does not attempt to predict the data in terms of imperfect knowledge of  direct and crossed-channel dynamics but rather determines the $s$-channel amplitudes from a simultaneous analysis of all the available data. The interpretation of these model-independent amplitudes in terms of specific crossed-channel dynamics is the subject of a separate paper.

\noindent Having partial wave amplitudes that automatically cover the full angular range are a key input into future dispersive analyses of hadronic light-by-light scattering that appear in contributions to the anomalous magnetic moment of the muon. Having precision information on the real photon amplitudes is a crucial step towards reducing the present uncertainties in such calculations by a factor of four, demanded by future experiments.

\noindent This paper is organized as follows. In Sect.~2 we focus on the formalism and the determination of the hadronic $T$-matrix elements.
In Sect.~3 we give the overall fit to both $\gamma\gamma\to\pi\pi$ and $\gamma\gamma\to K\overline{K}$ data, including integrated cross-section and angular distributions. In Sect.~4 we extract the $\gamma\gamma$ couplings. In Sect.~5 we compare the resulting radiative decay widths to that from different models, as well as discussing related analyses.
Finally we give our conclusions in Sect.~6.

\newpage
\baselineskip=5.5mm

\section{Formal definitions of amplitudes}\label{sec:1}
\subsection{$\gamma\gamma\to\pi\pi$ amplitudes}\label{sec:1;1}
\noindent We begin with the unpolarized cross-section in the two photon center of mass frame, which is related to the two helicity amplitudes $M_{+\pm}$, by:
\be
\frac{d \sigma}{d \Omega}=\frac{\rho(s)}{128\pi^2 s}\,\left[|M_{+-}|^2\,+\,|M_{++}|^2\right] \,\, , \label{eq:cs}
\ee
where for $\gamma\gamma\to MM$ (with $M$~=~meson) the phase-space factor
\be
\rho(s)\,=\,\sqrt{1\,-\,4m_M^2/s}\quad .
\label{eq:phasespace}
\ee
These amplitudes have partial wave expansions with only even $J$
\bea
M_{++}(s, \theta, \phi)&=&e^2 \sqrt{16\pi}\, \sum_{J\geq0}\,F_{J0}(s)\,Y_{J0}(\theta,\phi) \,\, , \nonumber\\
M_{+-}(s, \theta, \phi)&=&e^2 \sqrt{16\pi}\, \sum_{J\geq2}\,F_{J2}(s)\,Y_{J2}(\theta,\phi) \,\, . \label{eq:M}
\eea
Integrating over the full angular range the cross-section for individual partial waves with specific isospin $I$, spin $J$ and helicity $\lambda$ is given by
\be
\sigma^I_{J\lambda}(s)\;=\;\frac{2\pi \alpha^2}{s}\, \rho(s)\,|F^I_{J\lambda}(s)|^2 \,\, , \label{eq:cs;iso}
\ee
where $\alpha$ is the usual final structure constant $e^2/(4\pi)$ in units in which $\hbar$ and $c$ are 1.

\noindent If one had experimental data that covered the complete angular range, one would take moments of the differential cross-sections and then fit partial waves to these. With full angular coverage the moments are independent of each other, the spherical harmonics $Y_{J\lambda}$ being orthogonal. However, the two photon process is here determined in  $e^+e^-$ collisions in an environment in which the electron and positron are scattered at small angles in the center of mass frame: small angles because that is when radiating a virtual photon is closest to being massless and so has its highest probability. The small scattering angle means that not only are the scattered electron and positron  undetected, but neither are forward or backward going mesons. Consequently, in the two photon center of mass frame, the determination of the cross-section is only possible for $|\cos \theta| < 0.6$ for charged pions, and for neutral pions over a larger region out to $|\cos \theta|\, =\,0.8$. This means that the moments of the measured angular distribution are not independent, and interferences between partial waves are not readily separable. Thus the observed integrated cross-section is not just the sum of the squared moduli of partial wave amplitudes, but their interferences are just as critical in determining their magnitude and energy dependence.
\baselineskip=5.5mm
\subsection{Isospin decomposition}\label{sec:1;2}
The produced pions are in a combination of isospin 0 and 2, and the kaons in $I\,=\,$0 and~1.
The isospin decomposition of $\,\gamma\gamma\to\pi^+\pi^-, \pi^0\pi^0\,$ and $\,\gamma\gamma\to K^+K^-, K^0\overline{K}^0\,$ amplitudes is:
\bea
\mathcal{F}_\pi^{+-}(s)&=&-\sqrt{\frac{2}{3}}\,\mathcal{F}_\pi^{I=0}(s)\,-\,\sqrt{\frac{1}{3}}\,\mathcal{F}_\pi^{I=2}(s) \,\, , \no\\
\mathcal{F}_\pi^{00}\,(s)&=&-\sqrt{\frac{1}{3}}\,\mathcal{F}_\pi^{I=0}(s)\,+\,\sqrt{\frac{2}{3}}\,\mathcal{F}_\pi^{I=2}(s) \,\, ,\label{eq:F;amppi} \\[5mm]
\mathcal{F}_K^{+-}(s)&=&-\sqrt{\frac{1}{2}}\,\mathcal{F}_K^{I=0}(s)\,-\,\sqrt{\frac{1}{2}}\,\mathcal{F}_K^{I=1}(s) \,\, , \no\\
\mathcal{F}_K^{00}\,(s)&=&-\sqrt{\frac{1}{2}}\,\mathcal{F}_K^{I=0}(s)\,+\,\sqrt{\frac{1}{2}}\,\mathcal{F}_K^{I=1}(s) \,\, .\label{eq:F;ampK}
\eea
(Note our charged pion amplitude here is the negative of those presented in Refs.\cite{MRP91,MRP08}, and the normalization factor for the $\mathcal{F}_\pi^I(s)$ arising from the property of identical particles has been absorbed into the coefficients in Eq.~(\ref{eq:F;amppi})).
We will concentrate first on the amplitudes with $\pi\pi$ final states. The extension to ${\overline K}K$ will then be straightforward.

\noindent At low energies, di-pion production is dominated by the one pion exchange Born amplitude as a consequence of Low's low energy theorem. This means that, at least at low energies close to threshold, the $\,\gamma\gamma \to \pi^0\pi^0\,$ cross-section is very much smaller than that for $\pi^+\pi^-$. Eq.~(\ref{eq:F;amppi}) means that the $I=2$ $\,\gamma\gamma\to\pi\pi\,$ amplitude is of comparable size to that with $I=0$ (aside from the factor of $\sqrt{2}$). This is unusual for a hadronic amplitude. The expected weakness of \lq\lq exotic'' channels is reflected in the fact that final state interactions only slowly change the two photon amplitude from its Born contribution for $I=2$, while in the isoscalar channel these differ appreciably within a few hundred MeV of threshold. The isospin 0 and 2 amplitudes interfere in the individual charged and neutral cross-sections and this interference helps to untangle these. Indeed,
to be able to separate amplitudes into their isospin components requires that comparable data on both neutral and charged meson pairs are available.
The Belle two photon experiment provides access to $\pi^+\pi^-$ and $\pi^0\pi^0$ channels in overlapping regions of $\cos \theta$.

\begin{figure}[h]
\centering
\includegraphics[width=0.8\textwidth,height=0.45\textheight]{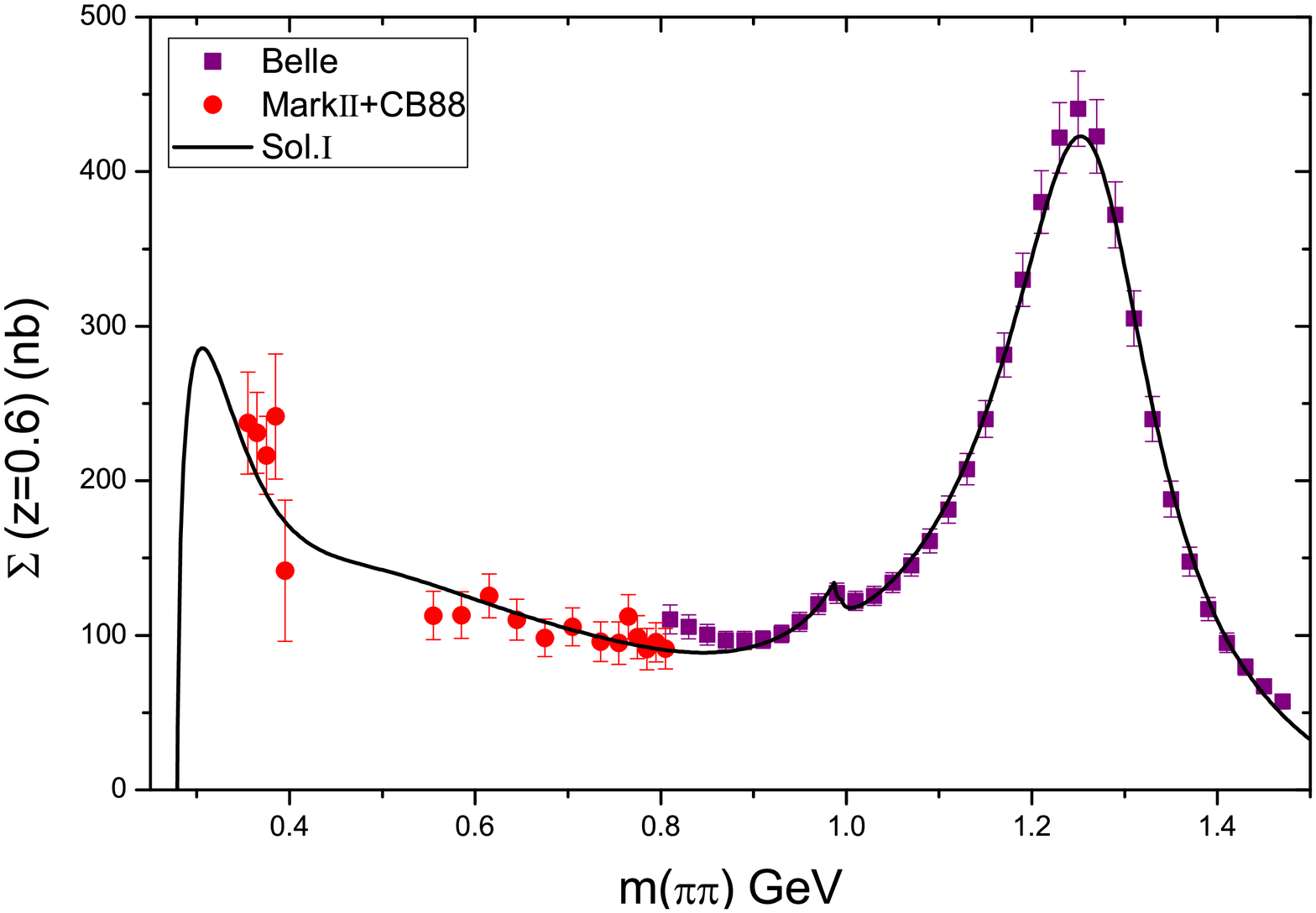}
\caption{\label{fig:Sigmacs}
The sum of the $\pi^+\pi^-$ and $\pi^0\pi^0$ integrated cross-sections, $\Sigma (0.6)$, Eq.~(\ref{eq:cssum}), with $Z\,=\,0.6$ from the Belle results of \cite{Belle-pm,Belle-nn}. At present the line is to guide the eye: it is our solution~I.}
\end{figure}

\noindent While interferences are critical to the amplitude analysis, these are not easy to convey in words or pictures.
However summing the cross-sections for charged and neutral pions integrated over the same region of $\cos \theta \equiv z$ up to $z\,=\,Z$,
which we can do for the Belle data, removes the $I=0$, $I=2$ interference. Thus
\bea
\Sigma(Z)&\equiv &\int_0^{Z}\,dz\;\left[\frac{d\sigma}{dz}\,(\gamma\gamma\to\pi^+\pi^-)\;+\;\frac{d\sigma}{dz}\,(\gamma\gamma\to\pi^0\pi^0)\right]\no\\[3mm]
&=&
\frac{2\pi\alpha^2}{s}\, \rho(s)\; \sum_{J,J',\lambda}\,\left[\mathcal{I}_{JJ'}^\lambda(Z)\,\left({\mathcal F^*}^0_{J\lambda}\,{\mathcal F}^0_{J'\lambda}\,+\,{\mathcal F^*}^2_{J\lambda}\,{\mathcal F}^2_{J'\lambda}\right)\right]\,,\label{eq:cssum}
\eea
where
\be
{\mathcal I}_{JJ'}^\lambda(Z)\;=\;\int_0^{Z}\;dz\, P_J^\lambda(z)\,P_{J'}^{\lambda'}(z)\quad .
\ee
In Fig.~\ref{fig:Sigmacs} we show $\Sigma(Z=0.6)$ from the Belle data~\cite{Belle-pm,Belle-nn} as a function of dipion mass $m(\pi\pi)=\sqrt{s}$.

\noindent Since the $I=2$ amplitude has not only no known direct channel resonances, but all indications are that it is smooth, it is natural to associate any structures in $\Sigma(Z)$ with dynamics in the $I=0$ $\pi\pi$ channel. Beyond the near threshold enhancement from the Born component,
the data in Fig.~\ref{fig:Sigmacs} show two clear peaks. The largest around 1250 MeV is associated with the spin-2 $f_2(1270)$ resonance. Two photon collisions favor the production of tensor mesons, and the $f_2(1270)$, having $\,\pi\pi\,$ as its dominant decay mode, appears very strongly. However, one sees that the position of the peak is shifted and the width larger for this enhancement, than the nominal PDG values~\cite{PDG12}. This is because with $Z <  1$ in Eq.~(\ref{eq:cssum}) there are important $S-D_0$ interferences within the $I=0$ channel, that we will discuss later. The second much smaller peak is seen just below 1 GeV. This is associated with the appearance of the $f_0(980)$. The $f_0(980)$ is an example of a particular type of resonance that is strongly coupled to a nearby opening channel. Many similar kinds of states are now being discovered in channels dominated by hidden charm and beauty~\cite{X3872,Y4260,eidelman}. As a consequence of its proximity to the ${\overline K}K$ threshold, the $f_0(980)$ appears as a dip in some processes and a peak in others. A peak means it couples through its hidden strange component, ${\overline s}s$ or ${\overline K}K$, as in $\,J/\psi\to \phi\pi\pi$, while a dip means its coupling is entirely through its ${\overline n}n$ or $\pi\pi$ component (where $n$ refers to the appropriate sum of $u,\,d$). Here in $\gamma\gamma$ we have a small peak implying a combination of both these non-strange and hidden strange components. These will be discussed later after we have performed our Amplitude Analysis.

\subsection{$S$-matrix and QED constraints}\label{sec:1;3}
Now to solve the problem of how to determine the partial wave amplitudes, we input in turn the three key properties of the $S$-matrix: unitarity, analyticity, crossing symmetry, together with the low energy theorem of QED.
In the energy region we study, below 1.4 or 1.5 GeV, there are a limited number of accessible $I=0, 2$ hadronic channels: $\pi\pi$, ${\overline K}K$,$\ldots$, before the $4\pi$ channels become important. In this region we presume that unitarity is saturated by the $\pi\pi$ and ${\overline K}K$ channels alone.
If we denote the hadronic scattering amplitudes by $T$, it is straightforward to show that coupled channel unitarity is fulfilled for the two photon reaction for each partial wave by
~\cite{AMP-FSI}:
\bea \label{eq:unitarity}
\mathcal{F}^I_{J\lambda}(\gamma\gamma\to\pi\pi;s)\;&=&\,{\alpha_1}^I_{J\lambda}(s)\,\hat{T}^I_J(\pi\pi\to\pi\pi;s)\;\;\, +\; {\alpha_2}^I_{J\lambda}(s)\,\hat{T}^I_J(\pi\pi\to{\overline{K}K};s)\; ,\no\\[3mm]
\mathcal{F}^I_{J\lambda}(\gamma\gamma\to{\overline K}K;s)&=&{\alpha_1}^I_{J\lambda}(s)\,\hat{T}^I_J(\pi\pi\to{\overline K}K;s)\;+\; {\alpha_2}^I_{J\lambda}(s)\,\hat{T}^I_J({\overline K}K\to{\overline{K}K};s)\; .
\eea
with the coupling functions $\alpha_i(s)$ real. This representation automatically embodies the final state interactions of the $\pi\pi$ and ${\overline K}K$ systems.
The hadronic amplitudes $\hat{T}$ represent what we call {\it reduced} amplitudes. As far as coupled channel unitarity is concerned, they are the same as the $T$-matrix amplitudes. However, to avoid right hand cut singularities in the functions $\alpha_i$, any real zeros in the $T$-matrix elements~\footnote{or their determinant. The determinant of the $T$-matrix elements can only have a zero below the inelastic threshold when the $\pi\pi\to {\overline K}K$ must have the elastic phase. {\it A priori} we do not know whether such a zero occurs. However, after making the new fits in Sect.~\ref{sec:2;1}, we check that no such zero exists, and so this complication can be ignored.} have to be removed. For the $S$-wave amplitudes, sub-threshold zeros are imposed by the Adler condition of chiral dynamics. We need to ensure that these zeros do not artificially transmit from one reaction to another, and allow their existence and position to be process-dependent. Thus with channel-labels defined by $1=\pi\pi$, $2={\overline K}K$ channels, we have for the hadronic process $i\to j$
\be
\label{eq:adler}
\hat{T}_{ij}(s)\;=\; T_{ij}(s)/(s-s_{0(ij)})
\ee
where $s\,=\,s_{0(ij)}$ is the position of the Adler zero in the hadronic channel $i \to j$.
For amplitudes with higher angular momentum, the hadronic amplitudes have a zero at threshold, and so reduced amplitudes are defined by dividing by factors of $(s- 4 m_M^2)^J$.
The $\gamma\gamma\to MM$ amplitudes, $\mathcal{F}$, and hadronic $MM\to M'M'$ amplitudes, $T$, each have right and left hand cuts. Unitarity requires their right hand cut structures to be the same. That is what Eq.~(\ref{eq:unitarity}) embodies with the coupling functions, $\alpha_i(s)$, real. However, their left hand cuts differ and the $\alpha_i(s)$ themselves have left hand cuts. These are generated by crossed channel exchanges, as we shall discuss. Along the right hand cut where the $\pi\pi$ and ${\overline K}K$ channels saturate unitarity, the functions ${\alpha_i}^I_{J\lambda}$ as defined by Eq.~(\ref{eq:unitarity}), are real. Thus Eq.~(\ref{eq:unitarity}) means that
the behavior of the hadronic partial wave amplitudes up to 1.5 GeV constrains the $\gamma\gamma$ amplitudes, even more so below the ${\overline K}K$ thresholds, when Eq.~(\ref{eq:unitarity}) imposes Watson's final state interaction theorem.
There in the elastic region, the hadronic and two photon amplitudes have the same phase for each and every value of $I,\,J$.
To implement unitarity up to 1.5 GeV  we therefore need to know the hadronic amplitudes for $\pi\pi\to\pi\pi$ and ${\overline K}K$ for each $I, J$. In the next subsection we will discuss how these are represented.

\noindent In principle, these inputs would be sufficient to describe the $\gamma\gamma\to\pi\pi$ data in the whole energy region where the hadronic channels we include saturate unitarity. However, it is really only above 800 MeV that we have data on the two photon reaction of any precision, which comes from the more recent Belle experiment.  While the Crystal Ball data~\cite{CB88,CB92} gives $\pi^0\pi^0$ results right down to threshold, the corresponding $\pi^+\pi^-$ results from CELLO~\cite{Cello1,Cello2} start at 800 MeV. Only the Mark II~\cite{MarkII} experiment from 25 years ago has charged differential cross-sections down to 600 MeV. With a special run Mark II also determined the charged cross-section at 5 energies between 300 and 400 MeV, but with 30\% error bars. Consequently, the low energy partial waves are poorly determined from the data alone. Fortunately, QED imposes a low energy theorem on the Compton scattering amplitude at threshold, which in turn fixes the $\gamma\gamma$ partial waves at $s=0$.  As one enters the $s$-channel physical region at $s = 4m_\pi^2$, these partial waves are modified by final state interactions. Knowledge of hadronic scattering in fact determines this modification rather precisely. Consequently, the partial waves in the low energy region are calculable from the hadronic scattering amplitudes. The tool for this calculation is a marriage of unitarity and analyticity. This we now discuss. It is important in following this analysis that while we use the representation of Eq.~(\ref{eq:unitarity}) for the $\gamma\gamma$ partial waves to fit the available data in the entire energy region from $\pi\pi$ threshold to 1.5 GeV, it is only in the very low energy region that we constrain these fits by calculations we now discuss. However, it is the same hadronic inputs that go into both. These will be  detailed in the next section.

\noindent Analyticity is imposed through the use of dispersion relations. The two photon amplitudes, $\mathcal{F}^I_{J\lambda}$, have both a right and a left hand cut. The left hand cut is generated by crossed-channel dynamics, and there have been several phenomenological efforts to model this with specific exchanges, as we will discuss later in Sect.~\ref{sec:4;1}. Here our intent is more general. We separate the left hand cut contribution to $\mathcal F$ into two parts: the known one pion exchange Born term ${\mathcal B}$ and the rest, which we denote by ${\mathcal L}$. Then for $ s\;<\;0$
\be
{\rm Im}\,\,{\mathcal F}^I_{J\lambda}\;=\;{\rm Im}\,{\mathcal B}^I_{J\lambda}(s)\;+\;{\rm Im}\,{\mathcal L}^I_{J\lambda}(s)
\ee
where the function $\mathcal B$ has a left hand cut starting at $s\,=\,0$, while the discontinutity in $\mathcal L$ starts at $s\;=\;s_L$, with $s_L\,\simeq\,-m_\rho^2$.
 The Born term is separated to ensure the low energy theorem for the Compton process $\gamma\pi\to\gamma\pi$ is satisfied. Thus with
\be
{\mathcal F}^I_{J\lambda}(s)\;\to\;{\mathcal B}^I_{J\lambda}(s)\quad{\rm as}\;s \to 0
\ee
with corrections of ${\mathcal O}(s)$, as shown by Abarbanel and Goldberger~\cite{AG}.

\noindent To implement these properties, we proceed as follows.
With the phase of the $\gamma\gamma\to\pi\pi$ partial wave amplitude, ${\mathcal F}^I_{J\lambda}(s)$ given by $\varphi^I_{J\lambda}$(s), we define the Omn$\grave{e}$s function~\cite{MRPhandbook}:
\begin{equation}\label{eq:Omnes}
\Omega^I_{J\lambda}(s)=\exp\left(\frac{s}{\pi} \int^\infty_{s_{th}} ds' \frac{\varphi^I_{J\lambda}(s')}{s'(s'-s)}\right) \,\, .
\end{equation}
Below the inelastic threshold (here effectively around 990 MeV), the phase $\varphi^I_{J\lambda}(s)\;=\;\delta^I_J(s)$, the $\pi\pi$ elastic phase-shift, as required by Watson's theorem.
This Omn$\grave{e}$s function contains (by construction) the right hand cut of ${\mathcal F}^I_{J\lambda}$. Using this we form the function $P^I_{J\lambda}(s)\,=\, {\mathcal F}^I_{J\lambda}(s)/\Omega^I_{J\lambda}(s)$ which only has a left hand cut. Then we can write a dispersion relation for $({\mathcal F}(s)\,-\,{\mathcal B}(s))\Omega^{-1}(s)$. For the $S$-wave amplitudes, this has two subtractions at $s=0$:
\bea\label{eq:F;ampS}
\mathcal{F}^{I}_{00}(s)\;=\;{\mathcal B}^I_{00}(s)+b^{I} s~\Omega^{I}_{00}(s)
 &+&\frac{s^2~\Omega^{I}_{00}(s)}{\pi}\int_L ds'\frac{{\rm Im}\left[ \mathcal{L}^{I}_{00}(s')\right]\Omega^{I}_{00}(s')^{-1} }{s'^2(s'-s)} \no \\[2.5mm]
              &-&\frac{s^2\;\Omega^{I}_{00}(s)}{\pi}\int_R ds'\frac{{\mathcal B}^I_{00}(s')\;{\rm Im}\left[ \Omega^{I}_{00}(s')^{-1}\right] }{s'^2(s'-s)}\, .
\eea
where the $\,b^I\,$ (with $I=0, 2$) are subtraction constants to be constrained below. For
$J\, >\, 0$ it is useful to take advantage of the known threshold behavior of the $\gamma\gamma$ partial wave amplitudes and their approach to the Born term as $s\,\to\, 0$ (which we detail later), and so write an unsubtracted dispersion relation for $({\mathcal F}(s)\,-\,{\mathcal B}(s))\Omega^{-1}(s)/s^n(s\,-\,4m_\pi^2)^{J/2}$ with $n\,=\,2\,-\,\lambda/2$. Then
\bea\label{eq:F;ampJ}
\mathcal{F}^I_{J\lambda}(s)\;=\;{\mathcal B}^I_{J\lambda}(s)
&+&\frac{s^n(s-4m_\pi^2)^{J/2}}{\pi}\,\Omega^{I}_{J\lambda}(s)\,\int_L ds'\frac{{\rm Im}\left[ \mathcal{L}^{I}_{J\lambda}(s')\right]\, \Omega^{I}_{J\lambda}(s')^{-1} }{s'^n(s'-4m_\pi^2)^{J/2}(s'-s)}\nonumber\\[3.5mm]
&-&\frac{s^n(s-4m_\pi^2)^{J/2}}{\pi}\,\Omega^{I}_{J\lambda}(s)\,\int_R ds'\frac{B^{I}_{J\lambda}(s')\,  {\rm Im}\left[ \Omega^{I}_{J\lambda}(s')^{-1}\right] }{s'^n(s'-4m_\pi^2)^{J/2}(s'-s)} \,\, .
\eea
These analytic representations, which we use  for $s > 4m_\pi^2$, automatically fulfill the low energy theorem of Low. The contribution from the nearby part of the left hand cut is contained in the one pion exchange Born term, and dominates the low energy integrals. As the energy, $\sqrt{s}$, increases above 400~MeV the more distant left hand cut contributions included in the function $\mathcal L$ start to become important and increasingly so. Whilst the contribution generated by $\rho$ and $\omega$ exchange can be reliably computed, heavier single particle and multi-particle contributions are more problematic. A number of studies have included $a_1$ and $b_1$ contributions~\cite{yumao09}, and more recently those from $f_2$ and $a_2$ exchanges have been added~\cite{Moussallam10}, in an attempt to make predictions for the $\gamma\gamma$ reaction up to 1.4 GeV. Here our aim is different. It is to determine from experiment what the two photon amplitudes are, not to predict them. The amplitude continued along the left hand cut is an output that we will discuss in a separate paper. Here the constraints of unitarity, analyticity and crossing symmetry are imposed as a general framework, within which amplitudes describing the experimental data are to be constructed.

\baselineskip=5.4mm
\subsection{Hadronic Inputs}\label{sec:1;4}
We now discuss the inputs to this framework for each partial wave amplitude.
We start  with the coupled channel $\pi\pi$ and ${\overline K}K$ hadronic amplitudes with $I=0, 2$ and $J=0, 2$, and treat the higher waves later. It is worth recapping that these inputs are used in two distinct ways. Up to 1.5~GeV, they are the key ingredients in ensuring that our fitted amplitudes to $\gamma\gamma\to\pi\pi$, and $\to {\overline K}K$, satisfy coupled channel unitarity through Eq.~(\ref{eq:unitarity}). The second use is as inputs in the dispersive treatment of the low energy $\gamma\gamma$ amplitudes, Eqs.~(\ref{eq:F;ampS},\ref{eq:F;ampJ}). Though this is only used at two photon energies below 600~MeV, being dispersive integrals they require inputs up to high energy. Thus above 1.5~GeV, it is only the hadronic inputs \lq\lq on the average'' that matter and not the fine details. Indeed, as the dispersive integrals all converge sufficiently fast, different behavior above 1~GeV or so, is only considered to gain an idea of the uncertainties in the dispersive calculations.
\begin{figure}[htbp]
\centering
\includegraphics[width=1.0\textwidth,height=0.65\textheight]{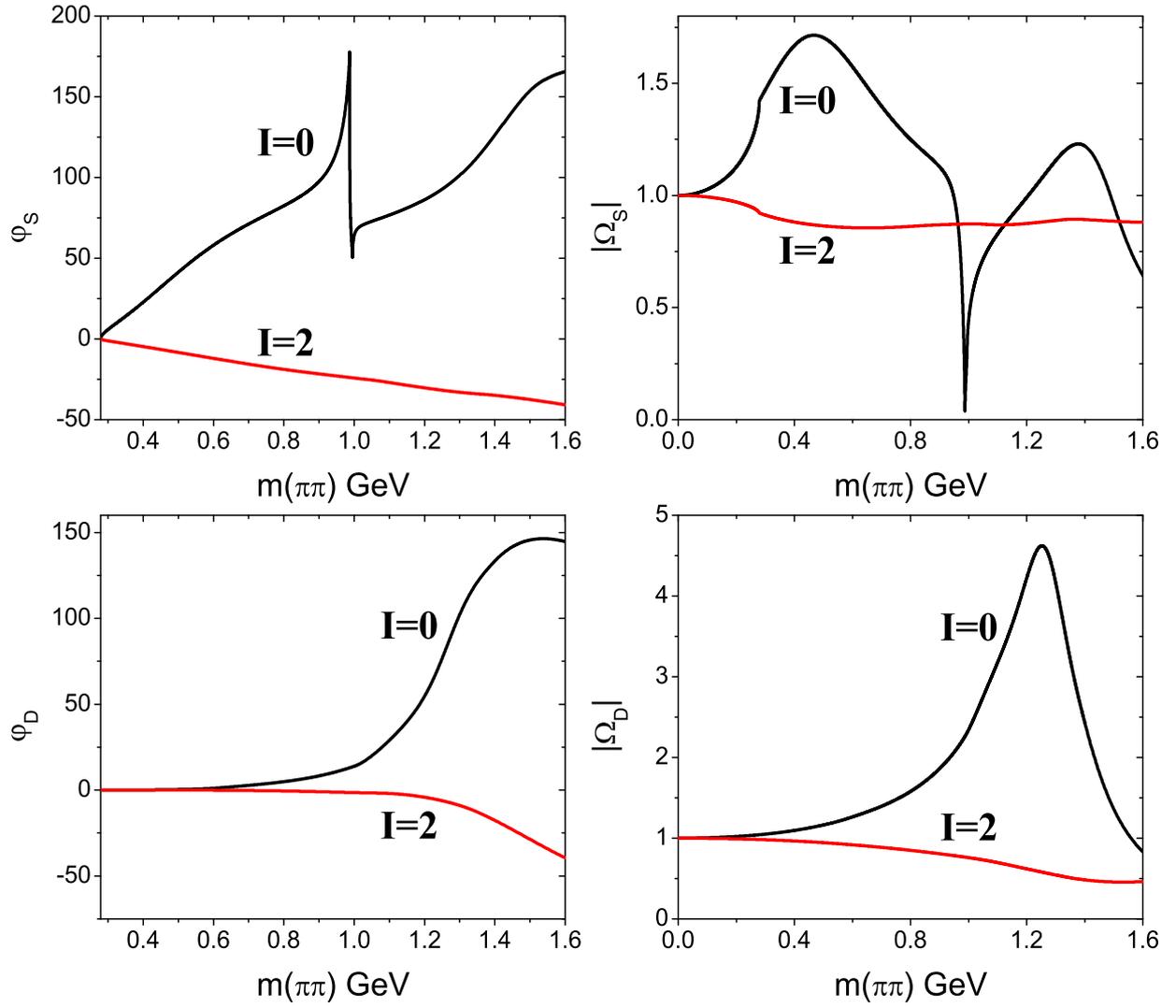}
\caption{\label{fig:Omnes} The phases and moduli of the Omn$\grave{e}$s functions of the process $\gamma\gamma\to\pi\pi$. Both $S$ and $D$-waves with
 $I=0$ and $I=2$  are presented.}
\end{figure}
\vspace{4mm}

\noindent $\mathbf{IJ=00\;channel}$\\[4mm]
As mentioned in the Introduction, we take advantage of the work of the
Madrid-Krakow collaboration~\cite{KPY} on updating knowledge of the $\pi\pi$  $T$-matrix element, for which we use their {\it Constrained Fit to Data IV (CFDIV)} parametrization.  In this reference, information  is expressed in terms of the phase-shift, $\delta^I_J$, and the inelasticity, $\eta^I_J$, from which we can define the $T$-matrix element in  the usual way:
\begin{equation}\label{eq:T}
T^I_J\;=\;\frac{1}{2i\,\rho_\pi(s)}\left[\eta^I_J\, e^{2 i \delta^I_J}\,-\,1\right]\;=\;|T^I_J|\, e^{i \varphi^I_J} \,\, .
\end{equation}
See ~\cite{KPY}, for the detailed expressions and for their range of applicability.
In the elastic region, when $\eta\,=\,1$, then of course we have $\,\varphi^I_J\,=\,\delta^I_J$\,.
In this $I=J=0$ channel we use a coupled channel $K$-matrix parametrization, imposing the $\pi\pi$ scattering amplitudes of Pel\'aez {\it et al.}~\cite{KPY} as a dataset,
up to $\sqrt{s}$~=~1.42~GeV . See Sect.~\ref{sec:2} for details. We also fold in the earlier dispersive results of Buttiker {\it et al.}~\cite{Descotes04} in the near threshold region of the $\pi\pi\to {\overline K}K$ channel.  Above that energy,  we use a Regge form for the full $\gamma\gamma\to\pi\pi$ amplitude
\bea\label{eq:Regg}
V_V(s,t)&=&\left(1-\exp[-i\pi\alpha_V(t)]\right)\,\Gamma[1-\alpha_V(t)]\,(\alpha's)^{\alpha_V(t)} \,\, , \no\\
V_P(s,t)&=&\left(1+\exp[-i\pi\alpha_P(t)]\right)\,\Gamma[\quad -\alpha_P(t)]\,(\alpha's)^{\alpha_P(t)} \,\,  , \no\\
\mathcal{F}^0(s,t)&=&V_P(s,t)+g_0\,h\,V_V(s,t) \,\, ,   \no\\
\mathcal{F}^2(s,t)&=&V_P(s,t)+g_2\,h\,V_V(s,t) \,\, ,
\eea
where $\,\alpha_V(t)=0.45+\alpha't$, $\alpha_P(t)=-0.017+\alpha't$ with $\alpha'=0.88$ GeV$^{-2}$, $h=3.71$, $g_0=-0.843$, $g_2=1.414$.
With this we predict the phase in the high energy region for the two photon reaction from 2~GeV to 5~GeV for each partial wave (not just $I=J=0$).
Making a smooth connection gives the phase and Omn$\grave{e}$s function in Fig.~\ref{fig:Omnes}.
It is important to emphasize that these phases are only imposed in the low energy region of our Amplitude Analysis. Phases up to 5~GeV are required only to limit the uncertainties in the dispersive calculations, using Eq.~(\ref{eq:F;ampS}).

\noindent In the low energy region below 600 MeV, we use the twice subtracted dispersion relation, Eq.~(\ref{eq:F;ampS}).
To keep the notation simple we here drop the indices $J= 0$, $\lambda = 0$. The left hand cut contribution is dominated by the one pion exchange Born amplitude, that we have included explicitly. Eq.~(\ref{eq:F;ampS}) encodes how the Born amplitude is modified by final state interactions in both the $I = 0,\,2$ channels.  To have an idea of how accurate this constraint on the near threshold two photon amplitude is, we estimate the effect of other particle exchanges in the $t$ and $u$-channels that contribute to $\mathcal{L}$. We approximate the left hand cut by individual exchanges $R =\, \rho,\,\omega,\, a_1,\, b_1$ and $T$ an effective exchange that \lq\lq sums'' the rest. With the mass squared of an exchange given by $s_R$, each generates a contribution to the discontinuity across the left hand cut for $s\, <\,-(s_R\,-\,m_\pi^2)^2/s_R\,\simeq -s_R$. Thus if their couplings to $\gamma\pi$ are of the same order of magnitude then their individual contributions to the $\gamma\gamma\to\pi\pi$ amplitudes only start to become important in the physical region when $s\,\ge s_R$. This  implies the Born term is dominant until the energy is a few hundred MeV above threshold. Including only simple exchange contributions to $\mathcal{L}\,=\,\mathcal{L}_R$, we have
\bea\label{eq:LR00}
\mathcal{L}^{I=0}_{R}(s)=-\sqrt{\frac{3}{2}}\mathcal{L}_{\rho}(s)-\sqrt{\frac{1}{6}}\mathcal{L}_{\omega}(s)-\sqrt{\frac{3}{2}}\mathcal{L}_{b_1}(s)
-\sqrt{\frac{1}{6}}\mathcal{L}_{h_1}(s)-\sqrt{\frac{2}{3}}\mathcal{L}_{a_1}(s)+\mathcal{L}_{T}(s) \,.
\eea
Here $\mathcal{L}_{T}(s)$ is the contribution of an \lq effective exchange' parametrized as in Appendix~\ref{app:A}.
This representation for $\mathcal{L}$ is just an approximation to allow us to assess the uncertainties in the calculated $\gamma\gamma$ amplitudes in the low energy region. In reality, there are, of course, contributions to the left hand cut from multi-meson exchange, pions and kaons, the calculation of which is more complicated. Nevertheless, this simple modeling will allow us to judge the range of the uncertainties in predicting the low energy $\gamma\gamma\to\pi\pi$ amplitudes that will anchor our partial wave analysis.
The subtraction constants $b_0,\, b_2$ can be fixed by the approach to
Low's theorem: $\mathcal{F}_\pi^{+-}(s)\rightarrow B(s)+\mathcal{O}(s^2)$,
and $\mathcal{F}_\pi^{00}(s)\,=\,0$ when $s\rightarrow 0$. The neutral amplitude has a chiral zero nearby, as
$\mathcal{F}_\pi^{00}(s_n)=0$ at $s=s_n=\mathcal{O}(m_\pi^2)$ ~\cite{MRP91}. Indeed, at lowest order in Chiral Perturbation Theory, the $\gamma\gamma\to\pi^0\pi^0$ cross-section is proportional to that for $\pi^+\pi^-\to\pi^0\pi^0$, as noted in ~\cite{Adler0-1loop}. The amplitude for the hadronic reaction having a zero at $s\,=\,m_{\pi}^{\,2}$ at tree level. Higher order corrections destroy this simple proportionality and shift the chiral zero in the the $\gamma\gamma$ reaction a little, see Eq.~(\ref{eq:zero}).
We then have
\bea\label{eq:F;bI0I2}
b^{I=0}&=&\sqrt{3}~\Delta/(\Omega^{I=0}(s_n)+2\Omega^{I=2}(s_n)) \,\, , \no\\
b^{I=2}&=&-\sqrt{2}~b^{I=0} \,\, ,
\eea
where
\bea\label{eq:F;Delta}
\Delta&=&-\sqrt{\frac{1}{3}}\frac{s_n\,\Omega^{I=0}(s_n)}{\pi}\left(
                                 \int_R ds'\frac{\sqrt{\frac{2}{3}}\, B(s')  {\rm Im}\left[\Omega^{I=0}(s')^{-1} \right] }{s'^2(s'-s)}
                                 +\int_L ds'\frac{{\rm Im}\left[\mathcal{L}^{I=0}_{R}(s')\right] \Omega^{I=0}(s')^{-1} }{s'^2(s'-s)}
                                  \right)   \nonumber\\
&&+   \sqrt{\frac{2}{3}}\frac{s_n\,\Omega^{I=2}(s_n)}{\pi}\left(
                                 \int_R ds'\frac{\sqrt{\frac{1}{3}} B(s')  {\rm Im}\left[ \Omega^{I=2}(s')^{-1}\right] }{s'^2(s'-s)}
                                +\int_L ds'\frac{{\rm Im}\left[\mathcal{L}^{I=2}_{R}(s')\right] \Omega^{I=2}(s')^{-1} }{s'^2(s'-s)}
                                  \right) \,\,\no
\eea
To determine the location of the Adler zero (and its uncertainty), we consider the values given by Chiral
Perturbation Theory at one loop~\cite{Adler0-1loop}~( $1.019m_{\pi^0}^{\,2}$ ), two loop~\cite{Adler0-2loop} ~( $1.175m_{\pi^0}^{\,2}$ ) and in the Muskhelishvili-Omn$\grave{e}$s analysis~\cite{Moussallam10}. These are all encompassed by taking:
\be\label{eq:zero}
s_n\,=\,(1\pm 0.2)\,m_{\pi^0}^{\,2}\,\quad,
\ee
which is within the range of ~\cite{MRP91}.


\vspace{4mm}
\noindent$\mathbf{IJ=02\;channel}$\\[4mm]
This amplitude for $\,\pi\pi\to\pi\pi\,$ and $\to {\overline K}K$ is dominated by the $f_2(1270)$ resonance. These hadronic amplitudes are suppressed below 1~GeV, by the $D$-wave angular momentum required to excite a tensor resonance. In contrast, in the $\gamma\gamma$ reaction, $J\,=\,2$ can be reached even in the $S$-wave in the helicity two channel. Consequently, even the spin two component of the pion exchange Born term is important from the lowest $\gamma\gamma$ energies.

\noindent Below 600 MeV we constrain the $D$-waves as we did the $S$-waves, but instead use the dispersion relation of Eq.~(\ref{eq:F;ampJ}). The Omn$\grave{e}$s function is computed using the following inputs.
For the $f_2(1270)$, the $\pi\pi$ channel contributes $\sim 84\%$, $K\overline{K}$ almost 5\% and the $4\pi$ channel 10\%, according to the PDG 2012 Tables~\cite{PDG12}. Since the shape of the $f_2$ is dominated by the $\pi\pi$ channel, the amplitude $T_{11}$ with  $I=0, \, J=2$ controls the pole position. To allow for the different shape in the two photon channels, the hadronic amplitude is modified by the appropriate coupling functions $\alpha(s)$. Thus we have for each helicity (dropping the $I=0,\,J=2$ labels):
\bea \label{eq:F02}
\mathcal{F}_1(s)&=&\alpha_1(s)\,\hat{T}_{11}\,+\,\alpha_2(s)\,\hat{T}_{21}\,+\,\alpha_3(s)\,\hat{T}_{31} \,\, ,
\no\\
                &\simeq&{\overline{\alpha}}_1(s)\,\hat{T}_{11} \,\, ,
\eea
where the subscripts $1,\,2,\,3\,$ label the $\pi\pi$, ${\overline K}K$ and $4\pi$ channels.
While $\alpha_1(s)$ is real, $\overline{\alpha}_1(s)$ could be complex  for $ s\,>\,s_{\rm{th2,th3}}$,
where these are the thresholds of the ${\overline K}K$ and $4\pi$ channels, respectively.
In practice there is a lack of sensitivity to the separation of these thresholds,
and $\overline{\alpha}_1$ is essentially real up to $\sqrt{s}=1.5$~GeV.
This is also so for the $IJ=20,~02$ channels discussed below.
Then we analyze the $T$-matrix element recalculated from Eq.~(\ref{eq:T}). The  phase-shift and inelasticity are given by~\cite{KPY} up to 1.4~GeV. At higher energies we choose a smooth connection up to 5~GeV given by the Regge representation of Eq.~(\ref{eq:Regg}). We use this result as the input to Eq.~(\ref{eq:F02}) and determine the Omn$\grave{e}$s function and phase for $\pi\pi\to\pi\pi$ shown in Fig.~\ref{fig:Omnes}.
Using a single channel dispersion relation gives a constraint on $\gamma\gamma$ amplitude, shown below 600~MeV in Fig.~\ref{fig:Flow}.
Indeed, the amplitudes with $J = 2$ behave~\cite{MRP88,Moussallam10} as:
\be \label{eq:F;pol}
\mathcal{F}^I_{20}-B^I_{20}\;\sim\; s^2 (s-4 m_\pi^2),\quad
\mathcal{F}^I_{22}-B^I_{22}\;\sim \; s (s-4 m_\pi^2),
\ee
Such behavior is built into the dispersion relation, Eq.~(\ref{eq:F;ampJ}), we use.

\vspace{4mm}
\noindent$\mathbf{IJ=20\;channel}$\\[4mm]
In isospin 2, inelasticity is produced by the opening of the $4\pi$ channel.
In a way similar to the isospin zero case just discussed, we relate the $\gamma\gamma\to\pi\pi$ partial waves to the hadronic amplitudes by
\bea\label{eq:F20}
\mathcal{F}_1(s)&=&\alpha_1(s)\,\hat{T}_{11}\,+\,\alpha_3(s)\,\hat{T}_{21}\;\simeq\;{\overline{\alpha}}_1(s)\,\hat{T}_{11} \,\, ,
\eea
where 1, 3 represents for 2$\pi$ and 4$\pi$ channel. We set $s_{\rm{th3}}\,=\,(1.05\ {\rm GeV})^2$ below which ${\overline{\alpha}}_1$ must be real.
In practice it is real up to $\sqrt{s}=1.5$~GeV.
The parametrization for the $\pi\pi$ $S$-wave amplitude ($T_{11}$) is given by the paper~\cite{KPY} below 1.35~GeV, and  above  we use a smooth Argand plot connection.
Then  we obtain the Omn$\grave{e}$s function shown in Fig.~\ref{fig:Omnes}. To determine the low energy behavior of the corresponding $\gamma\gamma\to\pi\pi$ partial wave, we still need the left hand cut contributions. For this we use for the non-Born term a form analogous to that of Eq.~(\ref{eq:LR00})
\bea \label{eq:LR20}
\mathcal{L}^{I=2}_{R}(s)=\sqrt{\frac{1}{3}}\mathcal{L}_{\omega}(s)+\sqrt{\frac{1}{3}}\mathcal{L}_{h_1}(s)-\sqrt{\frac{1}{3}}\mathcal{L}_{a_1}(s)
+\mathcal{L}_{T}(s) \,\, ,
\eea
and the $\mathcal{L}_{T}(s)$ is given in Appendix~\ref{app:A}.
With this input we use the subtracted dispersion relation given in Eq.~(\ref{eq:F;ampS}), where the subtraction constant $b^{I=2}$ is fixed by the chiral constraint in Eq.~(\ref{eq:F;bI0I2}).

\vspace{4mm}
\noindent$\mathbf{IJ=22\;channel}$\\[4mm]
When the energy is below 1.35~GeV, we use the parametrization set out in~\cite{KPY}.
Above we use an Argand plot connection making both the $T$-matrix element and its derivative continuous.  The corresponding Omn$\grave{e}$s function is calculated up to 5~GeV, and the result at lower energies is shown in Fig.~\ref{fig:Omnes}. We find that though the phase-shift is very small and inelasticity very close to unity, Eq.~(\ref{eq:T}), the phase will be quite different from the phase-shift, which means that the $4\pi$ channel cannot be ignored. Again we use a form for this $\gamma\gamma$ partial wave:
\bea \label{eq:F22}
\mathcal{F}^{IJ=22}_{\lambda,1}(s)&=&\alpha_{1}(s)\,{\hat T}_{11}\,+\,\alpha_{3}(s)\,{\hat T}_{31}\;\simeq\;{\overline{\alpha}}_{1}(s)\,{\hat T}_{11} \,\, ,
\eea
where $1,\,3\,$ represents for 2$\pi$ and 4$\pi$ channels (there being no coupling to ${\overline K}K$ with $I=2$ quantum numbers). Once again we set $s_{\rm{th3}}\,=\,(1.05 {\rm GeV})^2$, below that energy ${\overline{\alpha}}_1$ must be real. However, in practice it is real up to $\sqrt{s}=1.5$~GeV.
The low energy $\gamma\gamma$ amplitudes are then fixed using just the Born terms modified by final state interactions using dispersion relation, Eq.~(\ref{eq:F;ampJ}).

\noindent One may think these partial waves are too small to matter. However, though the $\lambda=0$ partial wave is small and the Born term plus other single particle exchanges\ref{eq:LR20} are good enough to represent its left hand cut, the $\lambda=2$ wave has a much bigger effect
because of its interference with the large $I=0,\, J\,=\,\lambda\,=\,2$ wave. Thus we parameterize this wave by
\be
\mathcal{F}^{\,2}_{D2}=\alpha^{\,2}_{D2}(s) B_{D2}(s) \exp[i \varphi^2_D],
\ee
where $\alpha^{\,2}_{D2}(s)$ is a polynomial function of s. Of course, the new amplitude should more or less reproduce the low energy Born amplitude.
Indeed, in this $I\,=\,J\,=\,2$ case, we demand it to be compatible up to 1.5 GeV, within the uncertainty  given by Eq.~(\ref{eq:F;ampJ}).


\vspace{4mm}

\noindent$\mathbf{J\geq4\; channel}$\\[4mm]
In the energy region we study here, final state interactions can be ignored for high spin waves. Importantly,  for charged meson production, these amplitudes contain the one meson exchange poles that sit just outside the physical region, close to $\cos \theta\,=\,\pm1$. Consequently, we set these waves equal to their Born terms to
be the $\gamma\gamma$ amplitudes, see Appendix~\ref{app:A} for their explicit formulae.

\subsection{The coupling functions, $\alpha(s)$}
It is important to repeat that the coupling functions that appear in Eqs.~(\ref{eq:unitarity},~\ref{eq:F02},~\ref{eq:F20},~\ref{eq:F22},~\ref{eq:Fk;1D})  not only carry channel labels $1$ for $\pi\pi$, $2$ for ${\overline K}K$ and $3$ for others, they each also have $I, J, \lambda$ labels that we have often suppressed  to make these equations transparent. Thus the $\alpha$ functions are different for $I\,=\,0,2$, $J\,=\,0, 2$ with $\lambda\,=\,0, 2$. These are parameterized as:
\begin{eqnarray}\label{eq:alphas}
{\alpha_i}^I_{J,\lambda}(s)\,=\, \exp(bs)\,\sum_n\,a_n\,X^n
\end{eqnarray}
with $X\,=\,(2s-s_1-s_2)/(s_2-s_1)$ where $s_1 = s_{th}$, $s_2 = 2.1$ GeV$^2$ and so in the range fitted $-1 \le X \le 1$. The exponential factor is an efficient way of taking account of the rapid change of the amplitudes from the near threshold Born term.  The parameters $b$ and $a_n$ depend, of course, on $I, J, \lambda$ and the channel label $i$.

\noindent All the different coupling functions $\alpha$ have in common that they are functions with only left hand cuts. Consequently, they are expected to be smooth along the right hand cut. Given the limited energy domain that is fitted here, we represent these by low order polynomials. Fits extending over a larger regime would need higher powers of $s$, and so a conformal representation may well be more economical. Here low order polynomials are sufficient. Typically, we have no more than 4 or 5 parameters per wave, in Eq.~(\ref{eq:alphas}). However, not having the correct analytic structure such parametrizations should not be continued far into the complex energy plane. We will consider that later in Sect.~\ref{sec:3}.

\newpage
\subsection{Dispersive constraints in the low energy region}\label{sec:1;5}
\begin{figure}[htbp]\label{fig:Flow}
\begin{center}
\includegraphics[width=0.48\textwidth,height=0.39\textheight]{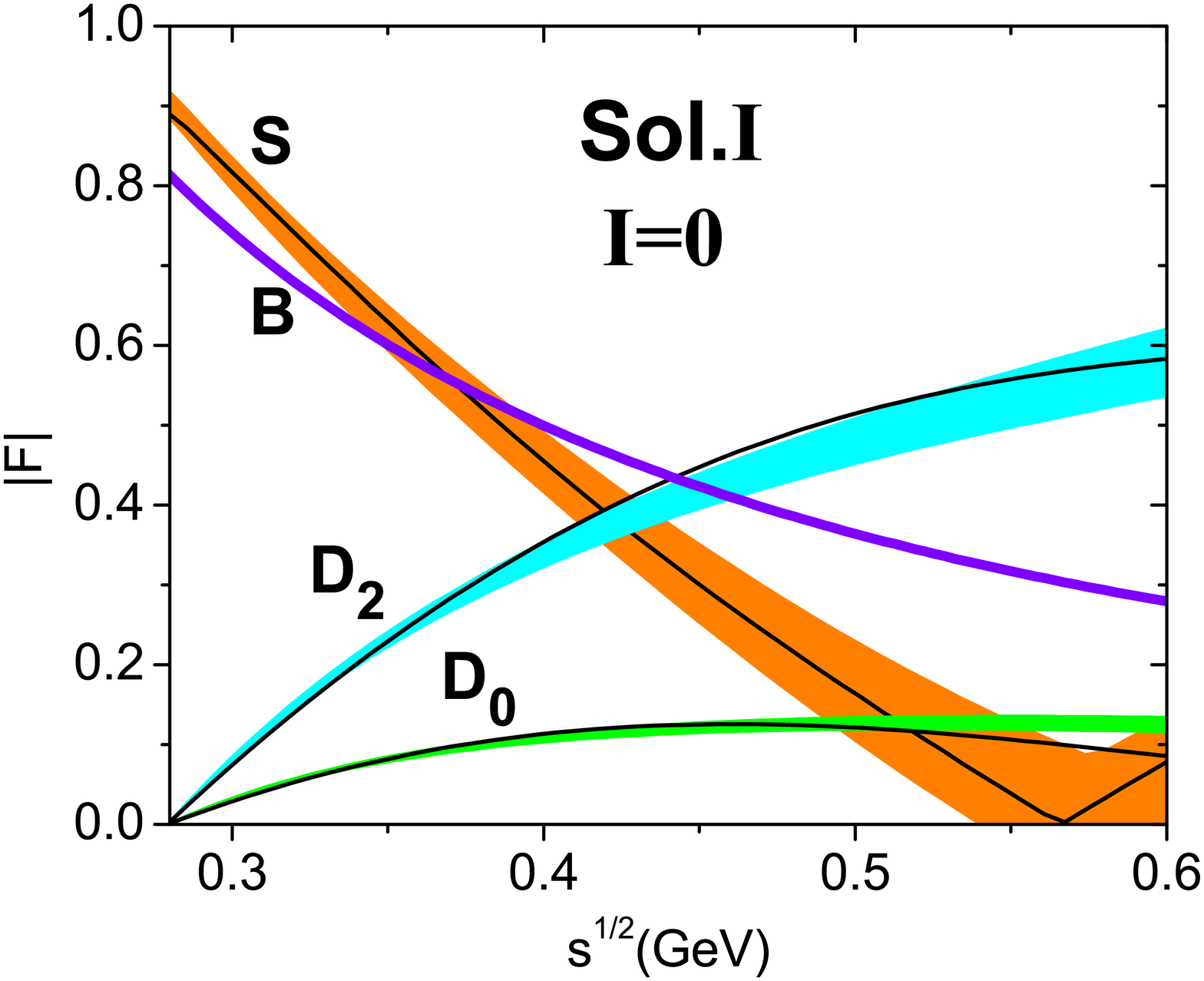}
\includegraphics[width=0.48\textwidth,height=0.39\textheight]{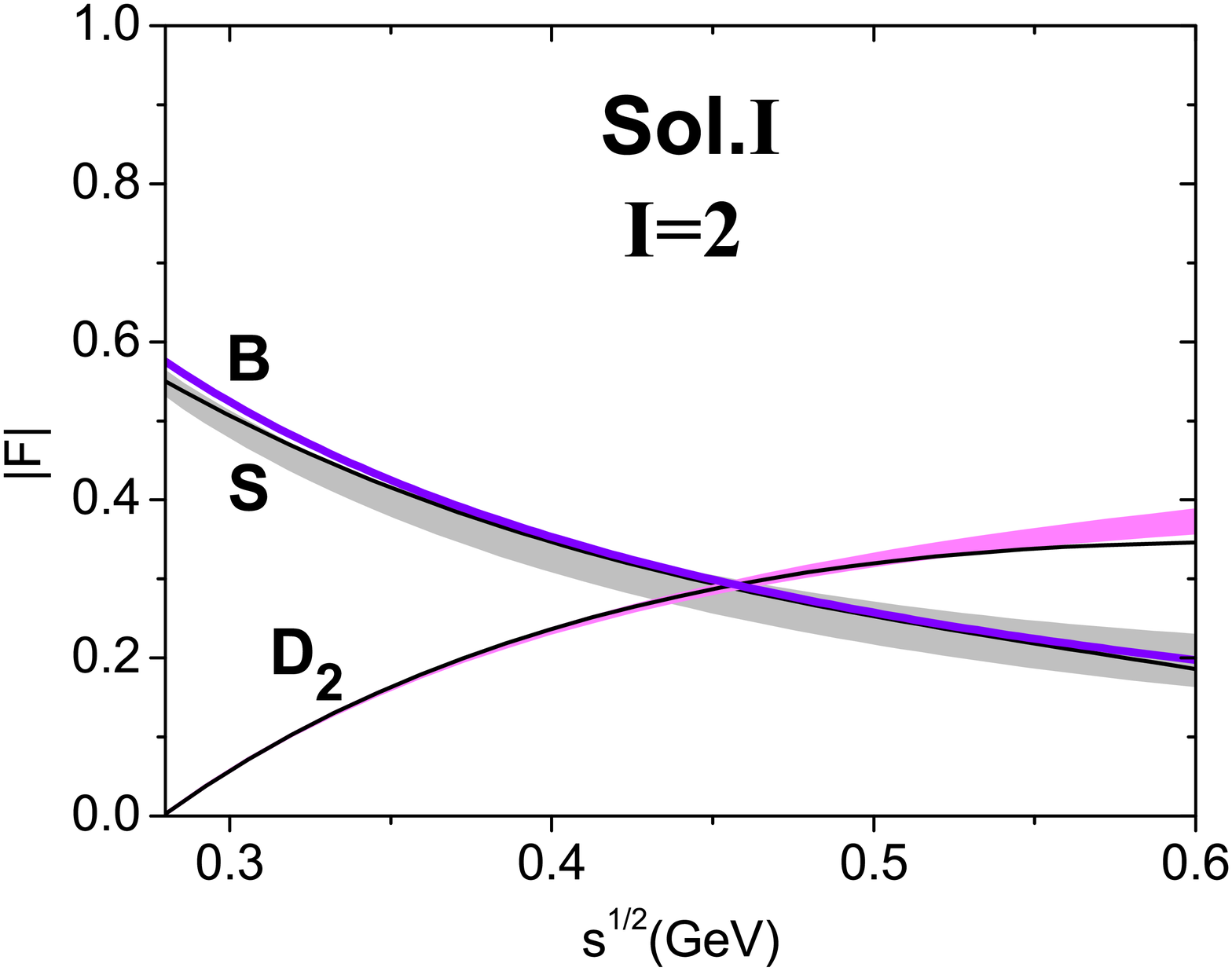}\\
\vspace{3mm}
\includegraphics[width=0.60\textwidth,height=0.36\textheight]{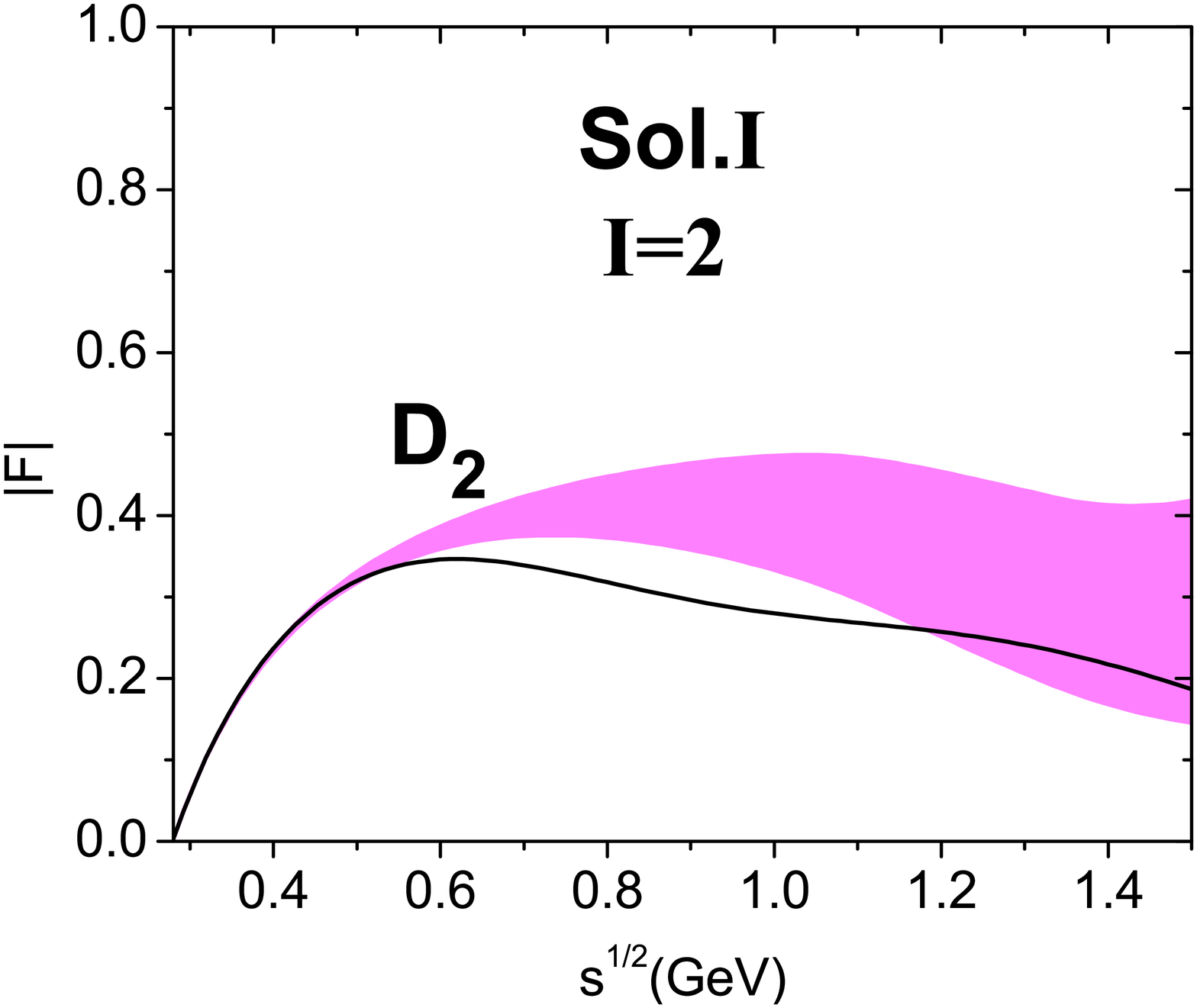}
\caption{\label{fig:Flow} Low energy amplitudes of $\gamma\gamma\to\pi\pi$ using Eqs.~(\ref{eq:F;ampS},\ref{eq:F;ampJ}). These are inputs to our partial wave determinations.
Only $F^2_{D2}$ wave has been input as a constraint up to 1.5~GeV as described in the text.
The bands indicate the uncertainties in these calculations, including the dispersive effect of the unknowns in the left hand cut  contributions and the phase of the relevant $\gamma\gamma$ partial waves above 1.5~GeV.  }
\end{center}
\end{figure}
With the inputs given above for the phase behavior of each partial wave and for the left hand cut amplitude, we can accurately compute the $I=0,\, 2$  $S$, $D_0,\, D_2$ waves in the low energy region. The uncertainties in the phases above 1.4~GeV or so contribute at most 10\% to the spread of the $\gamma\gamma$ partial waves below 600 MeV. A far larger contribution to this spread comes from the uncertainties in the modeling the left hand cut contributions by single particle exchanges (beyond one pion exchange), and for the $S$-waves in the position of the Adler zero location, Eq.~(\ref{eq:zero}).
These give rise to the error bands shown in Fig.~{\ref{fig:Flow}}.
For the ${\cal F}^2_{D2}$ wave, this is used up to 1.5~GeV, as described in Sect.~\ref{sec:1;5}.
The increase in uncertainties is why we only use the dispersive treatment for the partial waves below 600 MeV.
We will require our two photon amplitudes, described by the general unitary representation, Eq.~(\ref{eq:unitarity}), to lie within these bands.  Other calculational treatments~\cite{yumao09, Moussallam10} have imposed greater faith in knowledge of the crossed-channel exchanges and of the phases of the $\gamma\gamma$ amplitudes to make predictions even beyond ${\overline K}K$ threshold. Rather we will use  $\gamma\gamma$ data entirely to determine these amplitudes, and merely constrain the partial waves to lie within the calculated bands of Fig.~{\ref{fig:Flow}} in the energy ranges shown.

\vspace{4mm}
\subsection{$\gamma\gamma\to K \overline{K}\,$ amplitudes}\label{sec:1;6}
\vspace{4mm}
In principle the methodology adopted for the treatment of the $\gamma\gamma\to\pi\pi$ channels could be applied to that for ${\overline K}K$ production. Again coupled channel unitarity applies, and at the threshold for Compton scattering, $\gamma K\to\gamma K$, Low's low energy theorem equally holds and the amplitude is controlled by the one kaon exchange Born amplitude. However, the physical region for $\gamma\gamma$ production of kaon pairs is so far away from that for  Compton scattering that there is no part of the physical region where the left hand cut of this amplitude dominates its behavior. Very rapidly, $K$-exchange, gives way to $\kappa$ and $K^*$ exchange, or more generally correlated $K\pi$, $K\pi\pi, \cdots$ exchanges. Consequently, none of our previous dispersive machinery is useful in practice.  Moreover,
for the ${\overline K}K$ channels the older data are of much poorer statistics than for $\pi\pi$, particularly below 1.5~GeV with just 23 datapoints compared to nearly 3000 for the $\pi\pi$ channels. The recently published Belle results on $\gamma\gamma\to K_sK_s$ have good angular coverage above 1.1~GeV and add a further 315 datapoints to our analysis. While the ${\overline K}K$ data may be expected to constrain the $I=0$ amplitudes through coupled channel unitarity~Eq.~(\ref{eq:unitarity}), they also have important $I=1$ components, which Bose symmetry does not allow in the case of the $\pi\pi$ final state. Having data on $K^+K^-$, ${\overline K}^0K^0$ and $K_sK_s$ modes helps to disentangle these, at least approximately. A complete amplitude analysis of the isovector channel would require the inclusion not only of data on $\gamma\gamma\to \pi^0\eta$, but detailed information on the purely hadronic $\pi^0\eta\to\pi^0\eta$, ${\overline K}K$ channels, which are not available. Consequently, we have to attempt a more limited separation of the isovector component, and so instead use a representation that is less detailed. Below we give our parametrization of the ${\overline K}K$ amplitudes, which is particularly simple and crude in the case of the $I=1$ component.

\vspace{4mm}
\noindent $\mathbf{IJ=00\;channel}$\\[4mm]
\noindent In this channel, complete information is provided by the final state interaction constraint, Eq.~(\ref{eq:unitarity}). This fixes the $\gamma\gamma\to K \overline{K}$ amplitude once the coupling functions  $\alpha_{1,2}$ have been determined.

\vspace{4mm}
\noindent $\mathbf{IJ=02\;channel}$\\[4mm]
\noindent The isospin 0 $D$-waves are dominated by the $f_2(1270)$ resonance, for which we use a simple parametrization:
\be\label{eq:Fk;0D}
F^0_{D\lambda}(\gamma\gamma\to K \overline{K})=F^0_{D\lambda}(\gamma\gamma\to \pi \pi)
\frac{\rho_2^2(s)}{\rho_1^2(s)}\sqrt{\frac{\rho_1^5(M_{f_2}^2)\Gamma(f_2(1270)\to K \overline{K})}{\rho_2^5(M_{f_2}^2)\Gamma(f_2(1270)\to\pi\pi)}}\exp[i \varphi^0_{D}]\; ,
\ee
where recall Eq.~(\ref{eq:phasespace}) $\rho_i\,=\,\sqrt{1 - 4 m_i^2/s}$ with $i=1$ for the $\pi$ and $i=2$ for the $K$ channels.
In reality our fit shows that $\varphi^0_{D}$, $\varphi^1_{D}$ (see for the latter Eq.(\ref{eq:Fk;1D})) only make the fit a bit better, with a  $\Delta\chi^2<5$.  Given  the large uncertainty in the $I=1$ components, that also contributes in the ${\overline{K}}K$ case, {\it e.g.} Eq.~(\ref{eq:Fk;1S}) in $IJ=10$ channel, we simply set these phases to zero.

\vspace{4mm}
\noindent $\mathbf{IJ=10\;channel}$\\[4mm]
\noindent
For the isospin 1 $S$-wave, we simply parameterize this as a complex function:
\bea\label{eq:Fk;1S}
F^{I=1}_{S}(\gamma\gamma\to K\overline{K})&=&f^K_1(s)\, +\,i~f^K_2(s)\;\;.
\eea
with $f^K_i(s)$ polynomials of s. We then let the data fix these.

\vspace{4mm}
\noindent $\mathbf{IJ=12\;channel}$\\[4mm]
\noindent
The isospin 1 $D$ waves are dominated by the appearance of the $a_2(1320)$. Since its shape is controlled by its large $\rho\pi$ decay mode, we simply parameterize this wave in the ${\overline K}K$ channel by:
\be\label{eq:Fk;1D}
F^1_{D2}(\gamma\gamma\to K \overline{K})={\alpha_K}^1_{D2}(s)\,T^1_{D}(\rho\pi\to \rho\pi)\,\frac{s-s_{th2}}{s-s_{th1}}\,\exp[i \varphi^1_{D}]\; ,
\ee
where  $s_{th1},\,s_{th2}$ are the $\rho\pi$ and ${\overline K}K$ thresholds, respectively, and $T^1_{D}$ is given by the simple Breit-Wigner parametrization:
\be\label{eq:T;1D}
T^1_{D}(\rho\pi\to \rho\pi)=\frac{g_1(s)^2}{ M^2-s-i \rho_1(s)g_1^2(s)-i \rho_2(s)g_2^2(s)-i \rho_3(s)g_3^2(s) }.
\ee
and the dimensionful functions $g_i(s)$ include the standard Blatt-Weisskopf barrier factors, here called $Q_i(s)$, with $i$ labeling the decay channel, so that
\bea\label{eq:T;1D;g}
g_i^2(s)&=&\frac{M_{a_2}~\Gamma_{a_2}~\text{BR}_i~\mathcal{Q}_i(M_{a_2}^2)}{\rho_i(M_{a_2}^2)~\mathcal{Q}_i(s)} \;,\;\;\;\;\;\; i=1,2,3\;.\nonumber\\
\mathcal{Q}_1(s)&=&1+\frac{q^2}{s-s_{th1}} \; ,\nonumber\\
\mathcal{Q}_j(s)&=&1+\frac{q^2}{s-s_{thj}}+\left(\frac{q^2}{s-s_{thj}}\right)^2 \;,\;\; j=2,3\;.
\eea
$\Gamma_{a_2} = 107$~MeV, and BR$_i$ are the individual channel branching ratios. We take the parameters $q$ to be 1~GeV.
Here 1, 2, 3 represents for $\rho\pi$, $K\overline{K}$ and remaining $n\pi$ channels respectively. Given the paucity of the data to be fitted we absorb all the contribution of $\omega\pi\pi$, $\eta\pi$, $\cdots$ into the $n\pi$ channel for simplicity. It is equal to setting this branching ratio to  $25\%$.
Moreover, we simply ignore the $IJ\lambda=1D0$ wave and higher spin partial waves. The coupling function ${\alpha_K}^1_{D2}$ is parameterized as in Eq.~({\ref{eq:alphas}).
Together with the $\gamma\gamma\to {\overline K}K$ data, these do act as a useful constraint in the coupled channel treatment we use, especially for the $S$-wave.

\baselineskip=5.5mm

\section{Fit results}\label{sec:2}
\subsection{$\pi\pi\to\pi\pi$ scattering amplitudes}\label{sec:2;1}
There are two main parts in our fit. One is the fit to hadronic data, which determines the  $T$-Matrix elements, the inputs for which we have described in detail in Sect.~2. The other focusses on
the two photon reactions, $\gamma\gamma\to\pi\pi$ and  $\gamma\gamma\to \overline{K}K$. Our strategy is to first fit the hadronic data, and use the resulting $T$-matrix elements  to fit the $\gamma\gamma$ data.

\noindent For $IJ=00$ $\pi\pi$ scattering amplitude we use a coupled channel $K$-matrix parametrization.  To fix the parameters we fit to the phase shift,
inelasticity as was done previously~\cite{MRP08}, including all the data from Refs.~\cite{CERN-Munich,Cohen80,Etkin82,NA48}. The {\it CFDIV} parametrization of $T^0_{0}$~\cite{KPY} are included as important new constraints, together with the dispersive results of Buttiker {\it et al}~\cite{Descotes04} on the $\pi\pi\to {\overline K}K$ amplitude. We include the effects of isospin breaking by taking into account the 8~MeV mass difference between the $K^+K^-$ and ${\overline K}^0K^0$ thresholds, rather than treating the kaons as having a common mass as in ~\cite{KPY}. Other than this mass difference, the $K$-matrix elements are treated as isospin invariant. While our fitting is only along the real axis, the parametrization does have the $\sigma$-pole at $\sqrt{s}\;=\;441\,-\,i\,272$~MeV.

\begin{figure}[bhtp]
\includegraphics[width=1.0\textwidth,height=0.38\textheight]{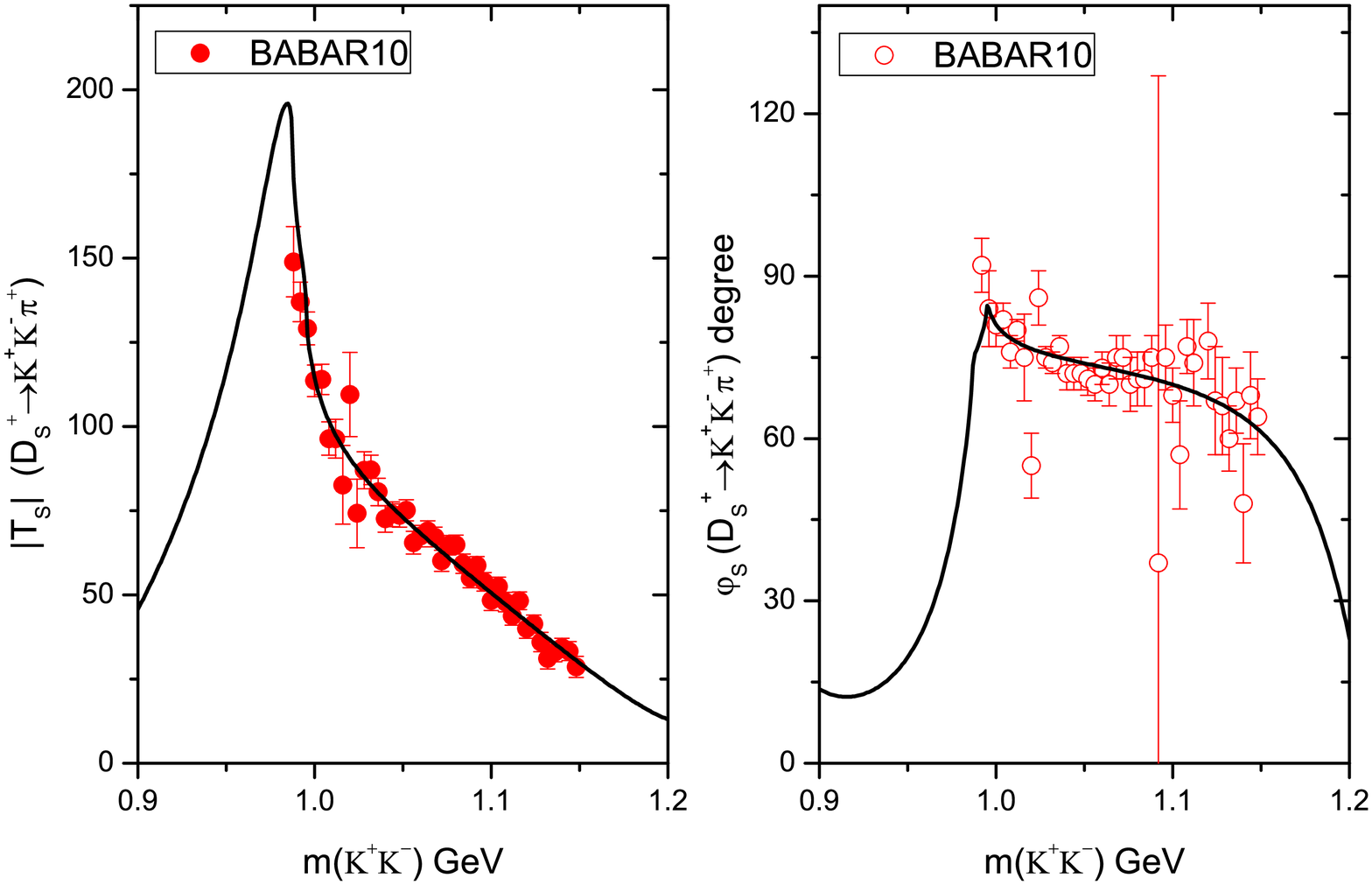}
\caption{\label{fig:BABAR} Fit to the $K^+K^-$ $S$-wave magnitude and phase in the decay $D_s^+\to (K^+K^-)\pi^+$ determined by BaBar~\cite{BABAR-K}. }
\end{figure}

\begin{figure}[htbp]
\includegraphics[width=0.5\textwidth,height=0.35\textheight]{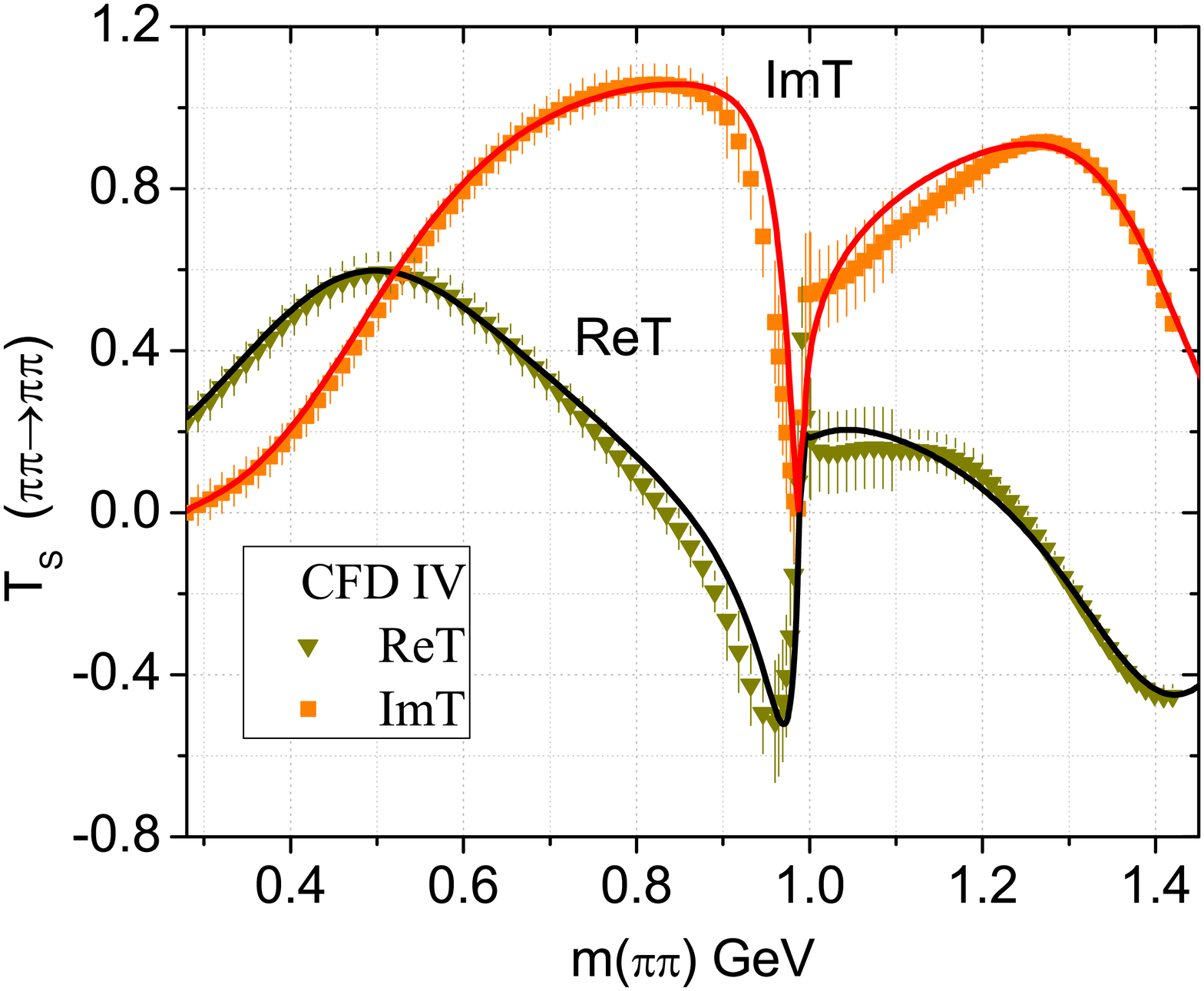}
\includegraphics[width=0.5\textwidth,height=0.35\textheight]{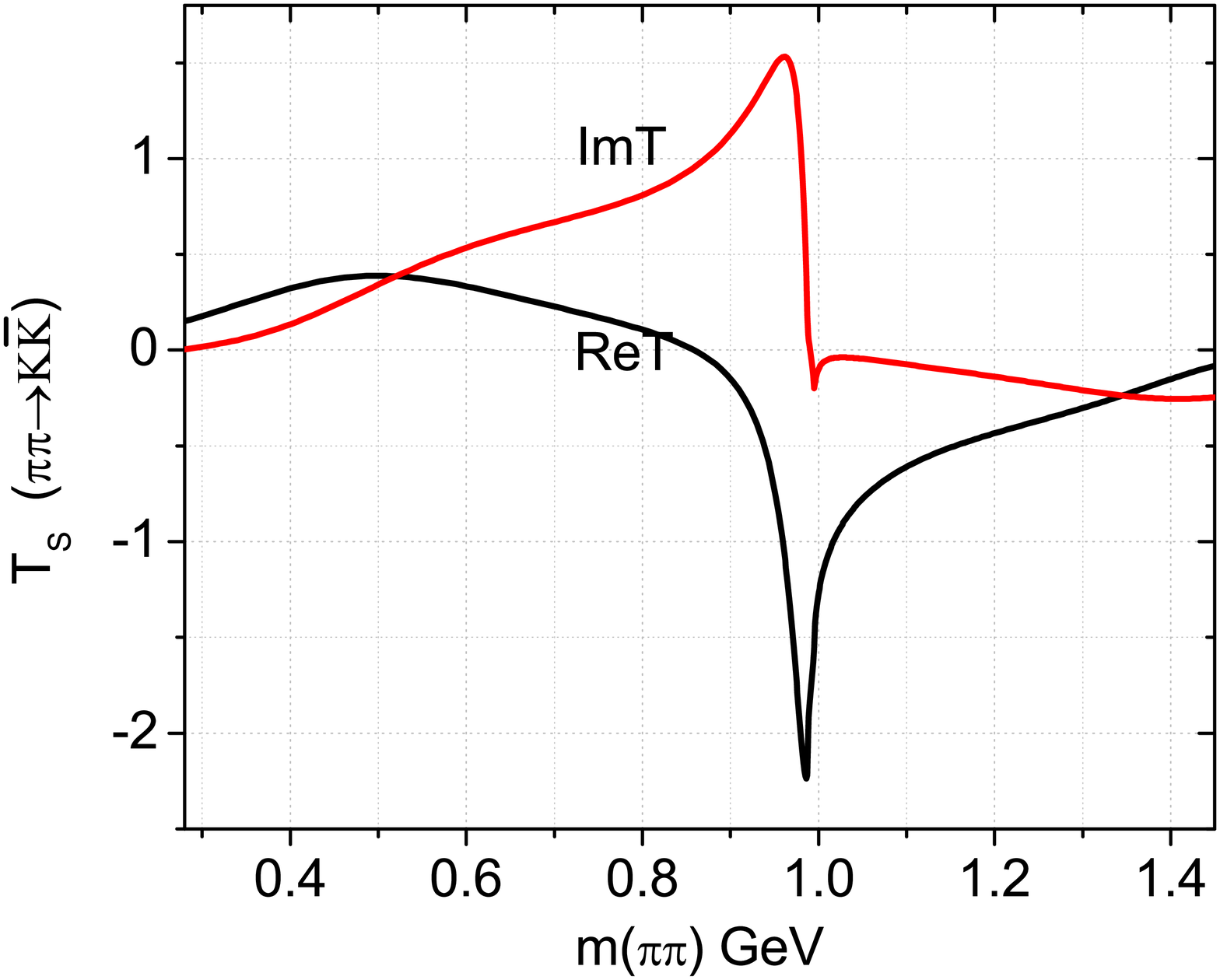}
\includegraphics[width=0.5\textwidth,height=0.35\textheight]{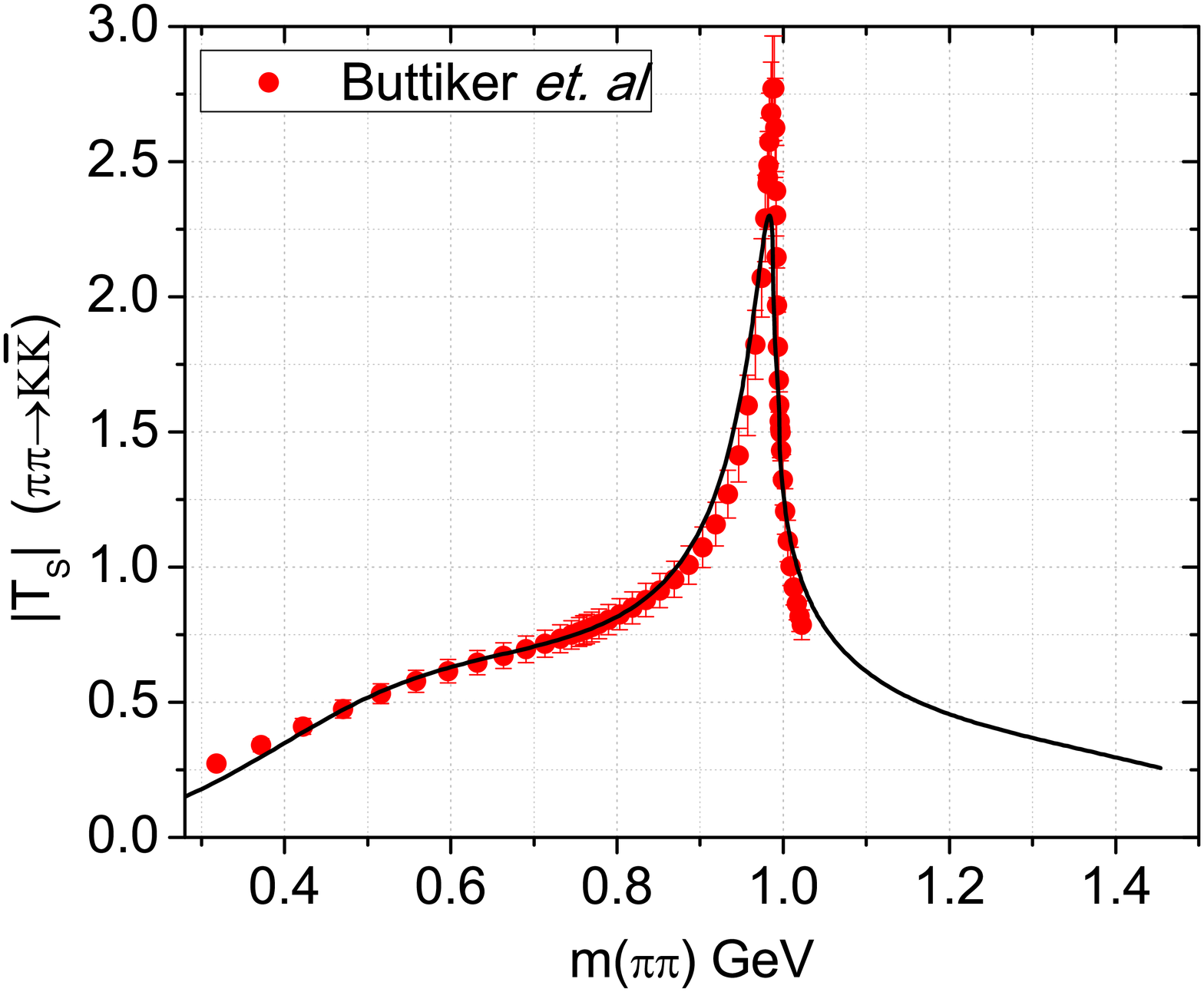}
\includegraphics[width=0.5\textwidth,height=0.35\textheight]{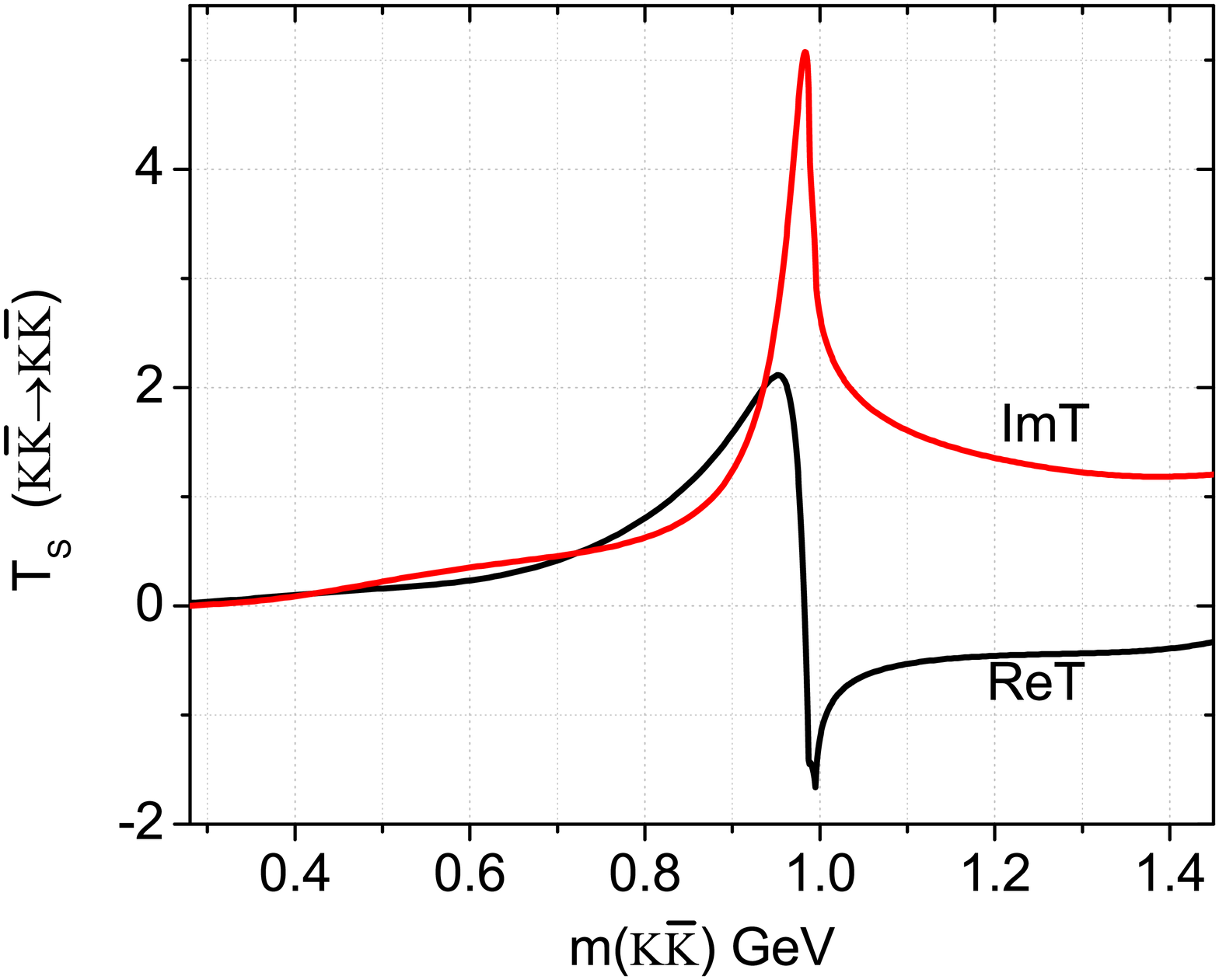}
\caption{\label{fig:T} The $I=J=0$ $\,T$-matrix elements for the hadronic processes: $\pi\pi\to\pi\pi$, $\pi\pi\to {\overline K}K$ and ${\overline K}K\to{\overline K}K$. The \lq\lq data'' labeled by {\it CFDIV} are from the dispersive analysis of Pel\'aez and  collaborators~\cite{KPY}.
The red points, representing for the modulus of $T_{12}$, are from the Roy-Steiner equation analysis of Buttiker {\it et al.}~\cite{Descotes04}.  }
\end{figure}

\noindent Since the ${\overline K}K$ threshold region features crucially in the $\gamma\gamma$ data, see Fig.~\ref{fig:csf980} later for instance, we also include  the latest BaBar Dalitz plot analysis of $D_s^+\to\pi^+\pi^-\pi^+$~\cite{BABAR-pi} and $D_s^+\to K^+K^-\pi^+$~\cite{BABAR-K}.
With a spectator $\pi$ in each case, the $\pi\pi$ and ${\overline K}K$ $S$-wave amplitude and phases have been determined~\cite{BABAR-pi, BABAR-K}, and for the latter in finer detail than in other reactions, as shown in Fig.~\ref{fig:BABAR}. According to Eq.~(\ref{eq:unitarity}), these decay amplitudes also constrain the $S$-wave hadronic amplitudes, of course, with appropriately different $\alpha$ functions.
The fitted $T$-matrix elements are shown in Fig.~\ref{fig:T}.
Compared to the precise description below 800~MeV and above 1.2~GeV, inconsistencies amongst the datasets around the ${\overline K}K$ threshold mean the description of any one dataset is not perfect there.

\noindent As seen in Fig.~\ref{fig:T} our fitted amplitudes are not identical to those of Pel\'aez {\it et al.}~\cite{KPY} though their amplitudes are an input, particularly in the region of 0.9-1.05~GeV. This is because we fit other information, such as that of \cite{Descotes04} and \cite{BABAR-pi,BABAR-K}. Important for the latter,  we treat the charged and neutral kaon pair thresholds with their actual (rather than a common) mass.

\noindent With the basic $T$-matrix elements that enter into the unitarity equation, Eq.~(\ref{eq:unitarity}), now fixed,  we focus on the analysis of the photon-photon amplitudes.

\subsection{$\gamma\gamma\to\pi\pi\,$ fits}\label{sec:2;2}
In this section we describe the fit to all the data on $\gamma\gamma\to\pi\pi$, both integrated cross-sections and
angular distributions. These datasets are listed in Table~\ref{tab:data}. The recent high statistics data
on $\gamma\gamma\to\pi^0\pi^0$ by Belle~\cite{Belle-nn} is the main addition to the earlier fit~\cite{MRP08}. The experimental binning, particularly in $\cos \theta$, are taken into account in the fitting. The fit forms are integrated over each bin, even though we show the fits as continuous lines. The charged meson data will, of course, have one meson exchange poles close to the forward and backward dierctions. With data only out to $|\cos \theta|\, \sim\, 0.6$, these poles are not obviously there in the plots we show. Nevertheless, these poles are there in our theoretical amplitudes used to fit the data, encoded in the $J\,\ge\,4$ partial waves. This is explicitly illustrated later in Fig.~\ref{fig:dcspredict}.

\begin{table}[t]
\centering
\begin{tabular}{|c|l|c|c|c|c|}
\hline
\rule[-0.4cm]{0cm}{12mm}
Experiment & {\hspace{3.5mm}}Process                              & Int. X-sect.  &  $|\cos \theta\, |_{max}$  & Ang. distrib.   &  $|\cos \theta\, |_{max}$   \\
\hline \hline
\rule[-0.6cm]{0cm}{12mm}
Mark II    & $\,\gamma \gamma \to \pi ^{+} \pi^{-}$ & 81            & 0.6                      & 63              & 0.6                       \\
\hline
\rule[-0.75cm]{0cm}{15mm}
Crystal Ball   & $\,\gamma \gamma \to \pi ^{0} \pi^{0}$ & 36            &\parbox{2.cm}{0.8~(CB88) \protect \\
                                                                         0.7~(CB92)}         & 90              & 0.8                       \\
\hline
\rule[-0.75cm]{0cm}{15mm}
CELLO      & $\,\gamma \gamma \to \pi ^{+} \pi^{-}$ & 28            & 0.6                      & \parbox{2.5cm}{104~(Harjes) \protect\\
                                                                                               201~(Behrend)}  & 0.55 - 0.8                \\
\hline
\multirow{2}{*}{\rule{0cm}{1.1cm}Belle}
           & \rule[-0.6cm]{0cm}{12mm}$\,\gamma\gamma\to\pi^+\pi^-$
                                                 & 128           & 0.6                      & 1536            & 0.6                       \\
\cline{2-6}
          & \rule[-0.6cm]{0cm}{12mm} $\,\gamma\gamma\to\pi^0\pi^0$
                                                 & 36            & 0.8                      & 684             & 0.6                       \\
\hline
\end{tabular}
\caption{\label{tab:data}~~Summary of datasets. Data in each experiment are fitted up to
1.44~GeV. Mark~II results are from Boyer {\it et al.}~\cite{MarkII},
CB88 (Crystal Ball 1988) from Marsiske {\it et al.}~\cite{CB88} and CB92 (Crystal Ball 1992) from Bienlein
{\it et al.}~\cite{CB92},  CELLO from Harjes {\it et al.}~\cite{Cello1} and Behrend {\it et al.}~\cite{Cello2}, $\gamma\gamma\to\pi^+\pi^-$ of
Belle from Mori {\it et al.}~\cite{Belle-pm}, and $\gamma\gamma\to\pi^0\pi^0$ of Belle from Mori {\it et al.}~\cite{Belle-nn}. }
\end{table}

\baselineskip=5.2mm
\noindent As the number of datapoints from different experiments are different, we weight the datasets to ensure each group contributes roughly the same in $\chi^2$.
Each experiment, except for CELLO, quotes separately a systematic uncertainty for normalizing their cross-sections. CELLO~\cite{Cello1,Cello2} folds this with the statistical uncertainties.
To fit the disparate datasets, we allow for  these global shifts. These are almost 7\% for Mark~II above 0.45~GeV, 2\% for Crystal Ball (CB88) in the whole energy region, 3\% for Crystal Ball (CB92), and 4\% for the Belle neutral pion data.
The normalization uncertainty of Belle charged pion data is energy dependent. However, it is dominated by a global shift of almost 5\%.
In the plots, we choose to renormalize our amplitudes rather than the datasets, so it is the published data that are plotted in Fig.~\ref{fig:cspic} and later figures.
\begin{figure}[htbp]
\vspace*{-0.0cm}
\includegraphics[width=0.5\textwidth,height=0.35\textheight]{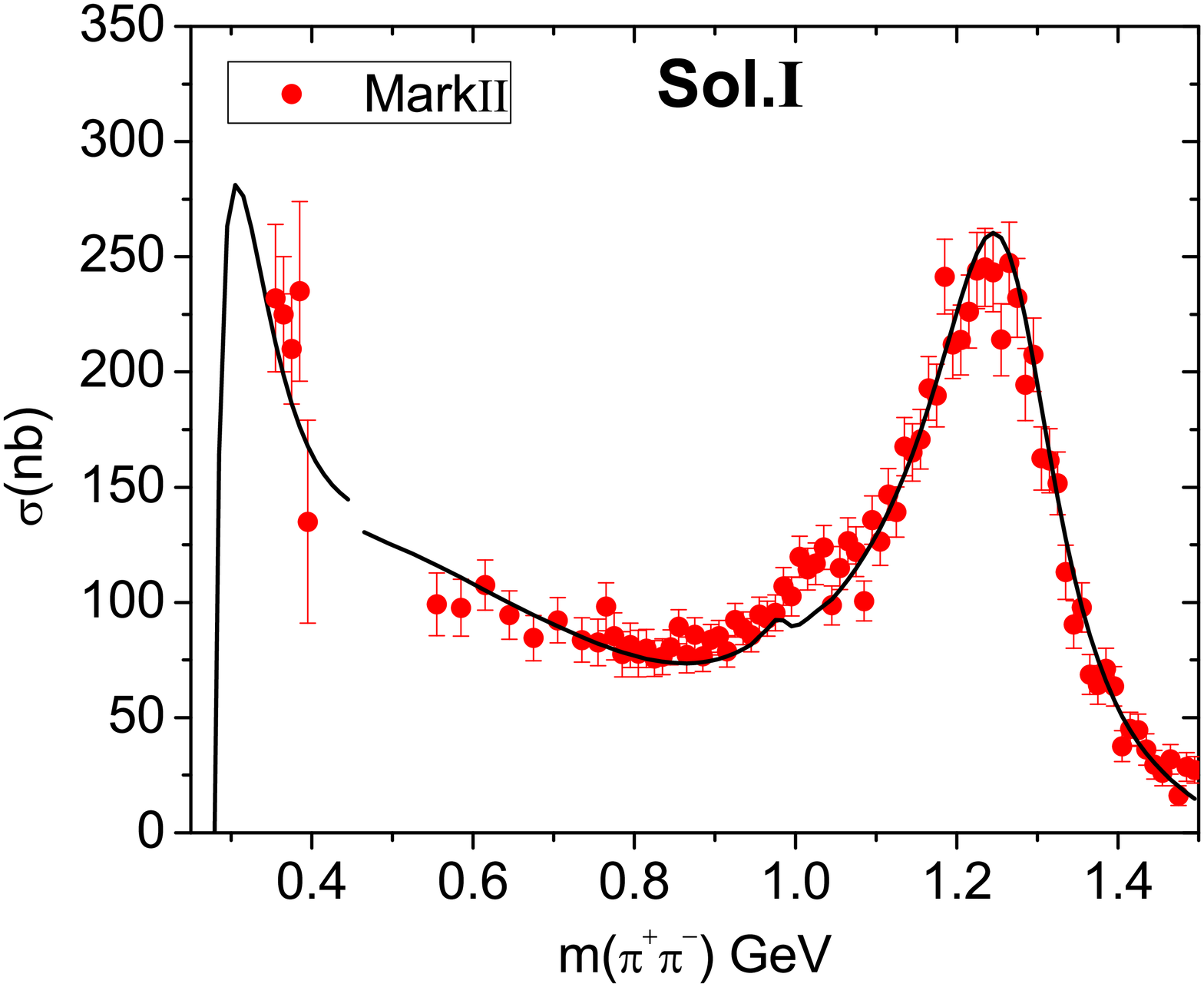}
\includegraphics[width=0.5\textwidth,height=0.35\textheight]{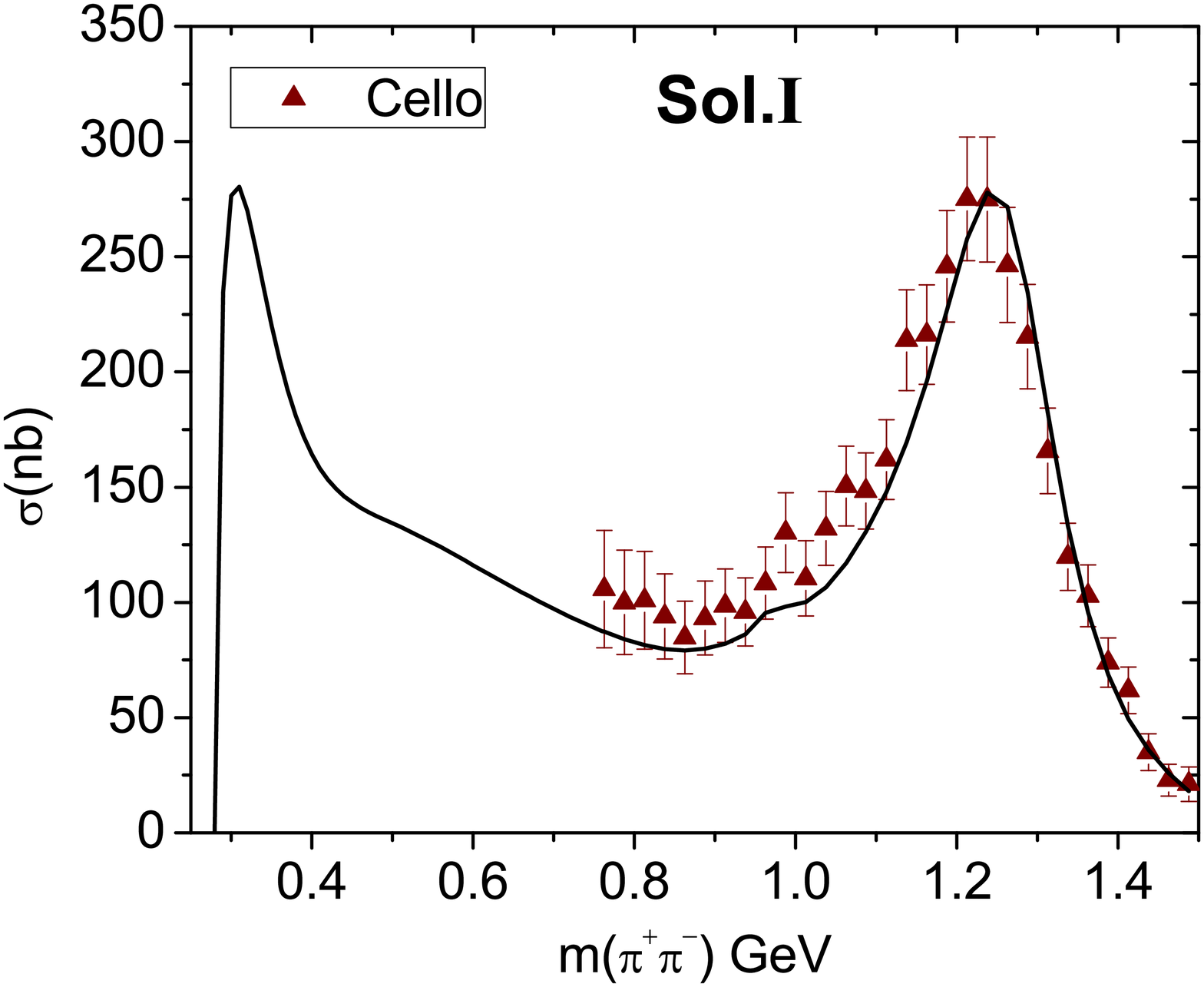}
\begin{center}
\includegraphics[width=0.5\textwidth,height=0.35\textheight]{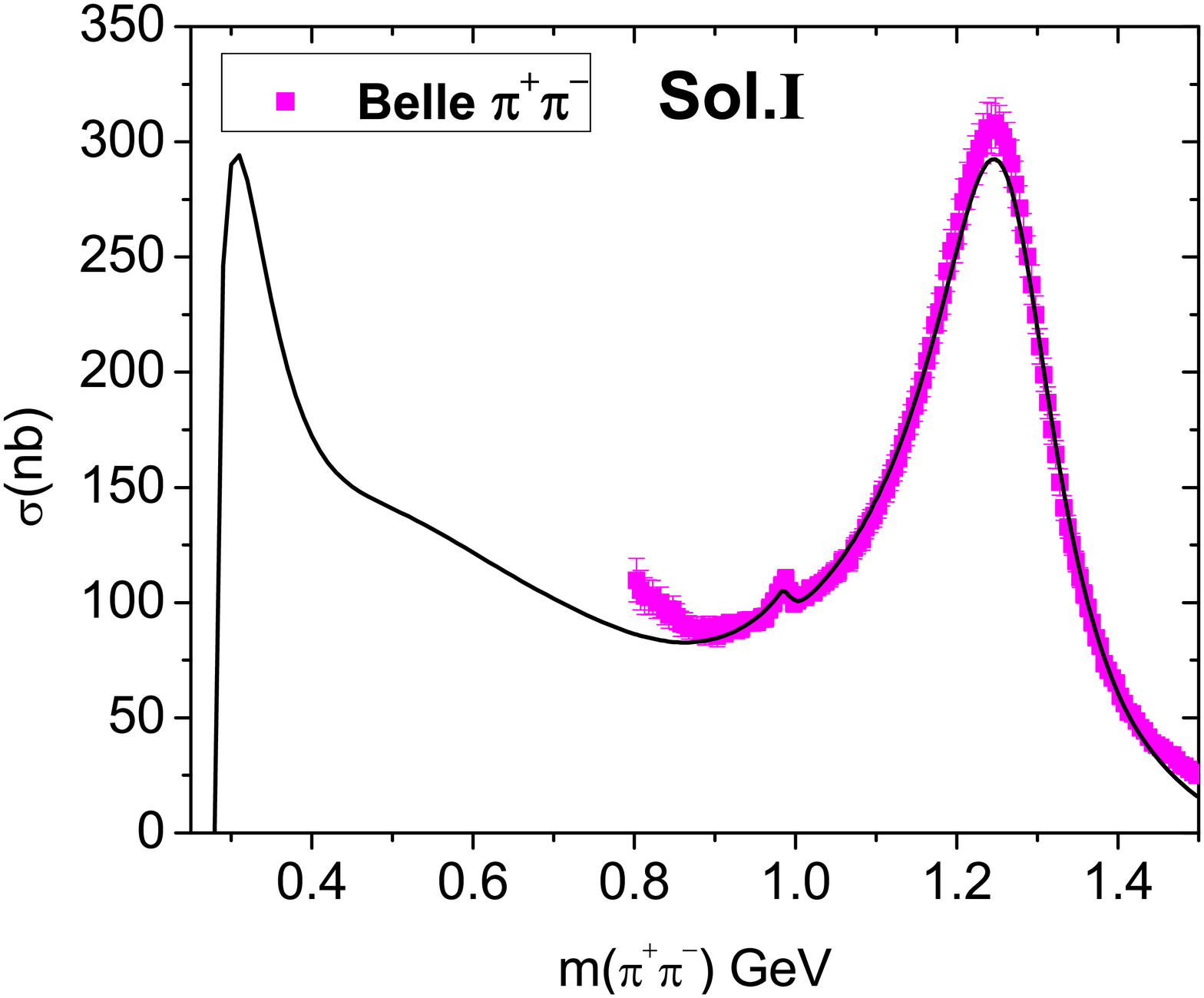}
\end{center}
\caption{\label{fig:cspic} Solution~I compared with the $\gamma\gamma\rightarrow\pi^+\pi^-$ datasets. The Mark~II~\cite{MarkII},
Cello~\cite{Cello1, Cello2} and Belle~\cite{Belle-pm} are all integrated over $|\cos \theta|\,\le\,0.6$.
The freedom to make a small systematic shift in normalization of Mark~II data above 0.45~GeV has been used to improve the fits.
While we make a systematic shift to all Belle data. Rather than shift the data, the solutions have been renormalized.
This results in the discontinuity in the solid curves at 0.45~GeV of Mark~II plots.
}
\end{figure}
\noindent Among the charged pion datasets only Mark~II has data below 700 MeV. We increase the weight of these data-points to give them some bite. In addition, we constrain the individual partial waves to lie within the dispersively evaluated bands of Fig.~\ref{fig:Flow}, as we have previously described.

\noindent We obtain a number of fits that almost equally well describe all the data.
Consequently, we show one representative solution (we call Solution~I) in these plots. Why this choice will become clear when we come to the ${\overline K}K$ data.  The differential cross-section plots for $\gamma\gamma\rightarrow\pi^+\pi^-$ are displayed in Figs.~\ref{fig:dcsMark}-\ref{fig:dcsBelle3}.
Fig.~\ref{fig:dcsMark} shows the fit to Mark~II, Fig.~\ref{fig:dcsCello} that to Cello and Figs.~\ref{fig:dcsBelle1},~\ref{fig:dcsBelle2},~\ref{fig:dcsBelle3} for Belle. The results for the integrated cross-section for $\gamma\gamma\rightarrow\pi^0\pi^0$ are shown in Fig.~\ref{fig:cspin}, and differential cross-sections in Figs.~\ref{fig:dcsCB},~\ref{fig:dcsBelle00}. The quality of the fits is good. This is quantified in Table~\ref{tab:fit}, where the $\chi^2$ for each dataset is listed.

\begin{figure}[t]
\includegraphics[width=1.0\textwidth,height=0.38\textheight]{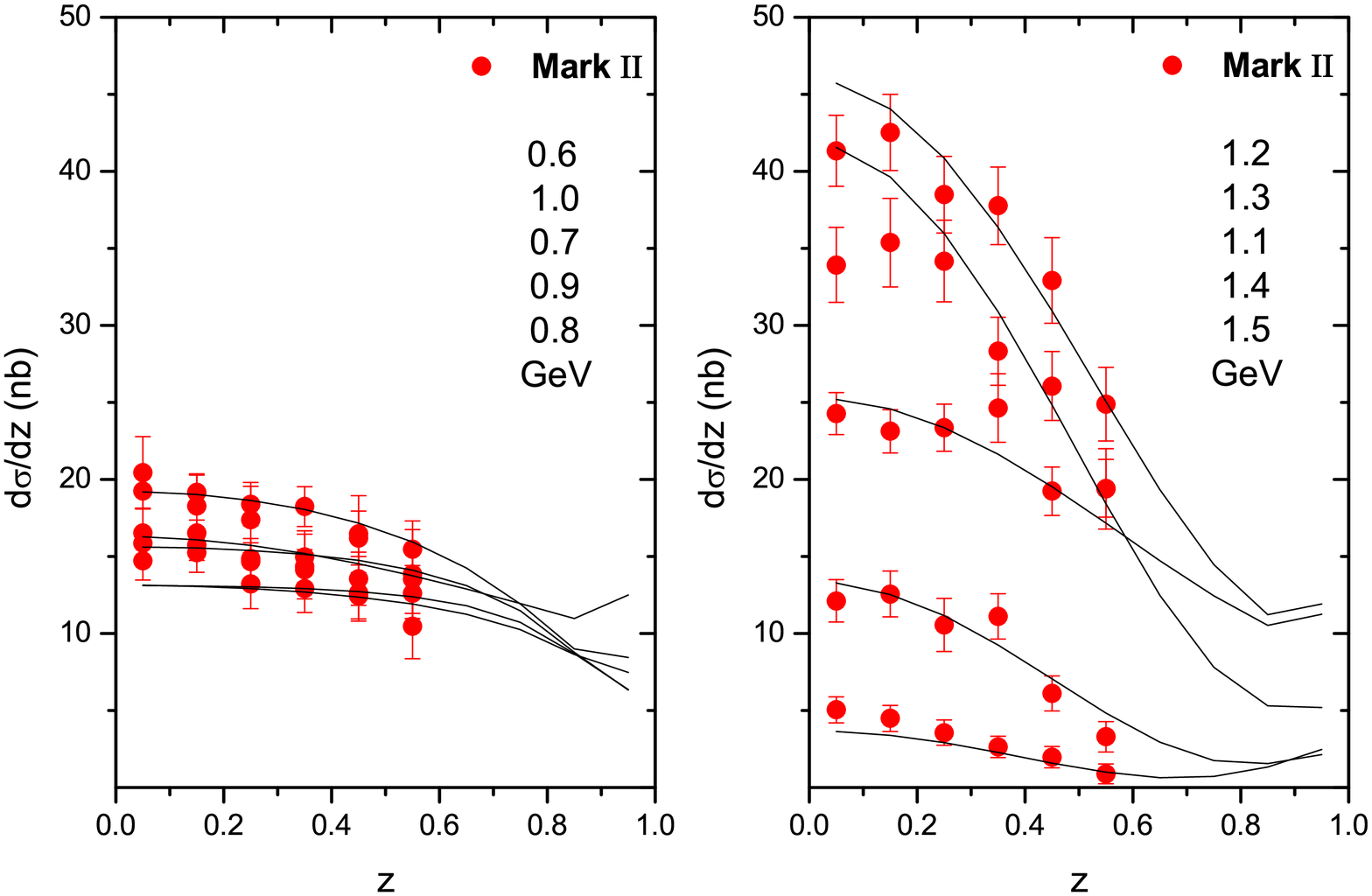}
\caption{\label{fig:dcsMark} Fit to $\gamma\gamma\to\pi^+\pi^-$ differential cross-section of the Mark~II experiment~\cite{MarkII}.
The fit shown is Solution~I. The numbers give the central energy in GeV of each angular distribution listed in order of the cross-section at $z=0$, where $z\,=\,\cos\,\theta$. The data are normalized so that the integrated cross-section is just a sum of the differential cross-sections in each angular bin.}
\end{figure}

\begin{figure}[htbp]
\includegraphics[width=1.0\textwidth,height=0.6\textheight]{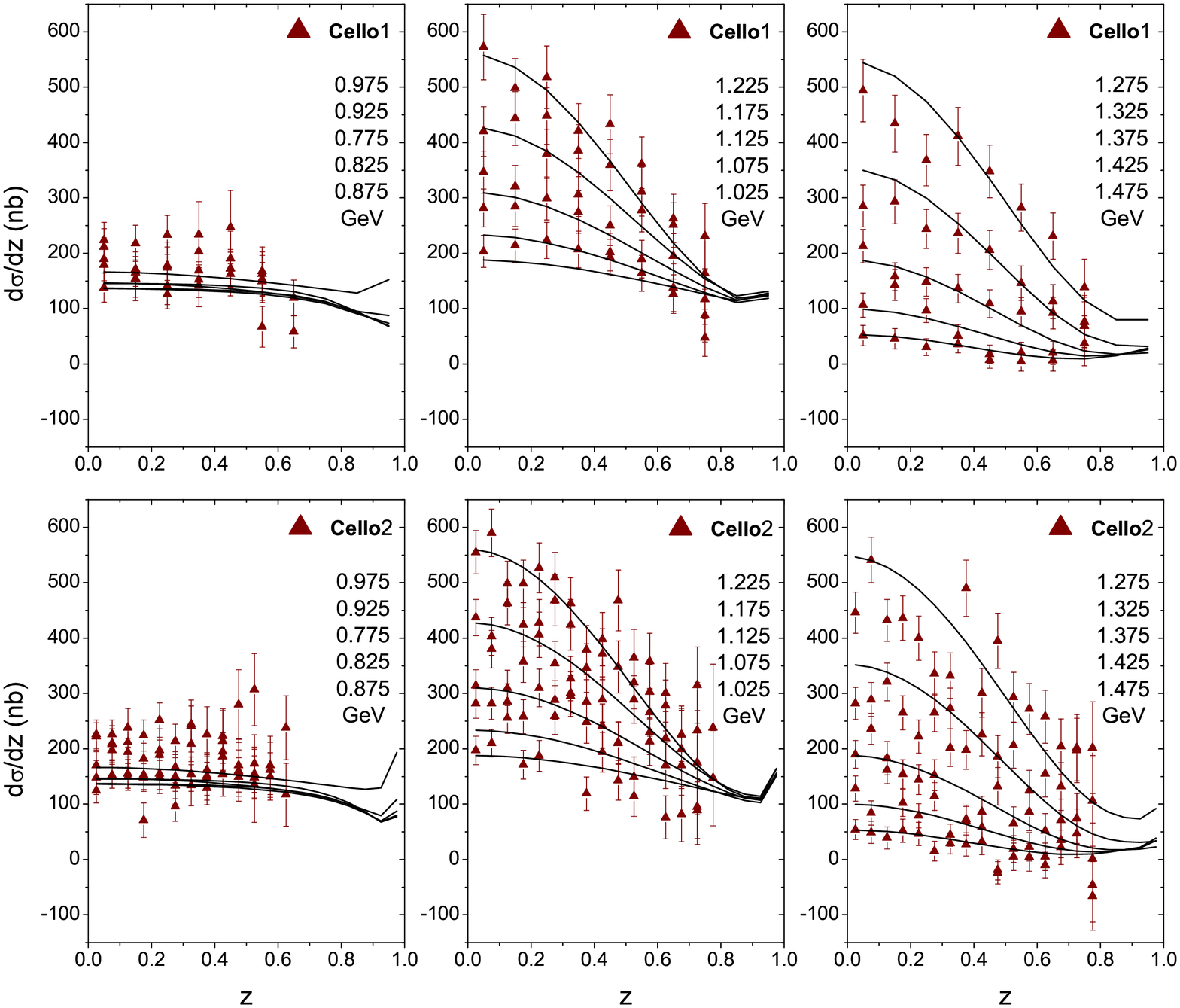}
\caption{\label{fig:dcsCello} Fit to $\gamma\gamma\to\pi^+\pi^-$ differential cross-section of CELLO experiment.
Here Cello1 is from Harjes~\cite{Cello1} and Cello2 from Behrend {\it et al.}~\cite{Cello2}.
The numbers give the central energy in GeV of each angular distribution listed in order of the cross-section at $z=0$, where $z\,=\,\cos\,\theta$.}
\end{figure}

\begin{figure}[htbp]
\vspace*{-2cm}
\includegraphics[width=1.0\textwidth]{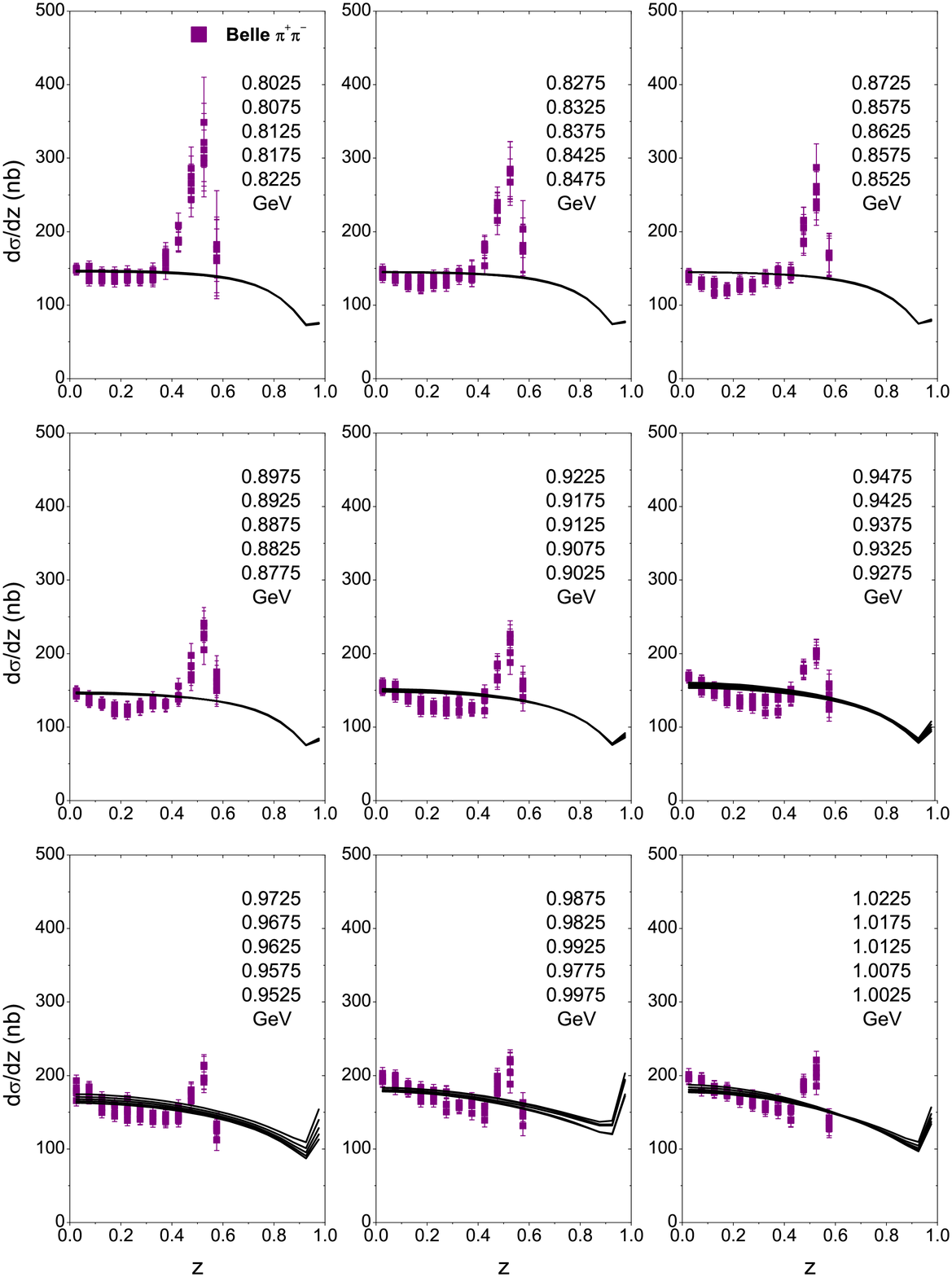}
\caption{\label{fig:dcsBelle1}Fit to $\gamma\gamma\to\pi^+\pi^-$ differential cross-section of Belle experiment~\cite{Belle-pm}.
The numbers give the central energy in GeV of each angular distribution listed in order of the cross-section at $z=0$, where $z\,=\,\cos\,\theta$.
The data are normalized so that the integrated cross-section is just a sum of the differential cross-sections in each angular bin.}
\end{figure}

\begin{figure}[htbp]
\vspace*{-2cm}
\includegraphics[width=1.0\textwidth]{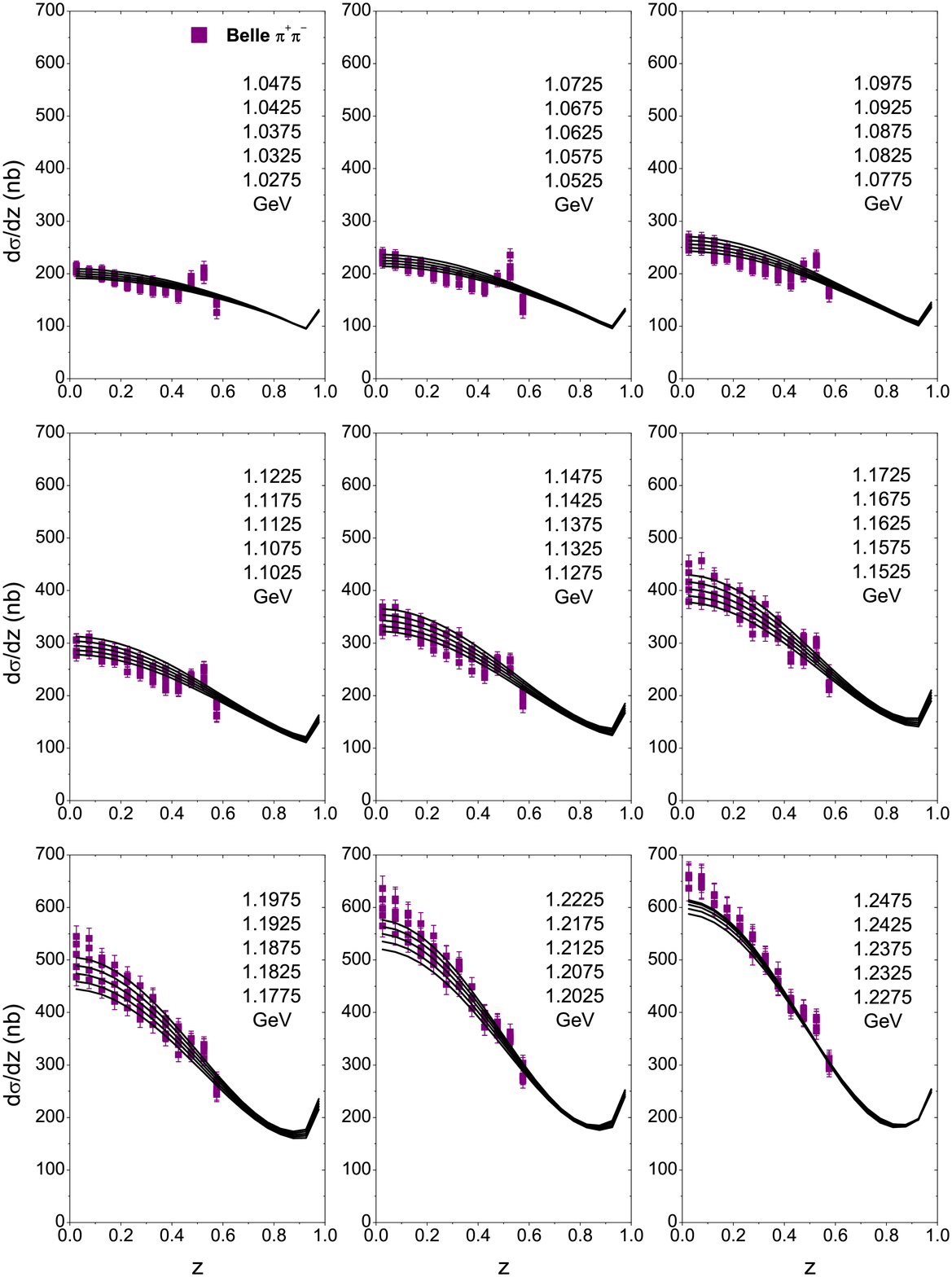}
\caption{\label{fig:dcsBelle2}Fit to $\gamma\gamma\to\pi^+\pi^-$ differential cross-section of Belle experiment~\cite{Belle-pm}.
The numbers give the central energy in GeV of each angular distribution listed in order of the cross-section at $z=0$, where $z\,=\,\cos\,\theta$. The data are normalized so that the integrated cross-section is just a sum of the differential cross-sections in each angular bin.}
\end{figure}

\begin{figure}[htbp]
\vspace*{-2cm}
\includegraphics[width=1.0\textwidth]{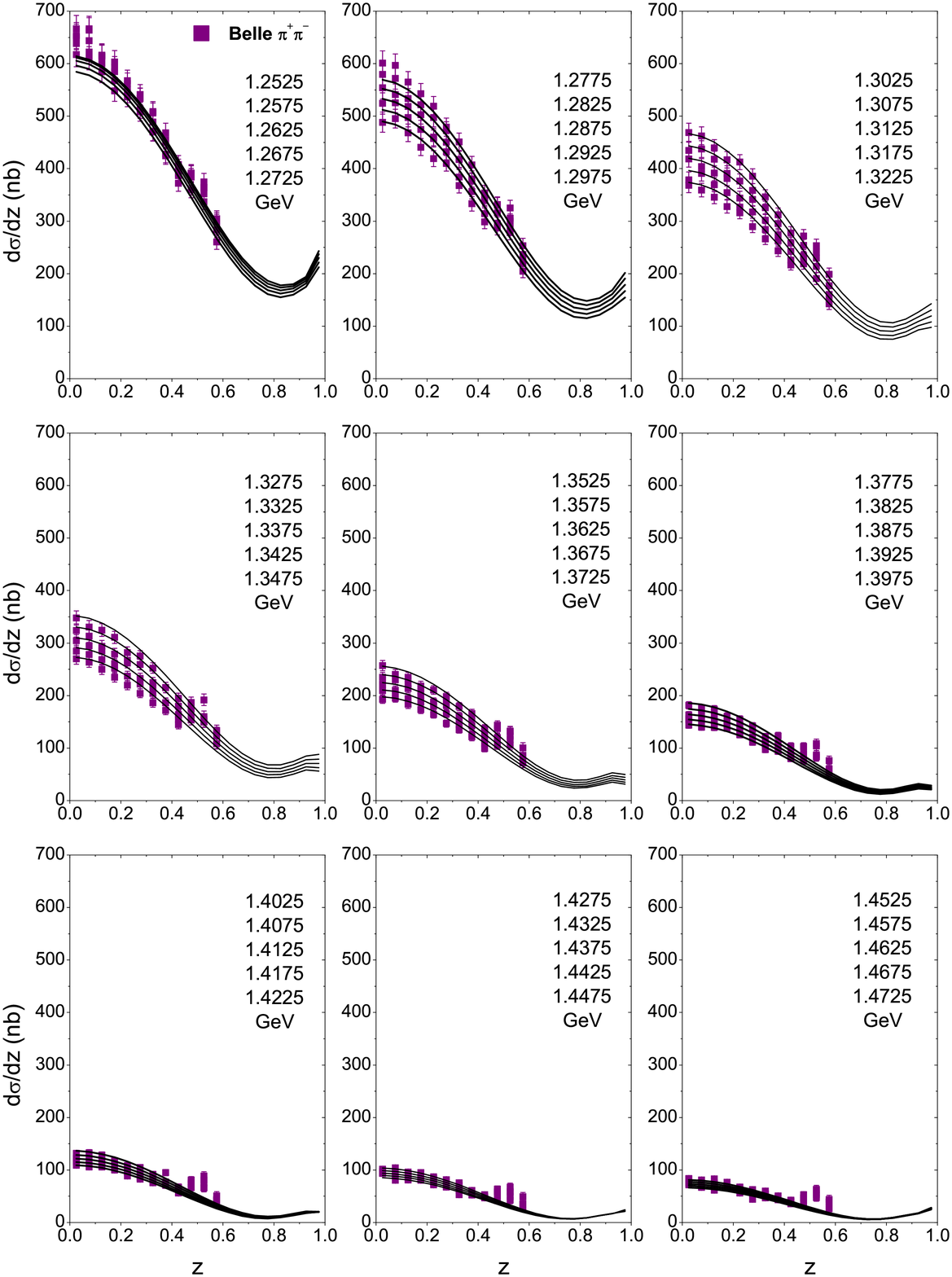}
\caption{\label{fig:dcsBelle3}Fit to $\gamma\gamma\to\pi^+\pi^-$ differential cross-section of Belle experiment~\cite{Belle-pm}.
The numbers give the central energy in GeV of each angular distribution listed in order of the cross-section at $z=0$, where $z\,=\,\cos\,\theta$.
The data are normalized so that the integrated cross-section is just a sum of the differential cross-sections in each angular bin.}
\end{figure}

\noindent We see that while overall the charged pion data are well described, there are zones of discrepancy. These are most apparent in the Belle data because of  their smaller statistical uncertainty. In the integrated cross-section we see that from 800 to 900 MeV, the trend of the Belle results is not well captured. This may be related to imperfections in the separation of the $\pi^+\pi^-$ component from the far larger $\mu^+\mu^-$ production --- a wholly QED process. This also appears to affect the angular distributions in this same mass domain seen in the strange upward blip of the distribution around $\cos \theta \simeq 0.5$. These are the main contributors to the average $\chi^2$ per datapoint of 2.2, as can be seen from Table~\ref{tab:fit}. These structures, which are difficult to reconcile with anything in the $\gamma\gamma\to\pi^+\pi^-$ channel, explain why our amplitudes give cross-sections below the Belle data in the energy region 0.8-0.9 GeV, as shown in the enlarged plots in Fig.~\ref{fig:csf980}.
This makes the Belle results on $\pi^0\pi^0$ production, which have no such contamination, particularly important. We also see, from Figs.~\ref{fig:Sigmacs}, \ref{fig:cspic}, that the Belle $\pi^+\pi^-$ cross-section is not well reproduced around 1.2-1.25~GeV, even allowing for a systematic normalization uncertainty of 12\% --- the fit being on the low side.
This amounts to a corresponding under-shooting of the differential cross-section in this mass range seen in the lower three plots of Fig.~\ref{fig:dcsBelle2}.
However, the fits to the other charged data from Mark~II and Cello are better, Figs.~\ref{fig:cspic}-\ref{fig:dcsCello}.

\noindent The integrated cross-sections for $\gamma\gamma\rightarrow\pi^0\pi^0$ and how well they are fitted is shown in Fig.~\ref{fig:cspin}.
Among these datasets only CB88~\cite{CB88} has data below 700 MeV. Consequently,  as discussed above, we have increased the weight of these low energy data in the fitting.
\begin{figure}[bhtp]
\vspace*{0.2cm}
\includegraphics[width=0.5\textwidth,height=0.35\textheight]{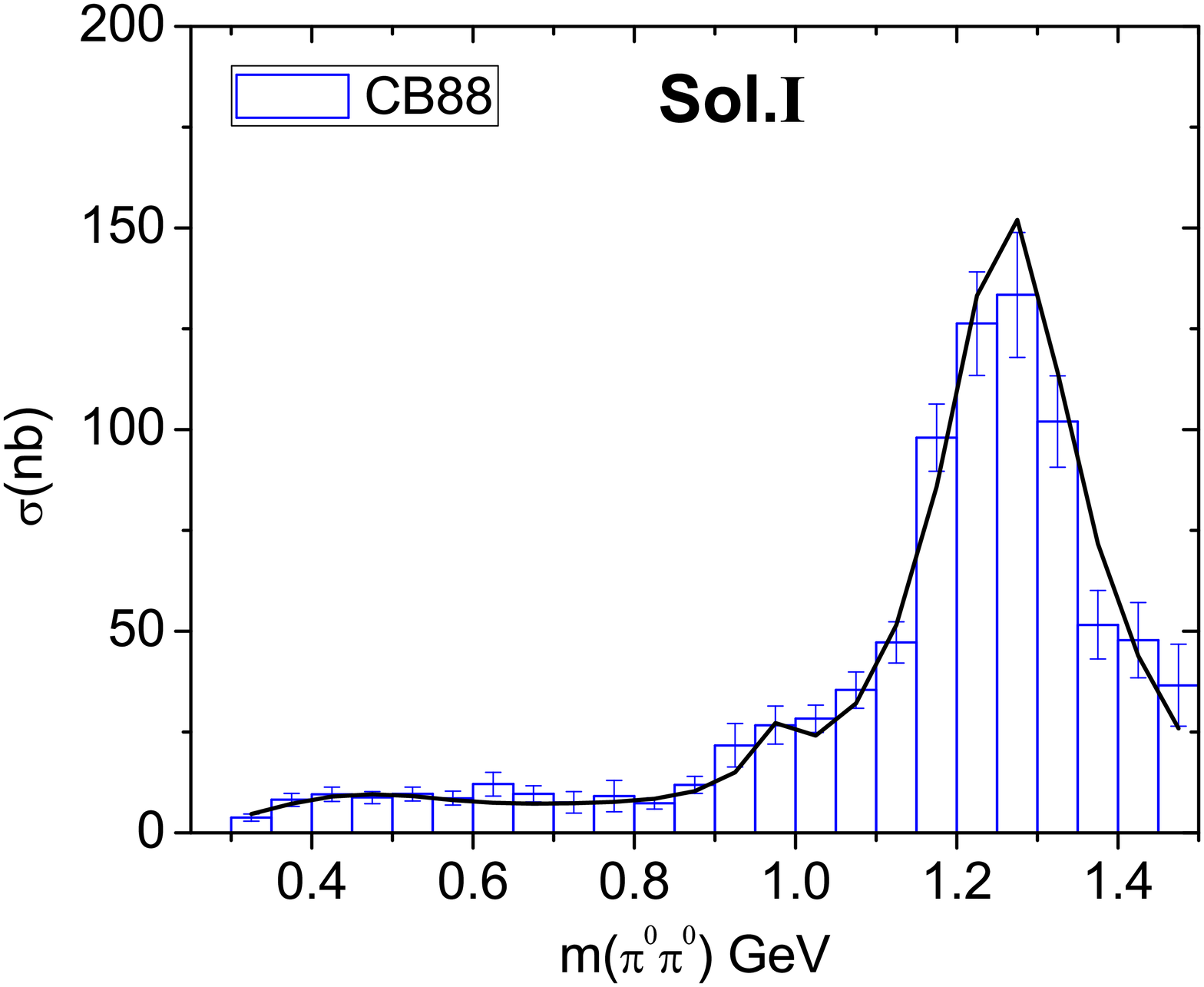}
\includegraphics[width=0.5\textwidth,height=0.35\textheight]{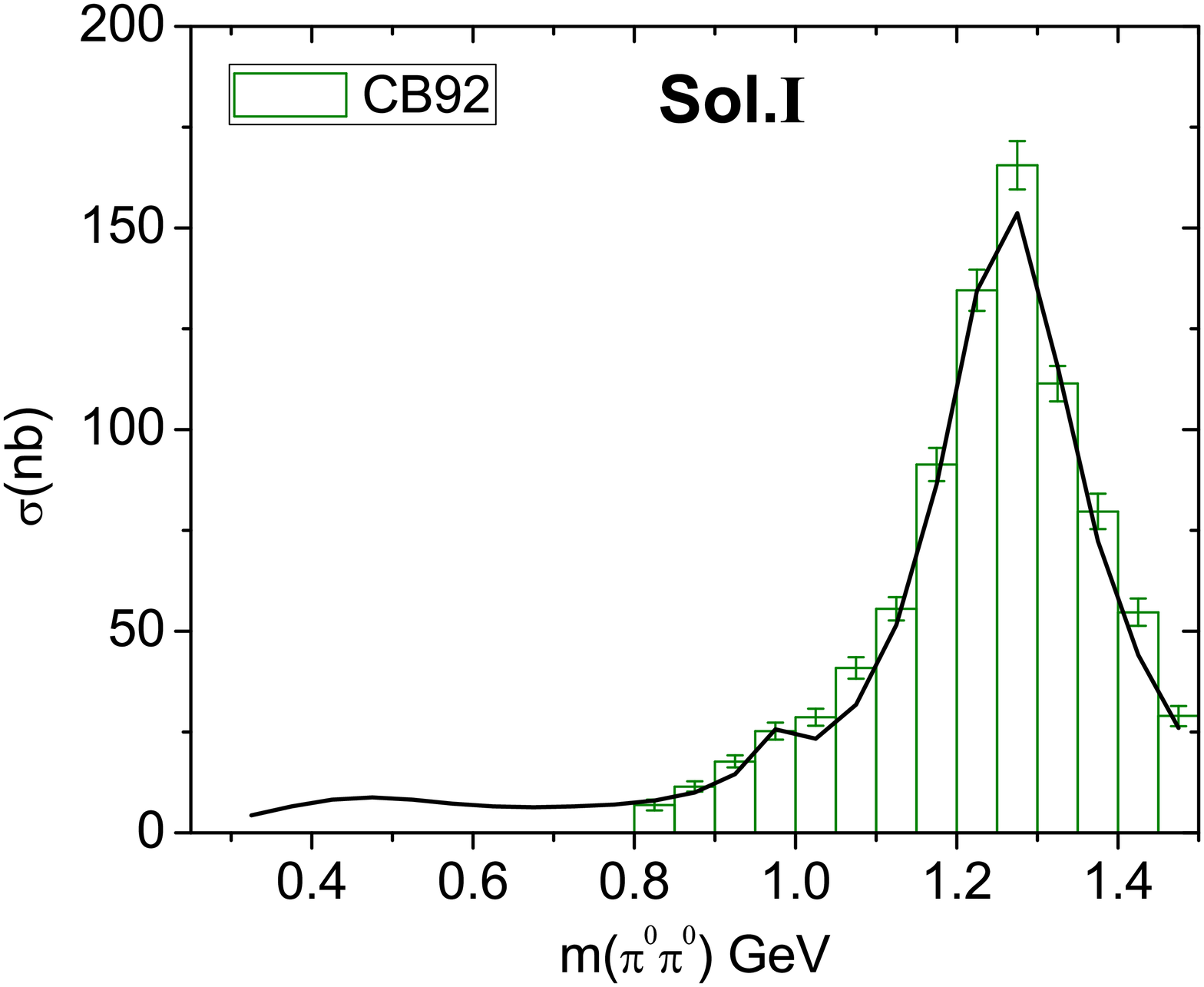}
\begin{center}
\includegraphics[width=0.5\textwidth,height=0.35\textheight]{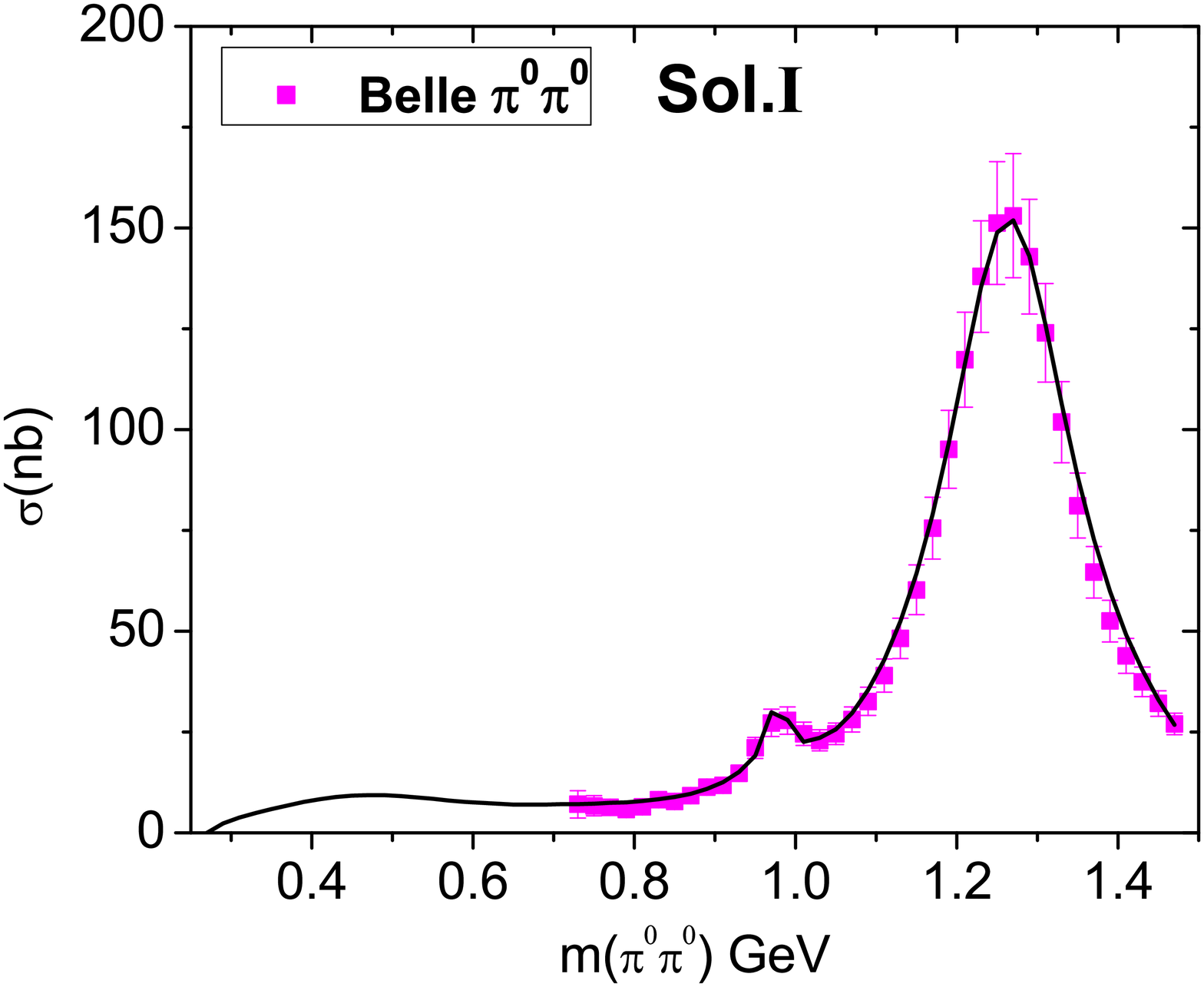}
\end{center}
\caption{\label{fig:cspin} Solution~I compared with $\gamma\gamma\rightarrow\pi^0\pi^0$ datasets. The CB88~\cite{CB88} and Belle~\cite{Belle-nn} data
are integrated over $|\cos \theta|\,\le\,0.8$, while the CB92 data~\cite{CB92} with increased statistics cover $|\cos \theta|\,\le\,0.7$.}
\end{figure}
The differential cross-section plots for $\gamma\gamma\rightarrow\pi^0\pi^0$ are shown in Figs.~\ref{fig:dcsCB},~\ref{fig:dcsBelle00},
which fit the data from  Crystal Ball and Belle, respectively.
The $\chi^2$'s are given in Table~\ref{tab:fit}.

\begin{figure}[ht]
\includegraphics[width=1.0\textwidth,height=0.4\textheight]{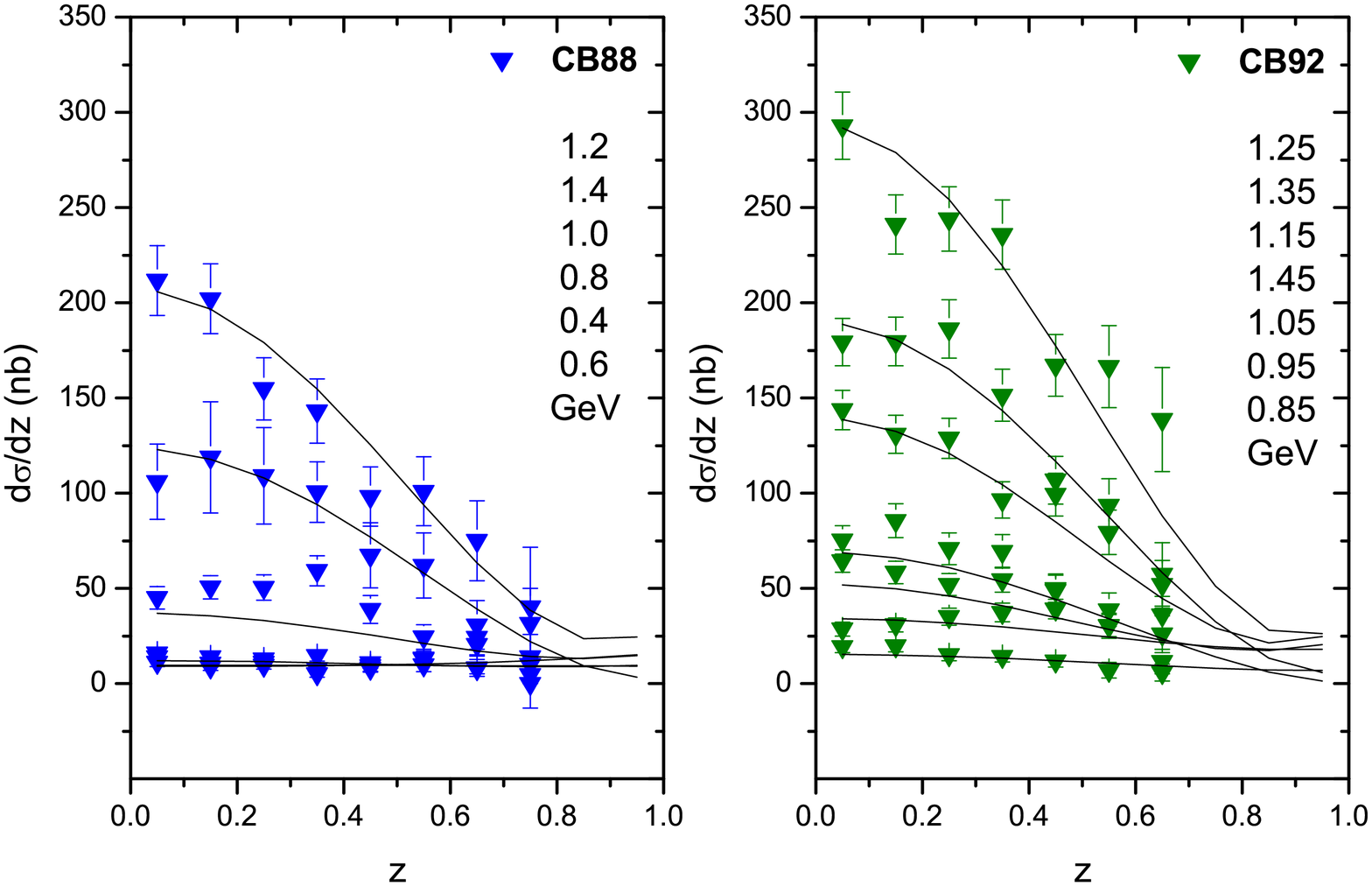}
\caption{\label{fig:dcsCB}Fit to $\gamma\gamma\to\pi^0\pi^0$ differential cross-section from the Crystal Ball experiment.
Here CB88 is from Marsiske {\it et al.}~\cite{CB88} and CB92 from Bienlein {\it et al.}~\cite{CB92}.
The numbers give the central energy in GeV of each angular distribution listed in order of the cross-section at $z=0$, where $z\,=\,\cos\,\theta$.
The data are normalized so that the integrated cross-section is just a sum of the differential cross-sections in each angular bin.}
\end{figure}

\begin{figure}[htbp]
\vspace*{-2cm}
\includegraphics[width=1.0\textwidth]{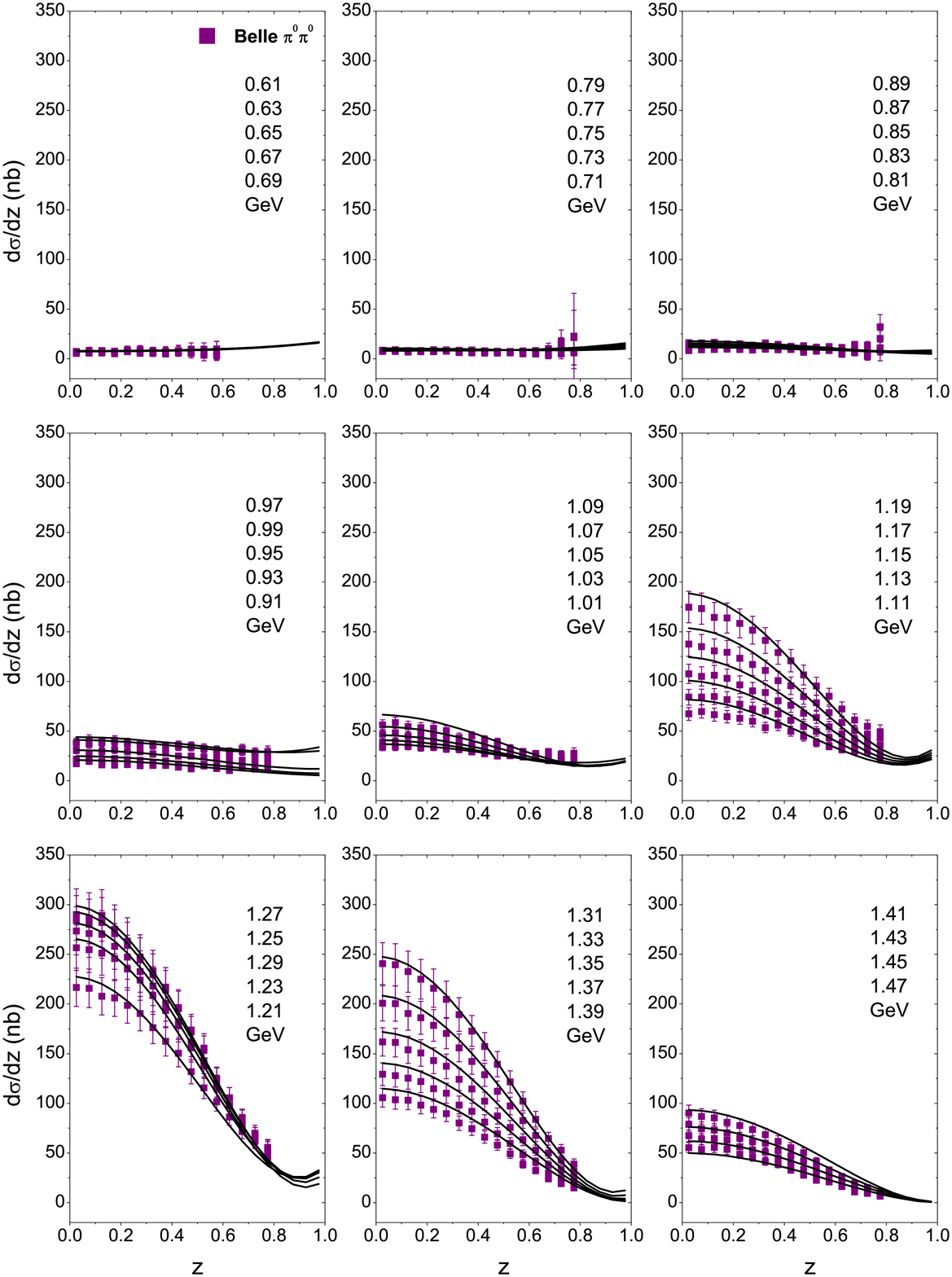}
\caption{\label{fig:dcsBelle00}Fit to $\gamma\gamma\to\pi^0\pi^0$ differential cross-section of the Belle experiment~\cite{Belle-nn}.
The numbers give the central energy in GeV of each angular distribution listed in order of the cross-section at $z=0$, where $z\,=\,\cos\,\theta$.
The data are normalized so that the integrated cross-section is just a sum of the differential cross-sections in each angular bin.}
\end{figure}
\begin{figure}[htbp]
\vspace*{-0cm}
\includegraphics[width=0.5\textwidth,height=0.32\textheight]{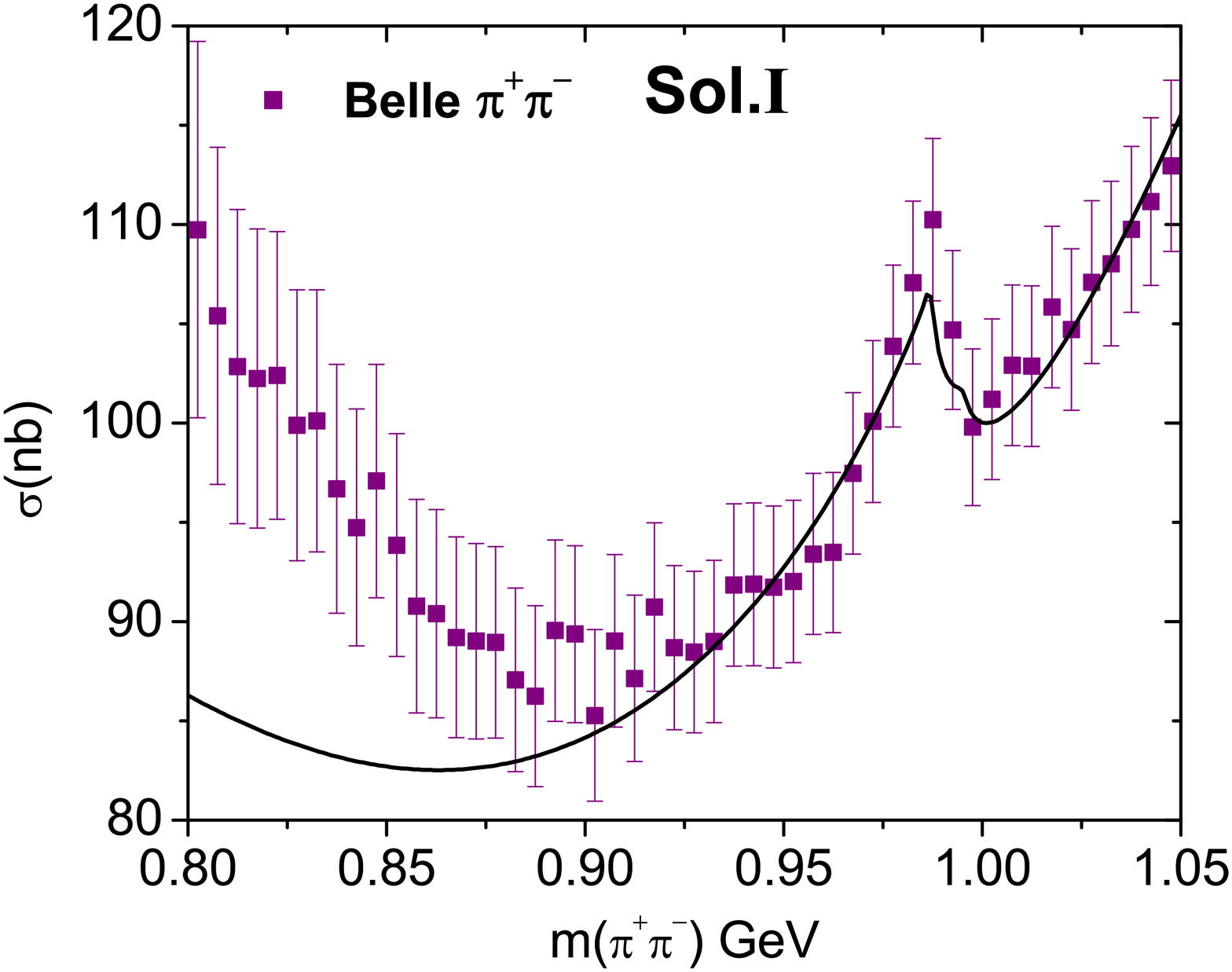}
\includegraphics[width=0.5\textwidth,height=0.32\textheight]{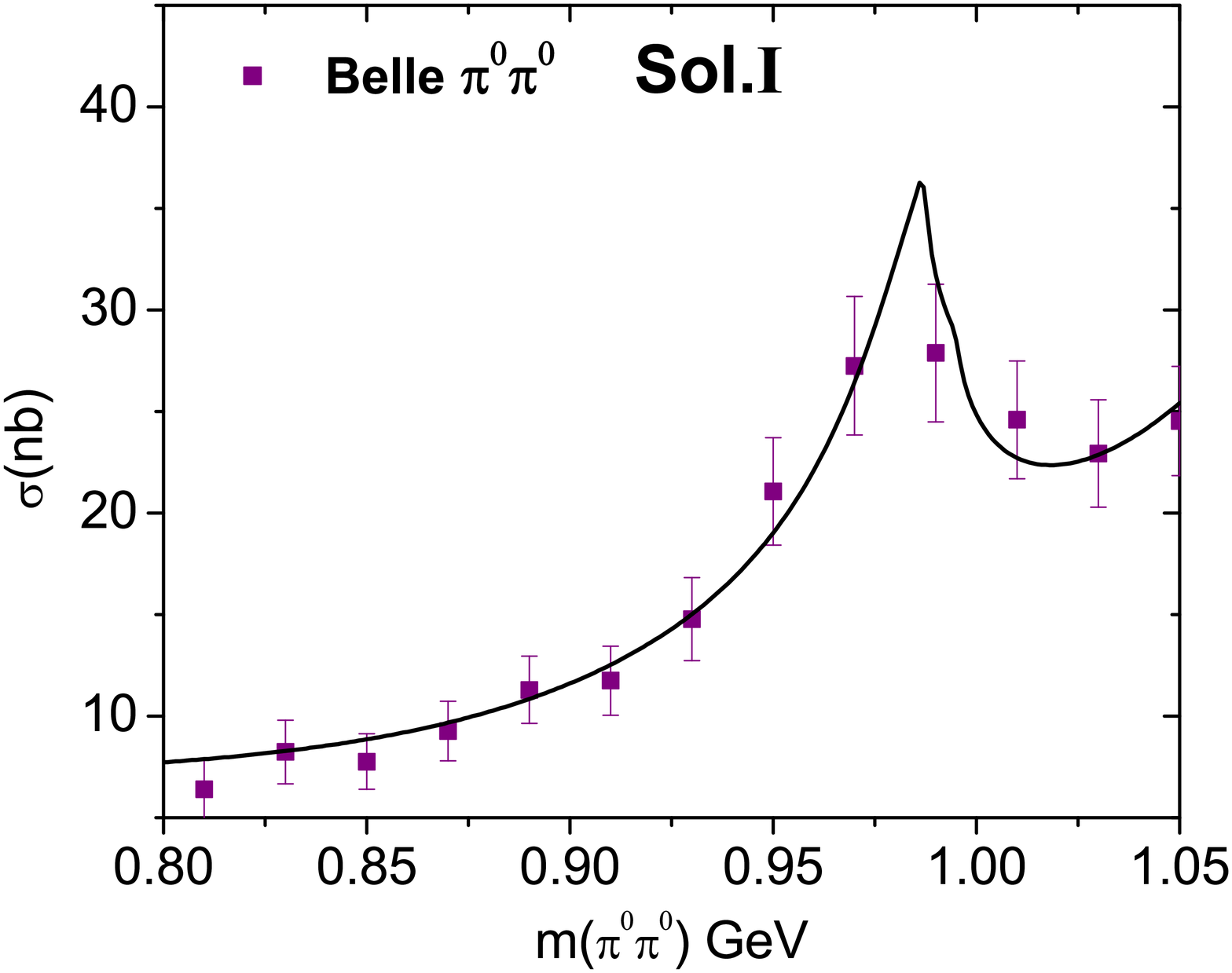}
\caption{\label{fig:csf980} Solution I compared with Belle results on $\gamma\gamma\rightarrow\pi^+\pi^-$~\cite{Belle-pm}, and $\pi^0\pi^0$~\cite{Belle-nn} in the energy region 0.8-1.05~GeV. The charged pion data is integrated over $|\cos \theta|\,\le\,0.6$ and the neutral one is integrated over $|\cos \theta|\,\le\,0.8$. This misfit of the Belle charged pion data below 0.9~GeV is perhaps due to $\mu^+\mu^-$ contamination, as discussed in the text. This is not seen in the Mark~II fits, Fig.~\ref{fig:dcsMark}. }
\end{figure}
\begin{figure}[htb]
\includegraphics[width=1.0\textwidth,height=0.40\textheight]{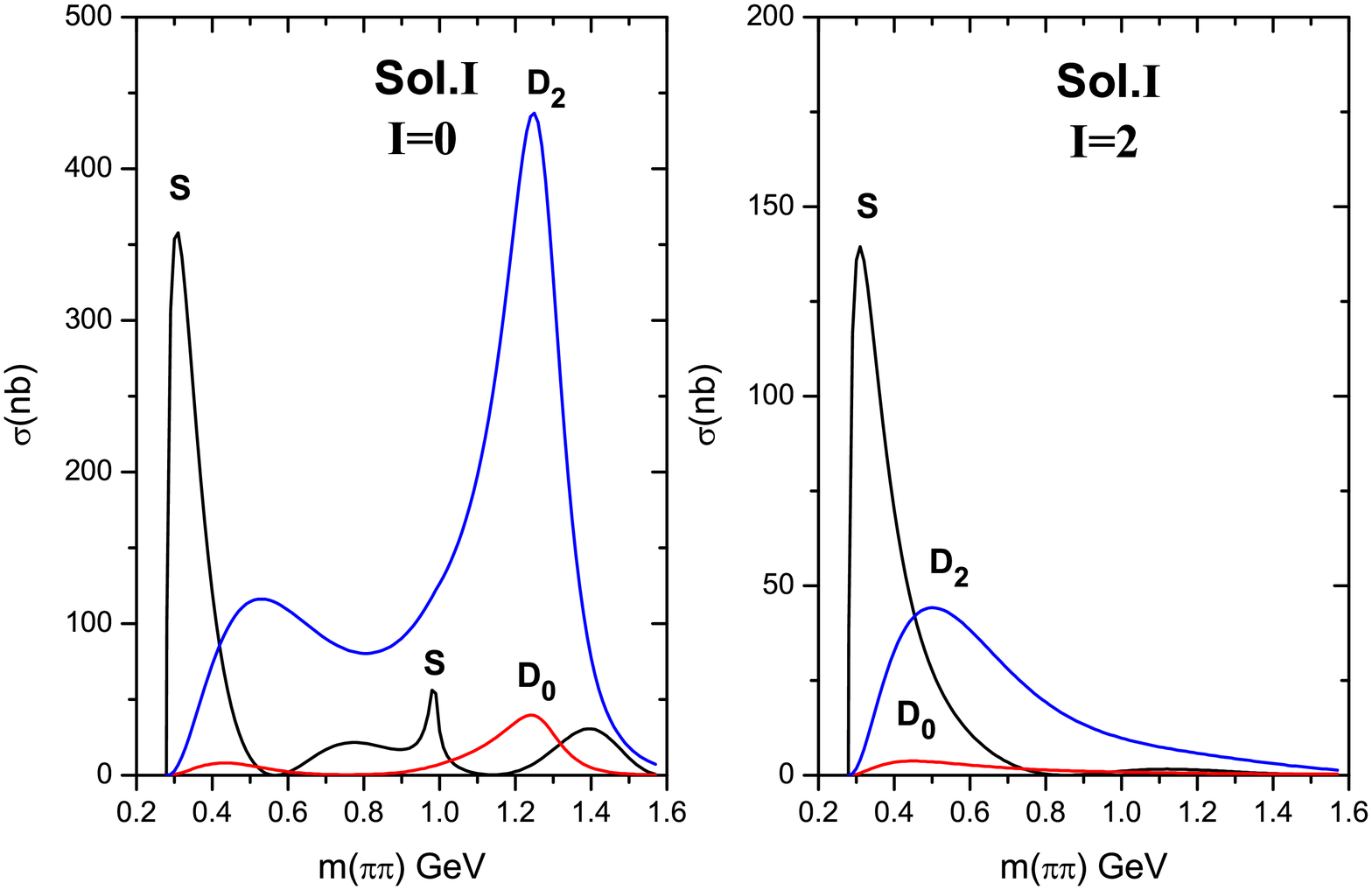}
\caption{\label{fig:csiso;pi} Individual partial wave components of the $\gamma\gamma\to\pi\pi$ integrated cross-section. }
\end{figure}
\begin{figure}[hp]
\centering
\includegraphics[width=0.78\textwidth,height=0.46\textheight]{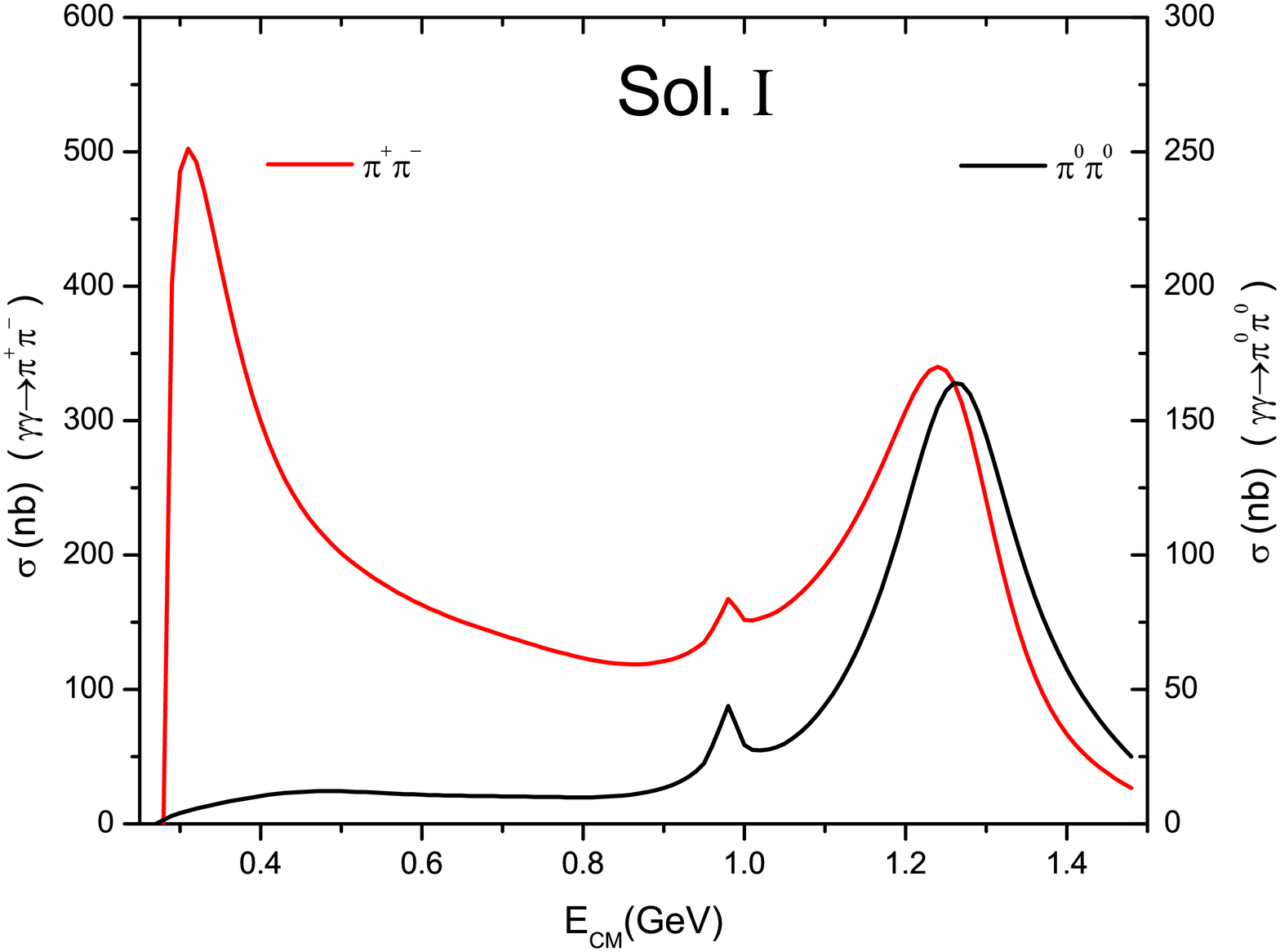}
\caption{\label{fig:cspredict}The cross-sections for $\gamma\gamma\rightarrow\pi^+\pi^-$ (with the scale on the left) and $\pi^0\pi^0$ (scale on the right) predicted by our Solution~I for the full angular range are shown. If the processes were pure $I=0$, the cross-sections would be equal everywhere.}
\vspace{4mm}
\centering
\includegraphics[width=0.6\textwidth,height=0.3\textheight]{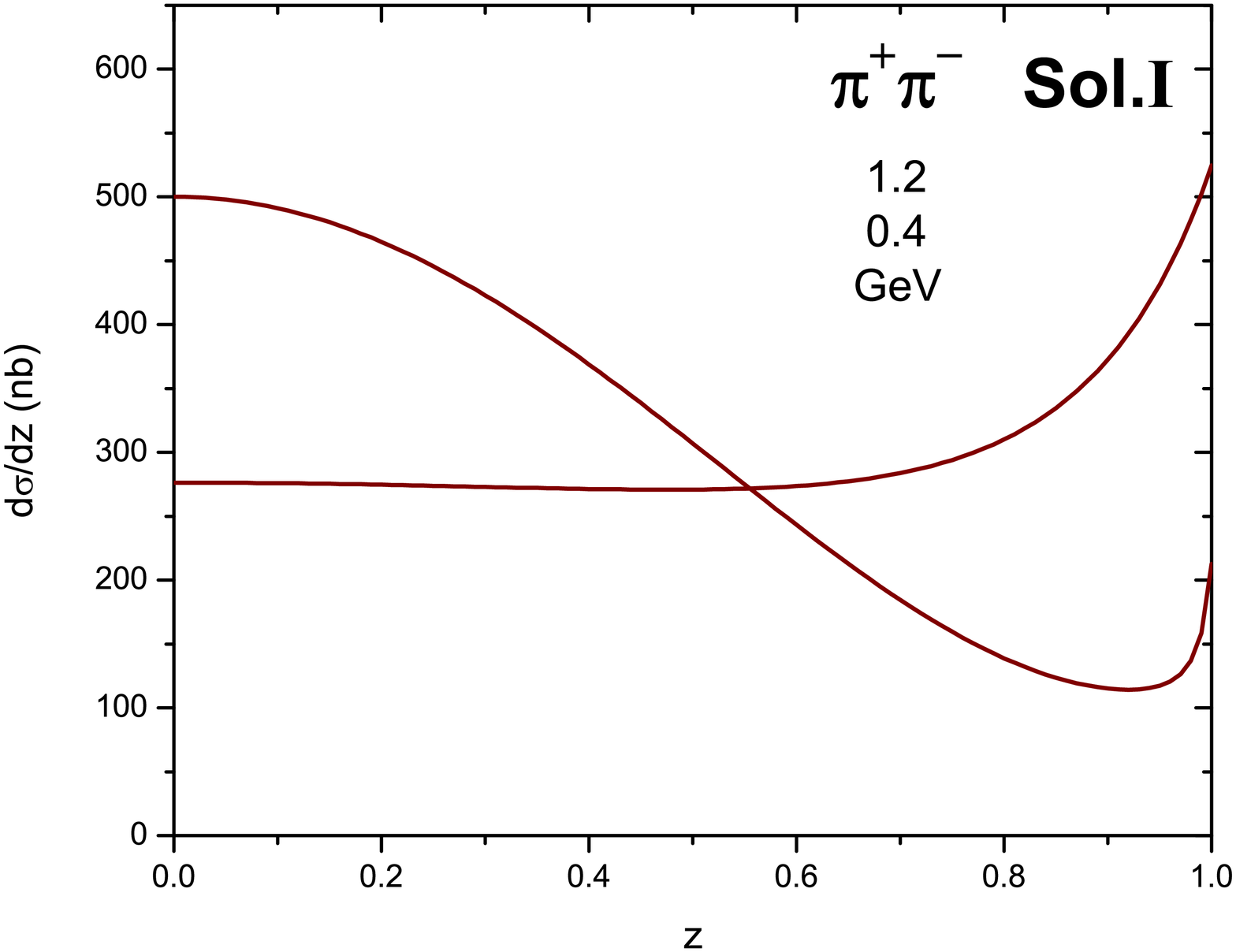}
\caption{\label{fig:dcspredict}The full angular distribution for $\gamma\gamma\rightarrow\pi^+\pi^-$ predicted by our Solution~I at 0.4 and 1.2~GeV. These energies are listed in order of the cross-section at $z=0$, where $z\,=\,\cos\,\theta$. While the fits to the angular distributions  in the preceding plots are integrated over each bin in $\cos \theta$, here we show the fits as continuous functions of angle to illustrate the effect of the pion poles at $|\cos \theta|\,>\,1$. These give the sweep up at the end of the angular range, even though this is not evident in data with limited angular coverage.}
\end{figure}

\noindent Having data of more comparable precision for  both $\gamma\gamma\to \pi ^{+} \pi^{-}$ and $\gamma\gamma\to \pi ^{0} \pi^{0}$ makes the separation of $I=0$ and $I=2$ components considerably more reliable. This is important for determining what of structures seen  around 950 MeV is the $f_0(980)$ and what is background, and so better constrain the isoscalar component. The angular distributions likewise help in separating the $S$ and $D$ waves that strongly interfere through the same mass region up to 1.4 GeV. We notice that there is a clear \lq peak' in the region of 0.95-1.025~GeV. In Fig.~\ref{fig:csf980}, we show a blow-up of this region for both the Belle charged and neutral pion cross-sections. We see that our Solution~I provides an adequate description of these data. However, as already remarked the $\pi^+\pi^-$ data below 900~MeV have a strange behavior, more easily seen in Fig.~\ref{fig:cspic}. This is likely an issue of the incomplete removal of the large $\mu^+\mu^-$ signal in this mass range. In Fig.~\ref{fig:csiso;pi} we plot the integrated cross section of each partial wave amplitude. This is one of our main result. It is a key input into future dispersive calculations of light-by-light scattering, as well as the basis for
determination of resonance two photon couplings as we discuss later in Sect.~\ref{sec:3}.

\begin{table}[hp]
\begin{center}
\begin{tabular}{||c|l|c|c|c|c||}
\hline \hline
\multicolumn{6}{||c||}{\rule[-0.4cm]{0cm}{10mm} SOLUTION I\parbox{1.2cm}{~~~~} $\chi ^2 _{{\rm tot}} = 2.17$ } \\
\hline
\rule[-0.5cm]{0cm}{10mm}
Experiment & {\hspace{3.5mm}}Process    & data-points & $\chi ^2_{{\rm average}}$ & $\chi^2_{\,{\rm Int. X-sect.}}$  & $\chi^2_{\,{\rm Ang. distrib.}}$ \\
\hline \hline
\rule[-0.5cm]{0cm}{10mm}
Mark II    & $\,\gamma\gamma\to\pi^+\pi^-$ & 144  & 1.55        & 1.50             & 1.61                   \\
\hline
\rule[-0.5cm]{0cm}{10mm}
Crystal Ball   & $\,\gamma\gamma\to\pi^0\pi^0$ & 126  & 1.63        &1.88              & 1.53                   \\
\hline
\rule[-1.1cm]{0cm}{22mm}
CELLO      & $\,\gamma\gamma\to\pi^+\pi^-$ & 333  & 1.87        &1.03              &\parbox{2.5cm}{1.41 \protect \\
                                                                                              from~Harjes \protect \\
                                                                                                     2.23 \protect \\
                                                                                                      from Behrend}\\
\hline
\multirow{2}{*}{\rule{0cm}{1.0cm}Belle}
          &\rule[-0.5cm]{0cm}{10mm} $\,\gamma\gamma\to\pi^+\pi^-$
                                         & 1664 &2.85         & 1.16             & 3.00                        \\
\cline{2-6}
          &\rule[-0.5cm]{0cm}{10mm} $\,\gamma\gamma\to\pi^0\pi^0$
                                         & 684  & 1.19        & 0.43             & 1.24                        \\
\hline
\rule[-0.5cm]{0cm}{10mm}
TPC/Argus/Belle      & $\,\gamma\gamma\to K^+K^-$   & 18   & 2.78        & 2.78     & $-$                   \\
\hline
\rule[-0.5cm]{0cm}{10mm}
TASSO/CELLO      & $\,\gamma\gamma\to {\overline K}^0K^0$   & 5   & 1.77        & 1.77              & $-$                   \\

\hline
\rule[-0.5cm]{0cm}{10mm}
Belle            & $\,\gamma\gamma\to {\overline K}_sK_s$   & 315 & 1.03        & 0.73              & 1.13                   \\

\hline\hline
\end{tabular}
\vspace{0.5cm}
\caption{\label{tab:fit}~~Summary of contributions to the $\chi^2$ from each experiment for our Solution I.
Here $\chi^2 _{\rm tot}$ is calculated as follows:
we sum $\chi^2$ of all datasets, and divide it by the total number of data-points we are fitting, namely 2951.
$\chi^2 _{\rm average}$ is computed in the same way, but for each dataset separately. The number in the bracket of
the line for $\gamma\gamma\to K^+K^-$ means that we do not take into account the Belle's data, which lies above 1.4~GeV. }
\end{center}
\end{table}
\noindent In Fig.\ref{fig:cspredict}, we give the prediction
for the integrated cross-section  from our Solution~I.
We display the charged and neutral pion cross-sections on scales that differ by a factor of two.
If the processes were pure $I=0$, the curves for $\pi^+\pi^-$ and $\pi^0\pi^0$ would be on top of each other. In the lower energy region the effect of a sizeable $I=2$ component is obviously apparent. However this is not so in the region of the $f_2(1270)$, where a factor of two does indeed approximate their magnitudes. Nevertheless, their shapes are different, with the $f_2$-peak shifted, reflecting the different mix of $I=2$ amplitudes.
Finally in Fig.~\ref{fig:dcspredict}, we give the prediction for the differential cross-section for $\gamma\gamma\to\pi^+\pi^-$ from our solution at two energies. This is to illustrate the effect of the one pion exchange poles as $|\cos\,\theta|\, \to\, 1$, not seen in data with limited angular coverage.

\baselineskip=5.5mm
\subsection{$\gamma\gamma\to K \overline{K}\,$ fits}\label{sec:2;3}
As previously indicated we have a patch of solutions that fit all the $\gamma\gamma\to\pi\pi$ data with almost the same $\chi^2$ and very similar characteristics.
Compared to previous Amplitude Analyses, the Belle $\pi^0\pi^0$ results have limited this significantly. Nevertheless, being a coupled channel treatment,
each of these solutions makes a different prediction for the isoscalar $\gamma\gamma\to{\overline K}K$ cross-section and its energy dependence.
It is here that the older experimental data on the $K^+K^-$ and ${\overline K}^0K^0$ channels, and the much newer high statistics results on $K_sK_s$, narrow the continuum patch to essentially a single solution.
Of course, ${\overline{K}}K$ production involves an important isovector component too,
which we crudely model by a Breit-Wigner-like form for the $a_2(1320)$, as set out in Sect.~\ref{sec:1;5}.

\noindent From the $\gamma\gamma\to \pi\pi$ amplitudes fitted in the last section, the isoscalar $S$-wave of $\gamma\gamma\to K \overline{K}$ has automatically been fixed according to Eq.~(\ref{eq:unitarity}), meanwhile isoscalar $D$-waves are fixed by Eq.~(\ref{eq:Fk;0D}). What we need to know are the isovector waves, and they are parameterized by Eqs.~(\ref{eq:Fk;1S}, \ref{eq:Fk;1D}). With these we obtain the fit in Fig.~\ref{fig:csK}.
\begin{figure}[htbp]
\vspace*{-2cm}
\centering
\includegraphics[width=0.48\textwidth,height=0.30\textheight]{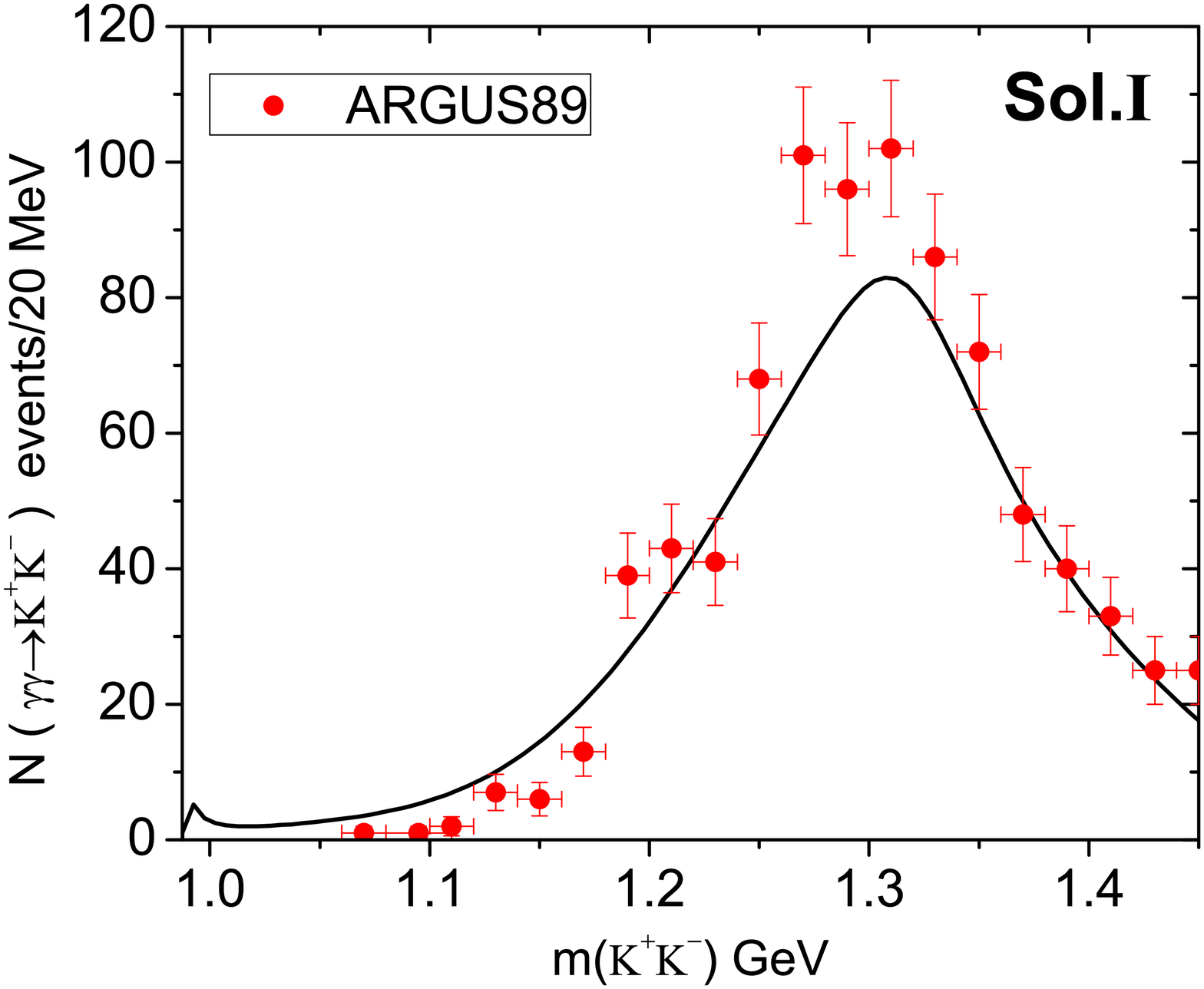}
\includegraphics[width=0.48\textwidth,height=0.30\textheight]{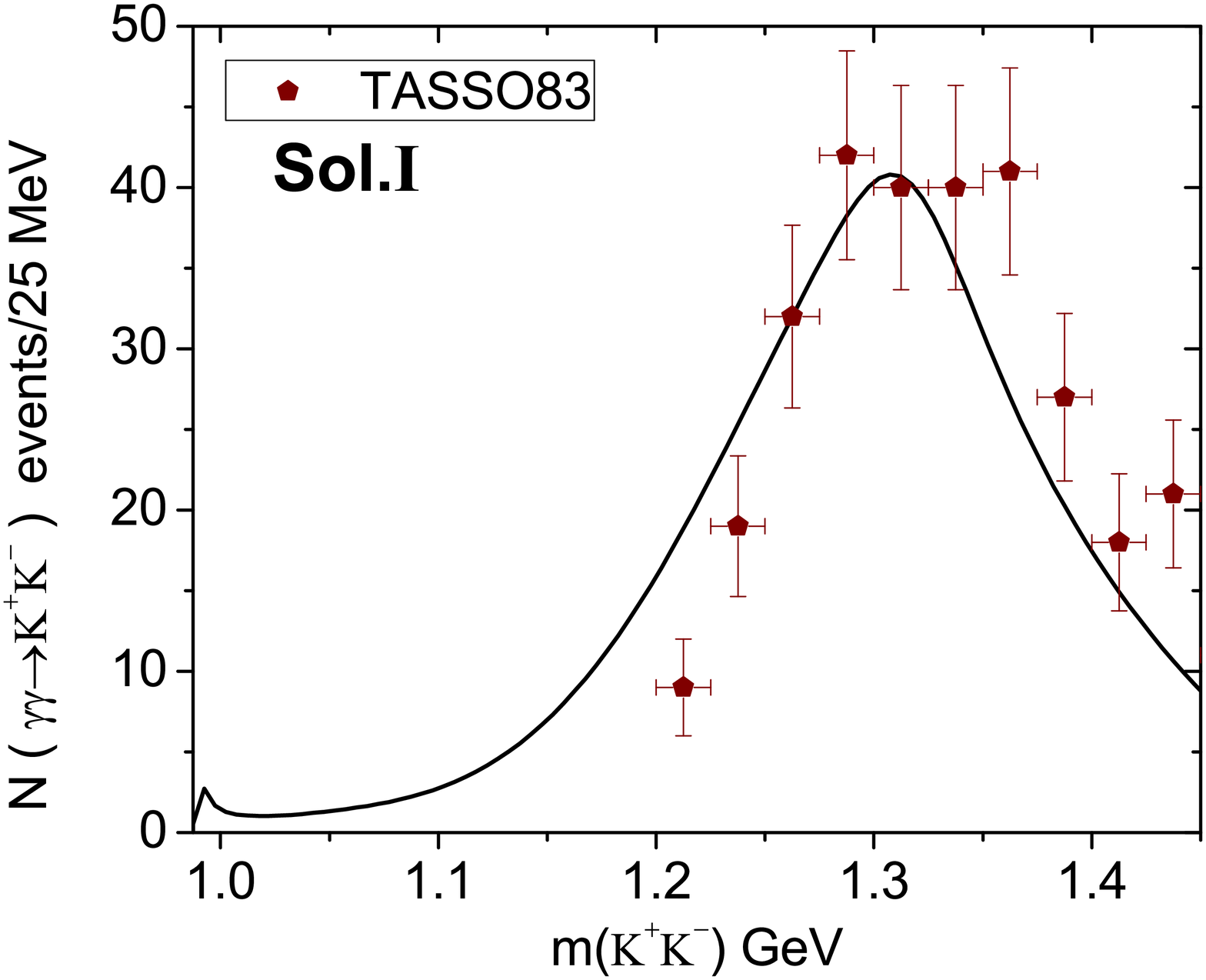}
\includegraphics[width=0.48\textwidth,height=0.30\textheight]{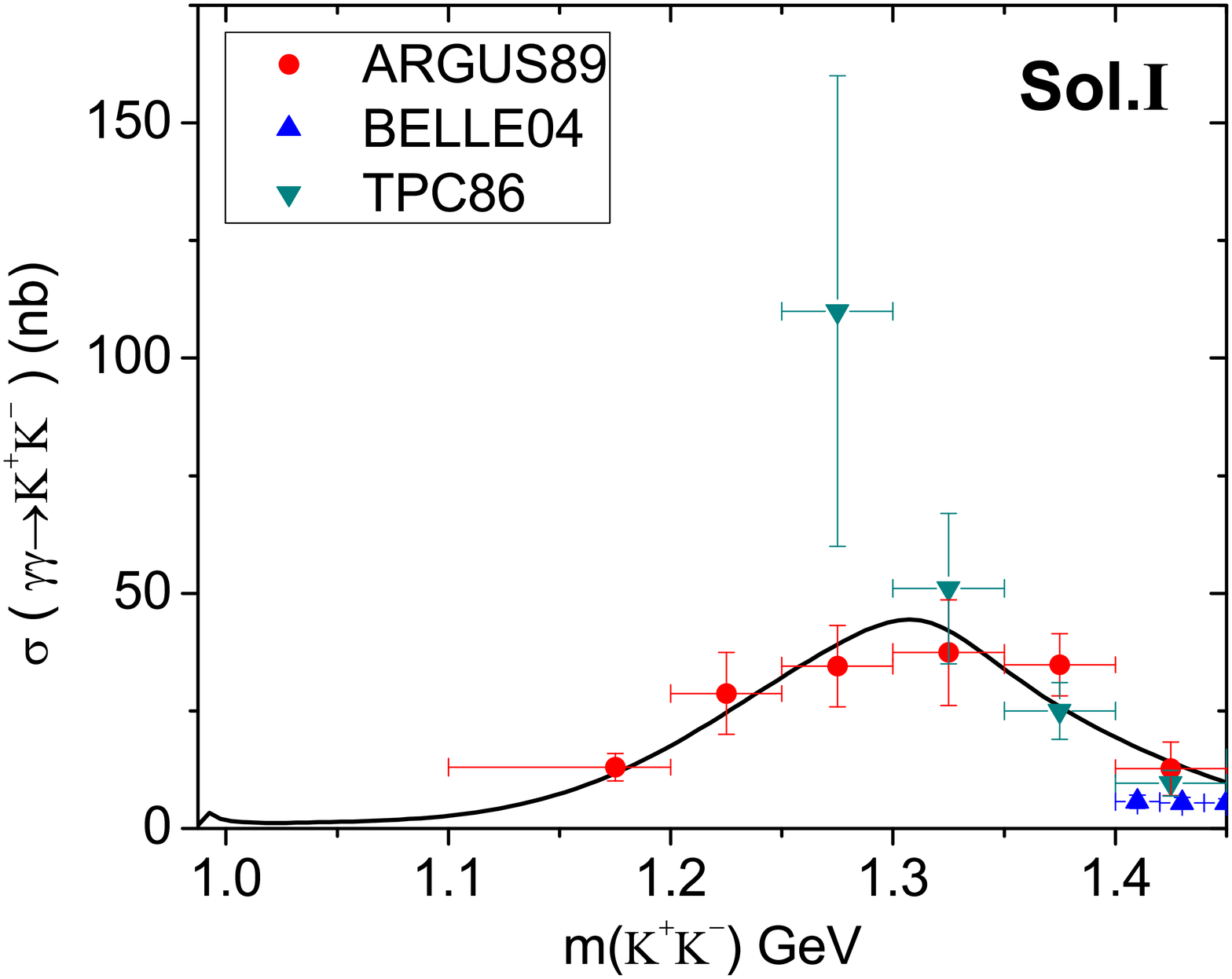}
\includegraphics[width=0.48\textwidth,height=0.30\textheight]{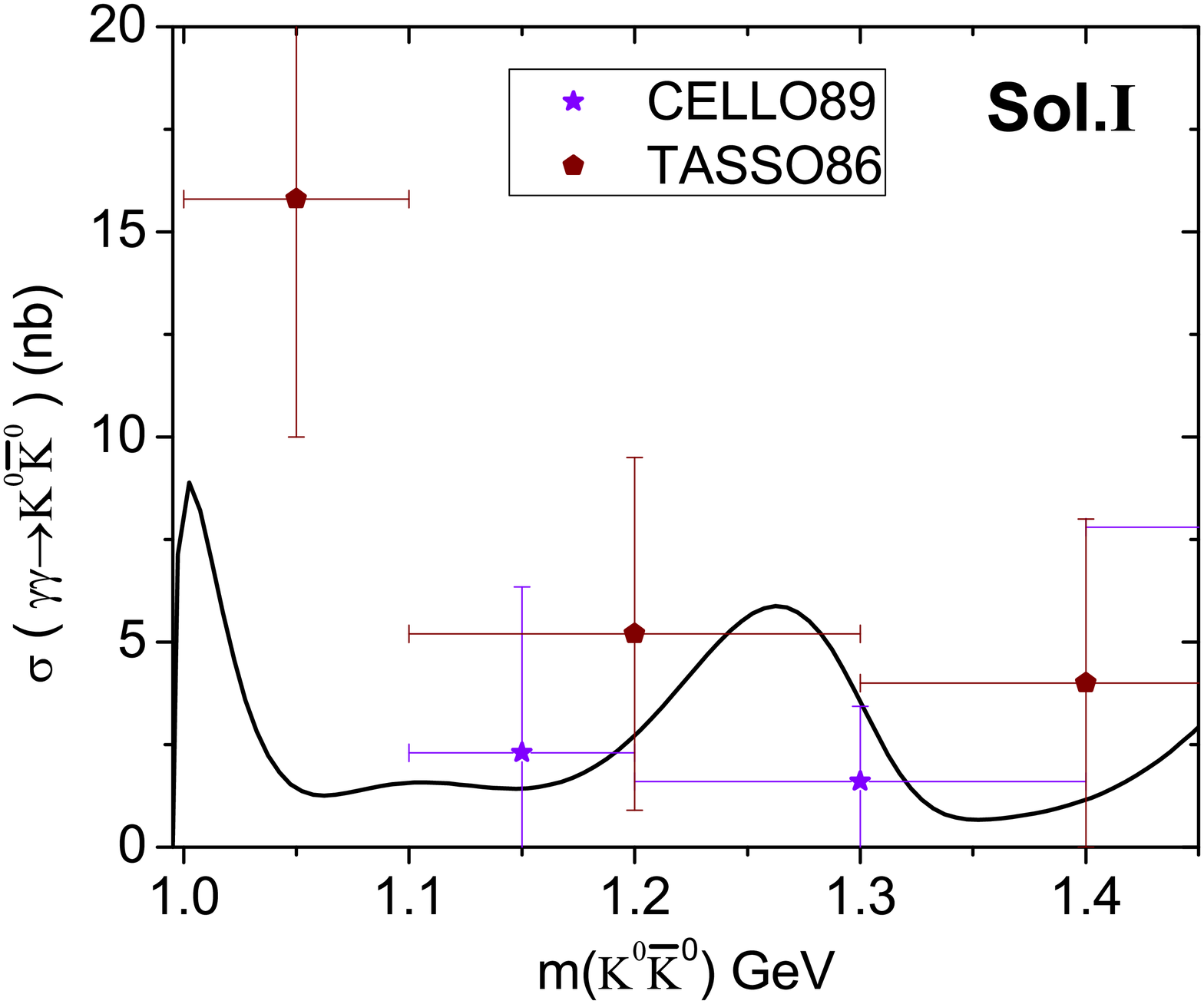}
\includegraphics[width=0.48\textwidth,height=0.35\textheight]{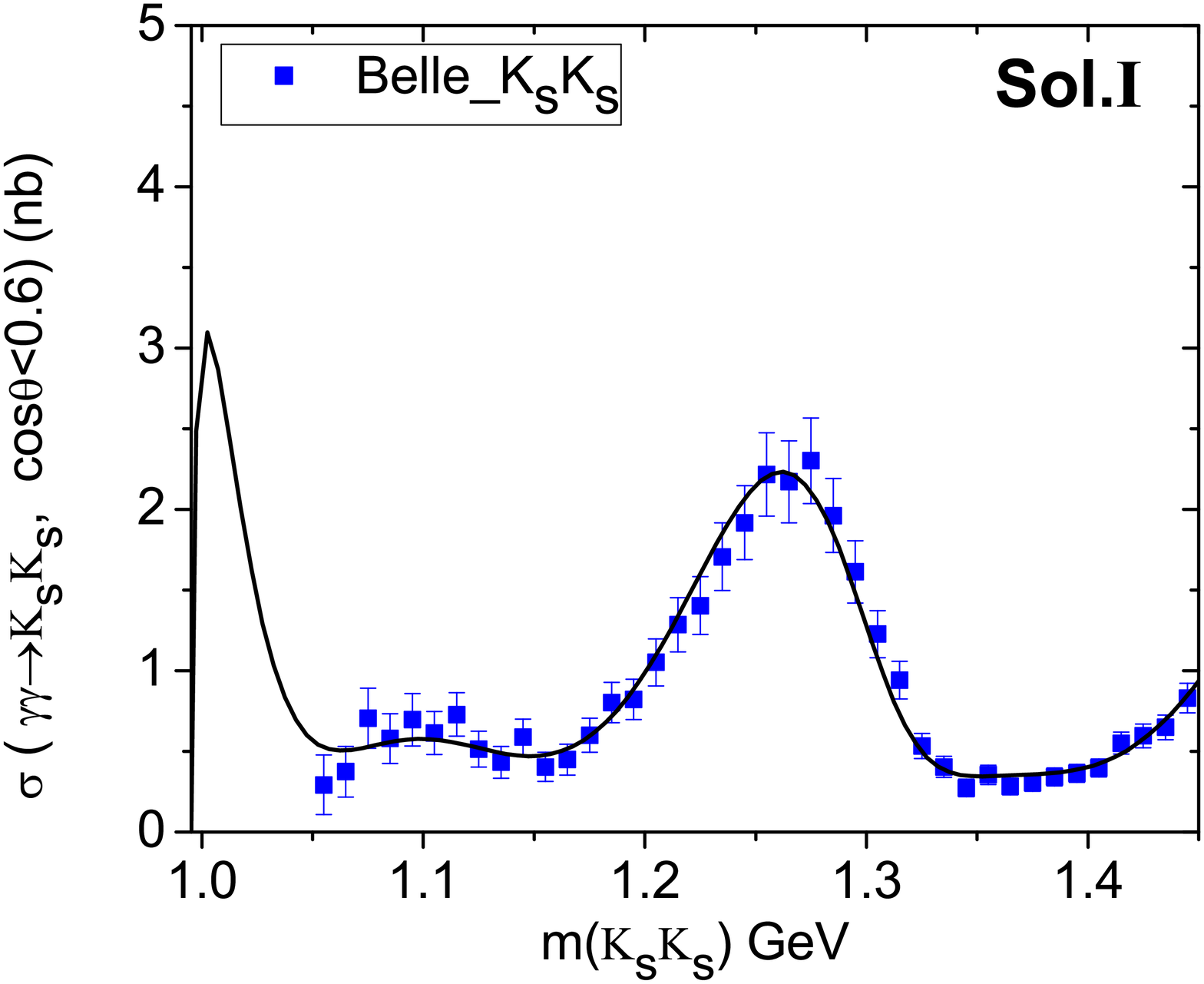}
\includegraphics[width=0.48\textwidth,height=0.35\textheight]{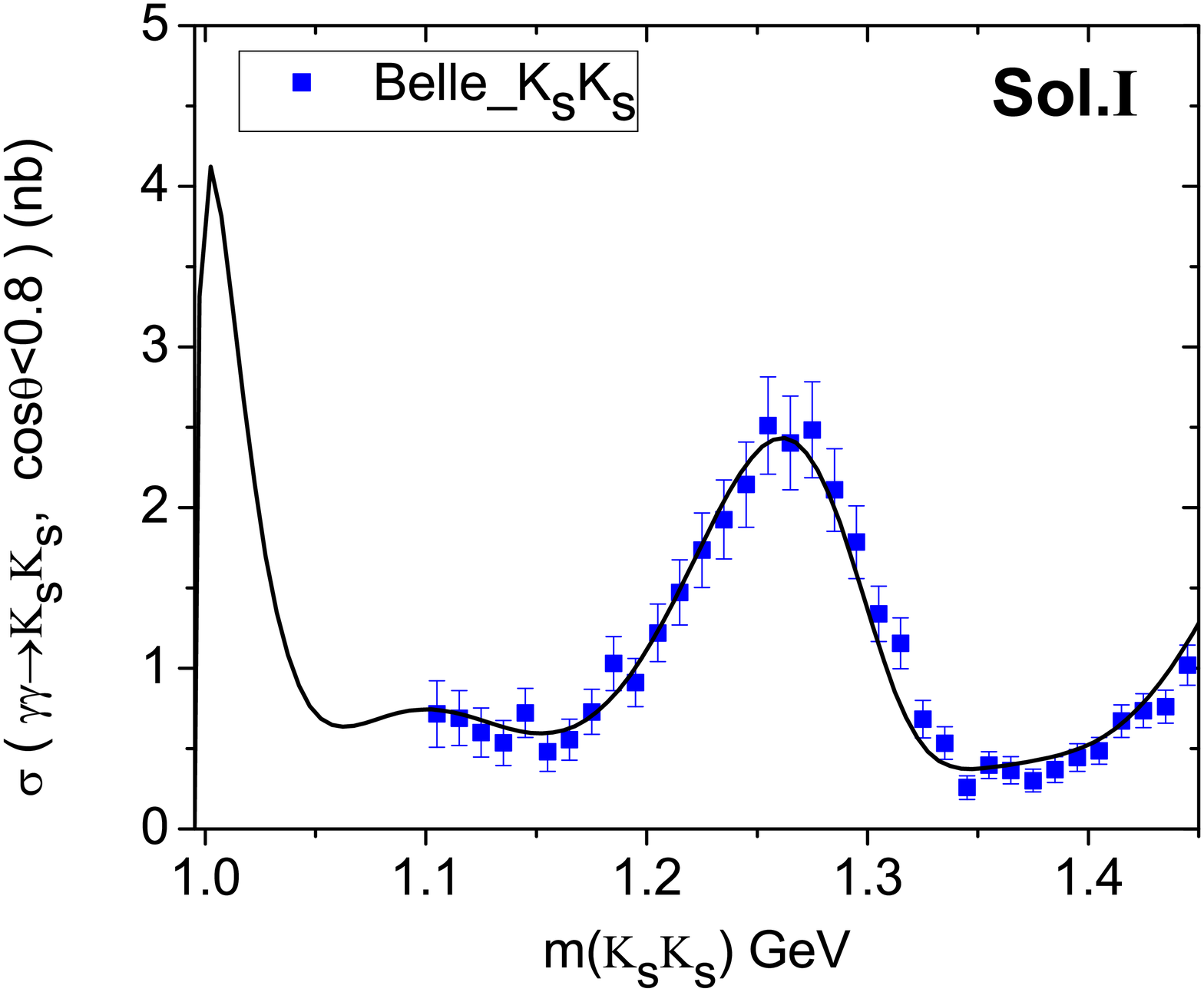}
\caption{\label{fig:csK} Solution I  compared with the $\gamma\gamma\rightarrow K \overline{K}$ data sets.
For $\gamma\gamma\rightarrow K^+ K^-$ process, ARGUS cross-section data~\cite{ARGUS89} are integrated over
$\cos \theta$. For their event distribution, the integral is over $|\cos \theta|\,\le\,0.7$. TPC~\cite{TPC86} and Belle~\cite{Belle04} are integrated up to $|\cos \theta|\,=\,0.6$. For $\gamma\gamma\rightarrow K^0 \overline{K}^0$ process, CELLO~\cite{Cello89} and TASSO~\cite{Cello89} are
integrated up to $|\cos \theta|\,=\,0.7$, $|\cos \theta|\,=\,0.87$ separately. The Belle
$K_sK_s$ data are from~\cite{Belle-KsKs}. }
\end{figure}

\begin{figure}[htbp]
\includegraphics[width=1.0\textwidth,height=0.66\textheight]{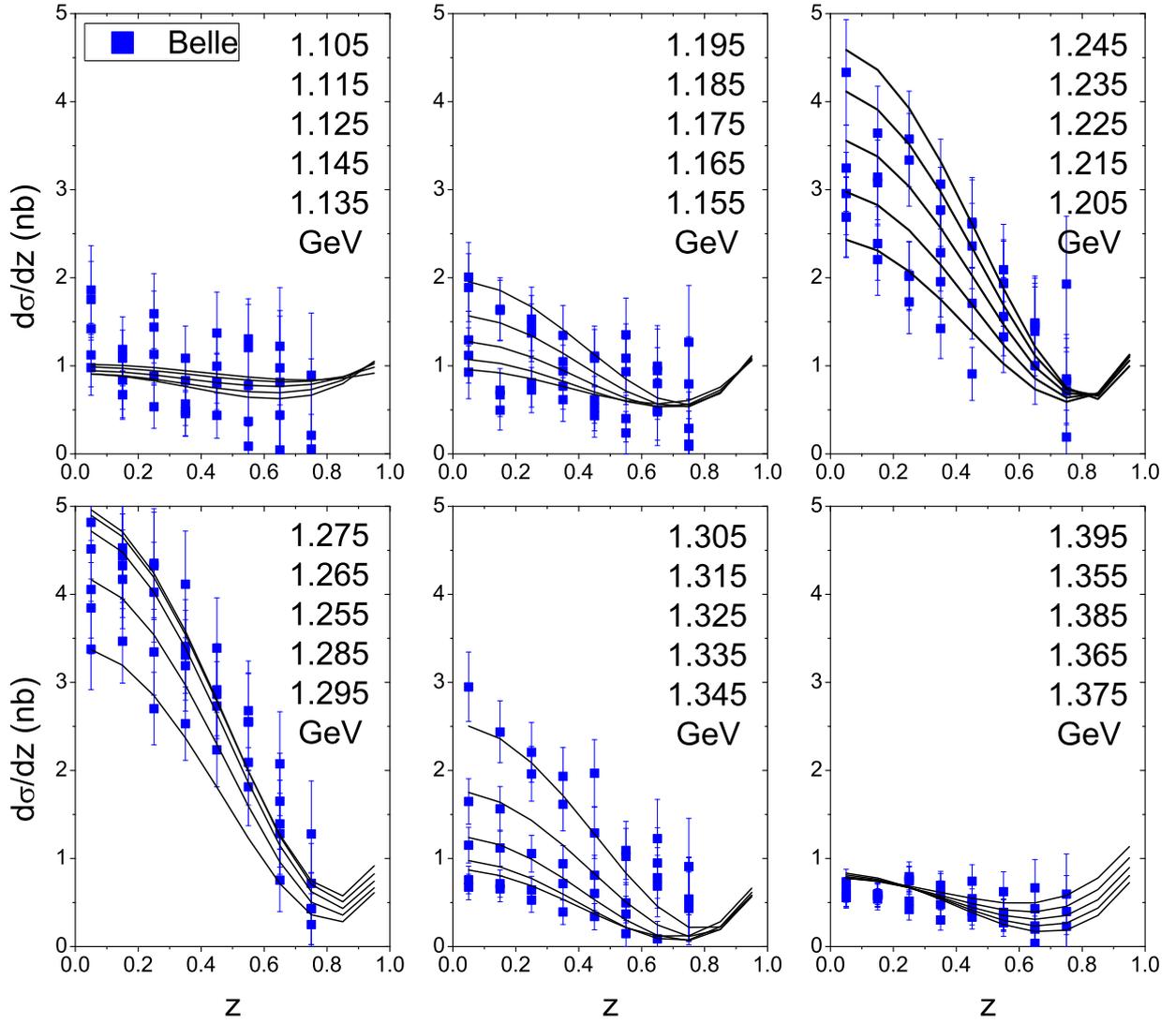}
\caption{\label{fig:csK} Solution I compared with the $\gamma\gamma\rightarrow K_s K_s$ differential cross-section from~\cite{Belle-KsKs}.}
\end{figure}

\begin{figure}[htbp]
\includegraphics[width=1.0\textwidth,height=0.35\textheight]{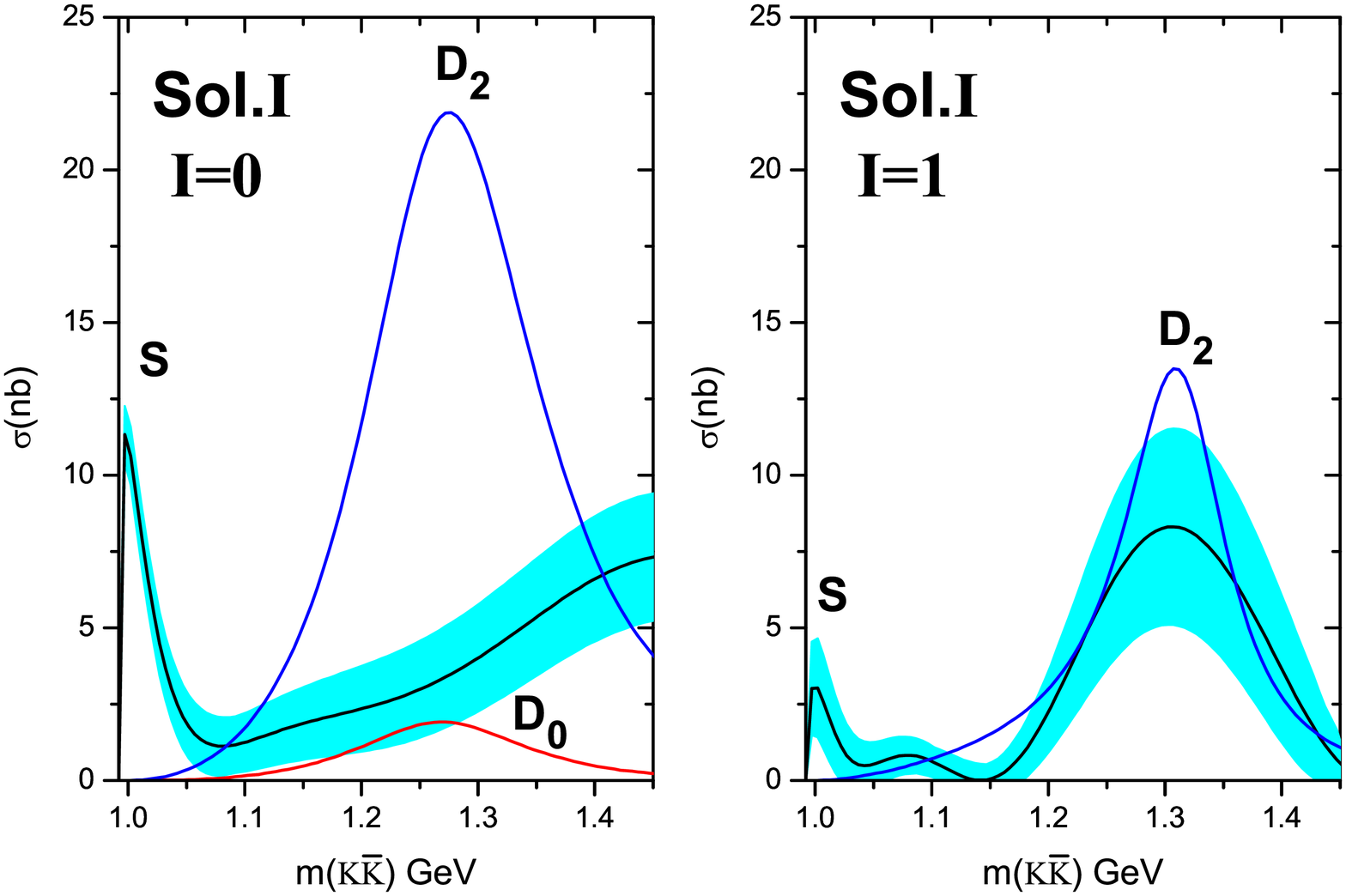}
\caption{\label{fig:csiso;K} Individual partial wave components of $\gamma\gamma\to K\overline{K}$ cross-sections. The lack of experimental data between 1.0 and 1.05~GeV contributes to the shaded (cyan) bands for the $I=0,\,1$ $S$-waves that reflect the systematic errors generated by different amplitude solutions.  }
\end{figure}

\baselineskip=5.7mm
\noindent For simplicity we only plot in Fig.~\ref{fig:csK} the $\gamma\gamma\to K^+K^-$ cross-section integrated over the whole angular range and
$\gamma\gamma\to K^0\overline{K}^0$ cross-section integrated over  $|\cos \theta|\,\le\,0.87$ , in accordance with the data from~\cite{ARGUS89,TASSO86}.
Since the ARGUS group do not give the errors for their event distribution, we include errors to make its $\chi^2$ comparable to that of other datasets.
By far the biggest constraint comes from the Belle $K_sK_s$ data. Indeed, these are the only data that cover a significant angular range ($|\cos \theta| \le 0.6$) down to 1.05~GeV.

\noindent Both the isoscalar and isovector $S$-waves are found to peak close to threshold. For $I=0$ the peak is 12~nb reflecting the appearance of the $f_0(980)$, while for $I=1$, which is to be expected of the $a_0(980)$, this is around 3~nb with a large uncertainty as indicated by the shaded band, Fig.~\ref{fig:csiso;K}. The isoscalar and isovector $S$-waves tend to cancel in the charged kaon channel rather than in the neutral kaon channel. This does not satisfy the model estimate of $\sigma_{K^0 \bar{K}^0}\,\leq\, 1$~nb calculated in ~\cite{Achasov11,Achasov12}.
The uncertainty band for the isovector $S$-wave, as shown in Fig.~\ref{fig:csiso;K}, is caused by a lack of information. A full coupled channel analysis including the $\eta\pi^0$ and multi-pion modes and scattering would reduce this. Without this here, recall we have  simply
parameterized this $S$-wave by Eq.~(\ref{eq:Fk;1S}). Of course, this uncertainty will affect the determination of $D$-waves too and this is reflected in the $\gamma\gamma$ couplings listed in Table~\ref{tab:poles}.

\newpage
\section{Two photon couplings}\label{sec:3}
Our analysis has determined the $I=0,\,2$ amplitudes for $\gamma\gamma\to\pi\pi$, and to a lesser extent the $I=0,\,1$ components for $\gamma\gamma\to {\overline K}K$. For the $\pi\pi$ channel the integrated cross-section for each partial wave is shown in Fig.~\ref{fig:csiso;pi}. For spin $\ge 4$ these are assumed to be given by the Born amplitude.

We see, as expected, the prominent peak for the $f_2(1270)$ in the $D$-waves, most obviously with helicity two. In the $S$-wave we see a smaller structure associated with the  $f_0(980)$, and at low energies masked by the peak from the Born amplitude the effects of the $\sigma/f_0(500)$. For the idealized case where a state is  well-described by a Breit-Wigner form with no background, its two photon width can be inferred from the height of the resonance peak through
\be\label{eq:Grr}
\Gamma(R\to\gamma\gamma)= \frac {\sigma_{\gamma\gamma}(\text{res.peak})~M_R^2~\Gamma_{tot}}{8\pi(\hbar c)^2(2J+1)~\text{BR}}\;.
\ee
In reality the states here are broad, overlapping with each other or with strongly coupled thresholds, and so the model-independent coupling of a resonance is given only by the residue of its pole. Knowing the underlying amplitudes we can determine these couplings to two photons as we discuss in Sect.~\ref{sec:3;2}.

\newpage

\subsection{Argand Plots}\label{sec:3;1}
An important outcome of our Amplitude Analysis is the behavior of the partial wave amplitudes, as complex functions of energy. For all the waves, but the $S$-wave, these are very simple and show no surprises. However, to learn about the structure of scalars, it is helpful to trace the variation of $S$-wave amplitudes from the Argand plots in Fig.~\ref{fig:argand}.
\begin{figure}[htbp]
\includegraphics[width=1.0\textwidth,height=0.36\textheight]{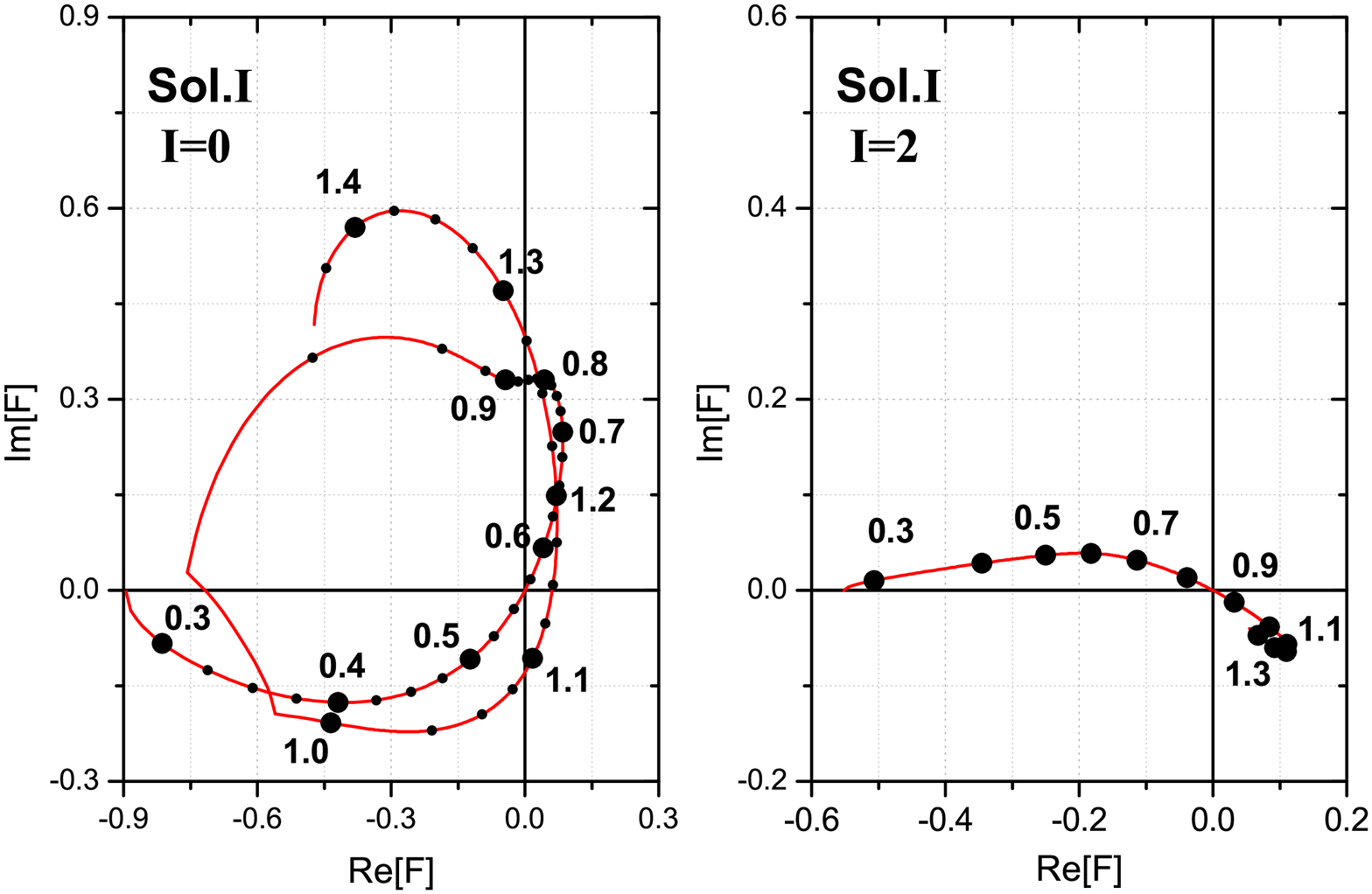}
\caption{\label{fig:argand}Argand plots for the $\gamma\gamma\to\pi\pi$ $I=0$ and $I=2$ $S$-wave amplitudes. For the $I=0$ $S$-wave, the bigger dots mark the energy every 0.1~GeV,  while the smaller dots are the intermediate energies every 25 MeV.  As seen the amplitudes move particularly fast between 950 and 1000 MeV because of the $f_0(980)$. The amplitude displays the expected \lq\lq kinks'' at $K^+K^-$ and $K^0{\overline K^0}$ thresholds. For the $I=2$ $S$-wave, the dots label the energies every 0.1~GeV.}
\end{figure}
The $I=0$ $S$-wave encodes the effect of $\sigma$, $f_0(980)$, $f_0(1370)$ poles. In the region from $\pi\pi$ threshold to 600 MeV, the real Born amplitude is increasingly being modified by strong final state interactions, generated by the $\sigma$-pole. These in fact decrease the amplitude from its simple Born value, and so produce the strong near threshold peaking seen in Fig.~\ref{fig:csiso;pi}. From Fig.~\ref{fig:argand} we see the amplitude varies fastest in the region of 0.95-1.025~GeV, corresponding to the narrow $f_0(980)$. The two thresholds, $K^+K^-$ and $K^0{\overline K}^0$, are clearly seen. Then once again above 1.1~GeV, the $S$-wave variation is generated by the deep pole of the $f_0(1370)$ and the approaching $f_0(1520)$.
We see the $I=2$ $S$-wave amplitude is smooth apart from a \lq kink' near 1.2~GeV reflecting where the 4$\pi$ (and $\rho\pi\pi$) channels become important.

\newpage
\subsection{Scalar couplings}\label{sec:3;2}
The couplings of a resonance are defined  by the residue of its pole on the nearby unphysical sheets. To determine these, we need to continue the amplitude shown, for instance, in Fig.~\ref{fig:argand} into the complex $s$-plane.
 It is the complex value of
$g_{\gamma\gamma}$ on the appropriate sheet that we quote in Table~\ref{tab:poles}. An intuitive feel for what this number means in terms of a two photon width is provided by the representation suggested by~\cite{MRP08,MRP98}
\be\label{eq:photonwidth}
\Gamma(R\to\gamma\gamma)= \frac {\alpha^2}{4~(2J+1)~m_R}\,|g_{\gamma\gamma}|^2\;,
\ee
where $\alpha$ is the usual QED fine structure constant.
This is, of course, {\bf not} a physical quantity, but merely a way to re-express $|g_{\gamma\gamma}|$.

\vspace{4mm}
\noindent $\mathbf{f_0(500)\to\gamma\gamma}$\\[4mm]
Since the existence of $\sigma$ (or $f_0(500)$) was established by dispersive analyses~\cite{zheng00,colangelo01},  a central issue has been what is its internal structure. This in turn is reflected in its two photon coupling. The representation we use, Eq.~(\ref{eq:unitarity}), which is well suited to fitting data on the real energy axis, can only be reliably continued a short distance into the complex plane, since our simple representation of the coupling functions, $\alpha(s)$, do not have the necessary analytic structure.  Consequently, to determine the residue of the $\sigma$-pole that is deep in the complex plane, and close to the left hand cut,
we use the dispersion relation of Eq.~(\ref{eq:F;ampS}) to continue the amplitudes far away from the real energy axis.
From this amplitude we can reliably extract its $\gamma\gamma$ coupling, together with its pole position, this is listed in Table \ref{tab:poles}, as well as its interpretation in terms of a radiative width, Eq.~(\ref{eq:photonwidth}).
\begin{table}[t]
\begin{center}
{\footnotesize
\begin{tabular}{|c|c||c|c|c|c|c|c|c|}
\hline
\rule[-0.5cm]{0cm}{1cm}\multirow{2}{*}{\rule[-1cm]{0cm}{2cm}State} & \multirow{2}{*}{\rule[-1cm]{0cm}{2cm}Sh} &  pole locations
& \multicolumn{3}{c|}{$g_{\gamma\gamma}=|g|e^{i\varphi}$}  & $\Gamma(f_J\to\gamma\gamma)$  &  $\lambda=0$  & \multirow{2}{*}{\rule[-1cm]{0cm}{2cm} C.L.}\\
\cline{4-6}
\rule[-0.5cm]{0cm}{1cm} & \multirow{2}{*}{}  & (GeV)  & $J_\lambda$ & $|g|~(GeV)$ & $\varphi$~($^\circ$) & (keV)   &  fraction~\%  &  \multirow{2}{*}{}   \\
\hline\hline
\multirow{4}{*}{\rule[-1.3cm]{0cm}{2.6cm}$f_2(1270)$} & \multirow{2}{*}{II} & \multirow{2}{*}{$1.270 -i0.081$}
   & \rule[-0.2cm]{0cm}{7mm} $D_0$ & 0.37$\pm$0.03 & 172$\pm$6       & \multirow{2}{*}{ 3.49$\pm$0.43 } & \multirow{2}{*}{8.4$\pm$1.4} & \multirow{2}{*}{ **** } \\
\cline{4-6}
\multirow{4}{*}{} & \multirow{2}{*}{}  &  \multirow{2}{*}{}
   & \rule[-0.2cm]{0cm}{7mm} $D_2$  &  1.23$\pm$0.08 & 176$\pm$5  & \multirow{2}{*}{}  & \multirow{2}{*}{} & \multirow{2}{*}{} \\
\cline{2-8}
\multirow{4}{*}{\rule[-0.3cm]{0cm}{8mm}}   & \multirow{2}{*}{ III}  &  \multirow{2}{*}{ $1.267 -i0.108$}
& \rule[-0.2cm]{0cm}{7mm} $D_0$& 0.35$\pm$0.03 &  168$\pm$6  &  \multirow{2}{*}{ 2.93$\pm$0.40 } & \multirow{2}{*}{ 8.7$\pm$1.7} & \multirow{2}{*}{****}  \\
\cline{4-6}
\multirow{4}{*}{}  & \multirow{2}{*}{} &  \multirow{2}{*}{}
& \rule[-0.2cm]{0cm}{7mm} $D_2$  & 1.13$\pm$0.08  & 173$\pm$6  & \multirow{2}{*}{}  & \multirow{2}{*}{}   &\multirow{2}{*}{}   \\
\hline
$a_2(1370)$
& \rule[-0.3cm]{0cm}{8mm} IV              & $1.313 -i0.053$   &$D_2$ &  0.72$\pm$0.08 & 174$\pm$3     &   1.04$\pm$0.22  & 0$^\dag$   &       **       \\
\hline
$f_0(500)$
    & \rule[-0.3cm]{0cm}{8mm} II          & $0.441 -i0.272$   & S    & 0.26$\pm$0.01  & 105$\pm$3     &   2.05$\pm$0.21  &     100    &       ****  \\
\hline
$f_0(980)$  & \rule[-0.3cm]{0cm}{8mm} II  & $0.998 - i0.021$  & S    & 0.16$\pm$0.01  &-175$\pm$5     &   0.32$\pm$0.05  &     100    &       ****  \\
\hline
\multirow{2}{*}{ \rule{0cm}{8mm}$f_0(1370)$}
    & \rule[-0.3cm]{0cm}{8mm} II          & $1.423 - i0.177$  & S    & 0.96$\pm$0.10  & 8$\pm$13      &   8.6$\pm$1.9    &     100    &       *     \\
\cline{2-9}
    & \rule[-0.3cm]{0cm}{8mm} III         & $1.406 - i0.344$  & S    & 0.65$\pm$0.15  & $-$146$\pm$15 &   4.0$\pm$1.9    &     100    &       *     \\
\hline
\end{tabular}
\caption{\label{tab:poles} The resonance poles and their two photon residues (both magnitude and phase) from Solution~I are listed. These residues can be interpreted in terms of  two-photon partial widths using Eq.~(\ref{eq:photonwidth}). These are tabulated in keV. For each the fraction of the width provided by helicity zero is given:
for the scalar resonances, it is, of course, 100\%. $^\dag$ Note that we assume $D_0$ waves to be zero for $a_2$.
In the first column \lq Sh' denotes the Riemann Sheets and in the last column \lq\lq C.L.'' (Confidence Level) indicates the reliability of the results. }
}
\end{center}
\end{table}
}
\noindent On the second sheet, our $\sigma$-pole in our hadronic amplitudes is at $E\,=\,0.441\,-i\ 0.272\,{\rm GeV}$ as found by~\cite{caprini06}, which is within the error of~\cite{zhou04}.
The more distant third sheet pole is located  at $E\,=\,0.386\, -\,i\ 0.108\,{\rm GeV}$, which is compatible with that of~\cite{DLY11}. Extracting first the coupling to $\pi\pi$ from the residue of the amplitude $T_{11}$, we then find the coupling $g_{\gamma\gamma}$ listed in the Table. For the pole on the second sheet this can be interpreted as a radiative width of $f_0(500)\to\gamma\gamma$  of $(2.05\ \pm\ 0.21)$ keV. The error is given by the band shown in Fig.~\ref{fig:Flow}. In Eq.~(\ref{eq:photonwidth}), the resonance mass $m_R$ is taken to be $|m_R|$. Other choices are included in the uncertainties.
The relation of these results to other works will be discussed in Sect.~\ref{sec:4;1}.

\newpage
\noindent $\mathbf{f_0(980)\to\gamma\gamma}$\\[4mm]
As shown in Fig.~\ref{fig:csf980}, our solution fits the peaks observed by Belle both in charged pion~\cite{Belle-pm} and neutral pion data~\cite{Belle-nn}.
Moreover, our input hadronic amplitude has only one pole located at $E\,=\,0.995\, -\, i\,0.042\,{\rm GeV}$ on the second sheet, see Table~\ref{tab:poles}.  Since this is quite close to the real axis, Eq.~(\ref{eq:unitarity}) with the functions $\alpha(s)$ represented by polynomials should be good enough to determine the couplings.
The way to extract the photon couplings on different Riemann sheets, numbered by $Sh$, is given in~\cite{MRP98,yumao09}. This can be expressed in a simplified way~\cite{zheng12} through:
\begin{eqnarray}\label{eq:residue}
&g^{Sh}_{\gamma\gamma}=\alpha_1(s_R)\,g^{Sh}_{\pi\pi}+\alpha_2(s_R)\,g^{Sh}_{KK}\;,
\end{eqnarray}
where the functions are evaluated at the pole $s\,=\,s_R$ on the appropriate sheet.
Here $\alpha_1(s)$, $\alpha_2(s)$ in Eq.~(\ref{eq:unitarity}) are simple functions, given by Eq.~(\ref{eq:alphas}), and $g^2_{\pi\pi}$, $g_{\pi\pi} g_{KK}$ are residues extracted from hadronic amplitudes $T_J(\pi\pi\to\pi\pi,\ K\overline K)$ with $J=0$.
In this way we find the $\gamma\gamma$ coupling of the  $f_0(980)\to\gamma\gamma$ listed in Table~\ref{tab:poles}.
This  can be interpreted as a width $(0.32\,\pm\,0.05)$ keV for our solutions. The error comes from the uncertainties in the $\alpha(s)$ and $T$-matrix
elements from MINUIT, together with the variation between $\gamma\gamma$ solutions.

\vspace{4mm}
\noindent $\mathbf{f_0(1370)\to\gamma\gamma}$\\[4mm]
As shown in Table~\ref{tab:poles}, there are two poles located at $E\,=\,1.423\, -\, i\,0.177\, {\rm GeV}$ on the second sheet and at $E\, =\,1.406\, -\, i\,0.344\, {\rm GeV}$ on the third sheet in our hadronic amplitudes. These poles are deep in the complex energy plane and close to the end of our fitted range, and have dominant couplings to 4$\pi$ channels that we have not treated uniformly well. Consequently, one must interpret their two photon couplings with a great deal of care. The couplings we find translate to a $f_0(1370)\to\gamma\gamma$ width of $(4.0\,\pm\,1.9)$ keV.
However, this number, even with its large error, does not have the credibility of our other results, and that is reflected in our associating {\it one star} with this result. Though it is near the upper limits of
$(3.8 \pm 1.5)$~keV and $(5.4 \pm 2.3)$~keV given in~\cite{Giacosa05}, in which the $f_0(1370)$ is treated as a mixed state of quarkonium and a glueball.

\baselineskip=5.5mm
\subsection{Tensor couplings}\label{sec:3;3}
\vspace{4mm}
\noindent $\mathbf{f_2(1270)\to\gamma\gamma}$\\[4mm]
Though the hadronic $\pi\pi$ amplitudes given by {\it CFDIV}~\cite{KPY} are quite precise along the real energy axis, they are not appropriate for continuing into  the complex plane.
Consequently, we use a coupled channel $K$-matrix parametrization to represent the phase and inelasticity given by {\it CFDIV}. To make it simple we absorb the 4$\pi$ channel contribution into the inelasticity for $K\overline{K}$.
As shown in Table \ref{tab:poles}, the resulting amplitude has two poles located at $E\,=\,1.270\, -\,i\,0.081\,{\rm GeV}$ on the second sheet and at $E\,=\,1.267\, -\,i\,0.108\,{\rm GeV}$ on the third sheet. The $f_2(1270)\to\gamma\gamma$ residue can then be interpreted, via Eq.~(\ref{eq:photonwidth}),
as a radiative width on the second sheet as $(3.49\,\pm\,0.43)$ keV, and   $(2.93\,\pm\,0.40)$ keV on sheet~III. It is the pole on Sheet~III that is the nearest to the real axis and produces the dominant physical effects on the cross-section.
These widths are compatible with the earlier \lq peak solution' of~\cite{MRP98} and Solution~A of~\cite{MRP08}.
The simple cross-section formula for a Breit-Wigner peak, Eq.~({\ref{eq:Grr}),  correspond to a width of $(3.32\,\pm\,0.37)$ keV from the cross-section seen in Fig.~\ref{fig:csiso;pi}.

\noindent The non-relativistic constituent quark model picture predicts that the coupling of the tensor mesons should be predominantly through helicity two~\cite{krammer} through an $E1$ transition.
As discussed in detail by Poppe~\cite{poppe} earlier data from SLAC and DESY could not rule out a significant $D_0$ component for the $f_2$ coupling to two photons: with limited angular coverage and poorer statistics $S$ and $D_0$ waves were interchangeable~\cite{MRP98,MRP08}. In ~\cite{Close91}, quark model calculations with relativistic corrections predict the helicity 0 component  to be 6\% of the total. Now finding a solution compatible with the Belle data in both the neutral and charged pion channels,
fixes the helicity zero fraction to be $(9\pm 2)\%$. This validates a simple ${\overline q}q$ picture for the $f_2(1270)$.

\vspace{4mm}
\noindent $\mathbf{a_2(1320)\to\gamma\gamma}$\\[4mm]
Our result tor the $a_2$ photon coupling does not have the precision of that for the $f_2$. While the isoscalar $D$-wave  is determined by the far greater $\pi\pi$ information we have fully discussed, the isovector $D$-wave has been crudely parametrized by a Breit-Wigner for the $a_2(1320)$.
The fit shown in Figs.~\ref{fig:csiso;K} gives a radiative width for the $a_2$ of $(1.04\,\pm\,0.22)$~keV.
The error reflects the systematic uncertainty produced by different solutions, as shown in Fig.\ref{fig:csiso;K}.
While the simple Breit-Wigner formula of Eq.~({\ref{eq:Grr}) for the peaks seen in Fig.~\ref{fig:csiso;K} correspond to a width of $(1.04\,\pm\,0.18)$~keV. These results are in agreement with the radiative width of $(1.00\,\pm\,0.06)$~keV that the PDG~\cite{PDG12} quotes.
The difference is tolerable as that width is deduced from the dominant 3$\pi$ decay mode of the $a_2$, which has a branching fraction of $\sim 70\%$,
while what enters here is the ${\overline K}K$ fraction of merely $5\%$.

\newpage
\section{Discussion}\label{sec:4;}
\subsection{Related Work}\label{sec:4;1}
In a series of papers, Achasov and collaborators~\cite{Achasov07}-\cite{Achasov12} have considered the two photon reactions we study here. Based on a $\pi\pi$, ${\overline K}K$ rescattering mechanism, they determine the two photon widths averaged over the resonance mass distributions. In particular, they have investigated the implication  of kaon loops for the appearance of the $f_0(980)$ in these reactions.
In ~\cite{Achasov07}, they fit the charged pion data in the energy range from 0.85 to 1.5~GeV, and the neutral pion results from threshold to 1.5~GeV. They do not fit the angular distributions, and presume the $D$-wave is pure helicity-two. They then find a width for $f_2(1270)\to\gamma\gamma$ of 3.68~keV.  Their model  separates the two photon couplings of resonances between a \lq\lq direct'' coupling and that from hadron loops. In ~\cite{Achasov08}, they find the direct components to be small. The radiative widths integrated over the resonances are then found to be 0.45~keV for the $\sigma$ and 0.19~keV for the $f_0(980)$. In ~\cite{Achasov12}, Achasov and Shestakov give a prediction for the scalar contributions to the ${\overline K}K$ channels from their kaon loop model. Their isoscalar $S$-wave is in agreement with ours, but the isovector $S$-wave is much bigger than ours in the range of 1.0-1.1~GeV, see Fig.~\ref{fig:csiso;K}.

\noindent In~\cite{Achasov94,AchasovPRL}, Achasov and Shestakov, using a linear realization of $SU(2)\times SU(2)$ $\sigma$ model~\cite{LSigma}, consider the $\sigma$ as a resonance in $\pi\pi\to\pi\pi$ and $\gamma\gamma\to\pi\pi$ scattering. They illustrate how the effect of the $\sigma$-pole, deep in the complex energy plane, on scattering on the real energy axis is shielded by multi-quark dynamics. In the AS treatment, a major part of the $\sigma$-meson self-energy is generated by intermediate $\pi\pi$, or $4q$, intermediate states. Unlike our two photon  discussion here, or that of $\pi\pi$ scattering using the Roy equations, the AS analysis has no crossing and so there is  no contribution from the nearby left hand cut in their bubble sum.  Nevertheless, the \lq\lq shielding''  effect AS noted was contemporaneously  demonstrated in the phenomenological description shown   in Fig.~3 of~\cite{MRP07Rev}.  A natural inference from this \lq\lq shielding of the $\sigma$'' is that the direct photon coupling cannot give an idea of the structure of the $\sigma$.

\noindent In ~\cite{Oller97}, Oller and Oset make a theoretical study of $\gamma\gamma\to MM$ reactions. They present a unified picture, for $\pi^+\pi^-$, $\pi^0\pi^0$, $K^+K^-$, ${\overline K}^0 K^0$, and $\pi^0\eta$ production up to 1.4~GeV. This model includes crossed-channel particle exchange and final state interactions. With this they fit the $\pi^+\pi^-$ data from Mark~II and Cello (including their angular distributions in three of the energy intervals), and the Crystal Ball results on the $\pi^0\pi^0$ channel. As a result, they find a radiative width for the $f_0(980)$ of 0.20~keV. Of course, this work pre-dates the high statistics Belle results.

\noindent In ~\cite{yumao09}, Mao~{\it et al.} perform a dispersive analysis of the $f_0(500)$ and $f_0(980)$ in $\gamma\gamma\to\pi\pi$. They find a width of 0.12~keV for the second sheet pole of the $f_0(980)$ and  quote 2.08~keV for the radiative width of the $\sigma$.

\noindent In ~\cite{Mennessier10}, Mennessier, Narison and Wang use an improved analytic $K$-matrix model for hadronic reactions, and extract the $\sigma$ and $f_0(980)$ radiative widths from studying the Crystal Ball and Belle data on the neutral pion channel. Their Feynman diagram based treatment allows \lq\lq direct'' and \lq\lq rescattering'' contributions to the $\gamma\gamma$ couplings to be separated. They find the direct $\sigma$ width to be 0.16~keV, while the total is $(3.08\pm 0.82)$~keV, and the direct $f_0(980)$ width is 0.28~keV with a total of only $(0.16 \pm 0.01)$~keV. However, they fit up to 1.09~GeV just the $\pi^0\pi^0$ mode. Their prediction for the  $\gamma\gamma\to\pi^+\pi^-$ process provides a rather poor approximation to the data.

\noindent In ~\cite{Moussallam10}, Moussallam and Garcia-Martin use the dispersive machinery used here supplemented by a coupled channel Mushkhevili-Omn\`es representation. Though the framework they use is similar to that we follow as set out in ~\cite{MRP87,MRP88}, we only attempt to determine the inputs to the dispersion relations, Eq.~(\ref{eq:F;ampS},\ref{eq:F;ampJ}), in as much they limit the partial waves from $\pi\pi$ threshold to 600~MeV, and even then within bands of uncertainty shown in Fig.~\ref{fig:Flow}. In contrast, Moussallam and Garcia Martin attempt to fix the inputs from distant energy regions along the left and right hand cuts to predict the $S$-wave through the $f_0(980)$ region. Thus, they assume the left hand cut is described by a combination of single particle exchanges: $\pi$, $\rho$, $\omega$, $a_1$, $b_1$, $h_1$, $a_2$ and $f_2$-exchanges.
No account is taken of multi-meson exchange contributions like $\pi\pi$, $\rho\pi$, $4\pi$, etc. Data on the two meson production are described up to 1.28~GeV.

\noindent In ~\cite{Phillips11}, Hoferichter. Phillips and Schat derive a system of Roy-Steiner equations for pion Compton scattering. These hyperbolic dispersion relations are then projected onto $s$ and $t$-channel partial waves with subtraction constants fixed in terms of pion polarizabilities. Focusing on the low energy region they find the two photon width of the $\sigma$ to be $(1.7 \pm 0.4)$~keV. The $f_2 (1270)$ is assumed to couple entirely through helicity-2 and is used to match parameters in their analysis below 1~GeV.

\noindent In ~\cite{Igor12}, Danilkin {\it et al.} consider $\pi\pi$, ${\overline K}K$, $\pi\eta$ and $\eta\eta$ production in a chiral Lagrangian model. They fit the $\pi\pi$ cross-sections up to 900~MeV, and $\pi^0\eta$ to 1.2~GeV. However, the model does not include tensor resonances and so is limited in applicability to 0.9~GeV and below.

\noindent All these papers have overlap with the study presented here. Many have used techniques close those that are the basis of this analysis. However, our treatment is the only one that is an Amplitude Analysis, using basic theoretical concepts to set the framework, and allowing the data to dictate the structure of the two photon partial waves. We now turn to the interpretation of our results.

\subsection{Scalar and Tensor Structure}\label{sec:4;2}
Having determined the two photon couplings of the scalar and tensor states, and having interpreted these in terms of a radiative width, we can use these to discuss the nature of the states.
The simplest is the relation of the tensors: $f_2(1270)$ and $a_2(1320)$. If their masses and annihilation probabilities were exactly equal, the two photon width measures the square of the mean square of the electric charges of their constituents. So if each is a ${\overline q}q$ system composed of $u$ and $d$ quarks, then
\bea
&&\Gamma\left(\gamma\gamma\to ({\overline u}u + {\overline d}d)/\sqrt{2}\right)\;:\;\Gamma\left(\gamma\gamma\to ({\overline u}u - {\overline d}d)/\sqrt{2}\right)\no\\[2mm]
&&{\hspace{35mm}}=\;\left[(2/3)^2\,+\,(-1/3)^2\right]^2\;:\;\left[(2/3)^2\,-\,(1/3)^2\right]^2\no\\[2mm]
&&{\hspace{35mm}}=\;25\;:\;9\; .
\eea
With $\Gamma(f_2\to\gamma\gamma)\,=\,(2.93\,\pm\,0.40)$~keV (the width on Sheet III), this quark model relation predicts $\Gamma(a_2\to\gamma\gamma)\,=\,(1.06\,\pm\,0.15)$~keV, compatible with our photon width for the $a_2$ presented in Table~\ref{tab:poles}.
This agreement suggests that the $f_2(1270)$  and $a_2(1320)$ are (not surprisingly) simple $\overline{q}q$ structures.
Of course, the determination of the $a_2$ radiative width is not on anything like the firm basis we have here for the $f_2(1270)$ with the partial wave separation and the careful treatment of its pole residue.

\noindent As discussed, for example, in Ref.~\cite{MRP06}, if the $\sigma$ were the quark model companion of the $f_2(1270)$ with a $({\overline u}u+{\overline d}d)/\sqrt{2}$ structure, it would satisfy
\be\label{eq:sigmaf2}
\frac{\Gamma(\sigma\to\gamma\gamma)}{\Gamma(f_2\to\gamma\gamma)}=\frac{15}{4}\left(\frac{m_\sigma}{m_{f_2}}\right)^n \;\; ,
\ee
in potential models. Here `${15}/{4}$' is reduced to 2 by relativistic effects, see~\cite{Close91}.}
The power $n$ indicates the shape of the potential, with $n=3$ in the Coulomb region and $n\,\to 0$ if there is linear confinement~\cite{Close91}.
To reproduce this relation with our radiative widths in Table~\ref{tab:poles}, $n$ should be around 0.7-1.2. Of course,
this is sensitive to what we choose the real parameter $m_\sigma$ in Eq.~(\ref{eq:sigmaf2}).

\noindent Meanwhile there are other models to classify state structure for the scalars. A simple $({\overline u}u+{\overline d}d)/\sqrt{2}$ gives $\sim 4$ keV as given by Babcock and Rosner~\cite{Babcock76}, while more recently Giacosa {\it et al.} predict it to be smaller than 1~keV in the case of the $\sigma$~\cite{Giacosa07}. An ${\overline s}s$ structure gives $\sim 200$ eV according to Barnes~\cite{Barnes}, and 62 eV from Giacosa {\it et al.}~\cite{Giacosa08}. A tetraquark, {\it i.e.} ${\overline {qq}}qq$, composition gives $\sim 270$~eV as calculated by Achasov {\it et al.}~\cite{Achasov82} through kaon loops. The prediction of largely $K\overline{K}$ composition for the $f_0(980)$ is more complicated.  Barnes~\cite{BarnesKK} gives $\sim 600$~eV  in the molecular model of Weinstein and Isgur~\cite{Weinstein}, while Hanhart {\it et al.}~\cite{Hanhart07} predict 220 eV. Alternatively Narison, and Ochs and Minkowski have proposed an intrinsic glueball  nature for the $\sigma$, with direct radiative widths between 0.2 and 0.6~keV~\cite{Narison}, which increase to $(3.9 \pm 0.6)$~keV when $\pi\pi$ rescattering is included~\cite{Ochs}. Here we list the radiative widths in Table~\ref{tab:radiativeG} predicted in these models.
\begin{table}[htbp]
\begin{center}
\begin{tabular}  {|c||c|c|}
\hline
\rule[-0.3cm]{0cm}{8mm} composition       &        prediction (keV)       &       author(s)                        \\
\hline\hline
\multirow{2}{*}{$(\overline{u}u+\overline{d}d)/\sqrt{2}$}  &   4.0 \rule[-0.3cm]{0cm}{8mm} & Babcock \& Rosner \cite{Babcock76}     \\
\cline{2-3}
                                          & $<1^\dagger$\rule[-0.3cm]{0cm}{8mm}   & Giacosa {\it et al.} \cite{Giacosa07}\\
\hline
\multirow{2}{*}{$\overline{s}s$}            &   0.2\rule[-0.3cm]{0cm}{8mm}  & Barnes \cite{Barnes}                   \\
\cline{2-3}
                                          & 0.062\rule[-0.3cm]{0cm}{8mm}   & Giacosa {\it et al.} \cite{Giacosa08}\\
\hline
$\overline{[ns]}[ns]$                    & 0.27 \rule[-0.3cm]{0cm}{8mm}  & Achasov {\it et al.} \cite{Achasov82}  \\
\hline
\multirow{2}{*}{\rule{0cm}{8mm}$\overline{K}K$}& 0.6\rule[-0.3cm]{0cm}{8mm}    & Barnes  \cite{BarnesKK}                \\
\cline{2-3}
                                          & 0.22\rule[-0.3cm]{0cm}{8mm}   & Hanhart {\it et al.}  \cite{Hanhart07} \\
\hline
$gg$ & 0.2--0.6\rule[-0.3cm]{0cm}{8mm}    & Narison  \cite{Narison}                \\
\hline
\end{tabular}
\end{center}
\caption{\label{tab:radiativeG}Radiative widths in different modellings of their composition.
Here $^\dagger$ means that the authors~\cite{Giacosa08} assume $M_\sigma~<~0.7-0.8~GeV$.
}
\end{table}
As discussed in Sect.~\ref{sec:4;1}, the radiative width for the $\sigma$  is dominated by the $\pi\pi$ rescattering effects regardless of its \lq\lq inner core", just as in the original calculation of ~\cite{MRP06} and updated here in Table~\ref{tab:poles}. The dominance of $\pi\pi$ effects shows how important hadronic final state interactions are.

\noindent For the $f_0(980)$, as shown in  Table~\ref{tab:radiativeG}, the radiative width of $\overline{[ns]}[ns]$, $\overline{s}s$, $\overline{K}K$
and $gg$ are quite close to each other, and close to the  $(260 \pm 40)$~keV we have determined from the latest data. Here too the ${\overline K}K$ rescattering component is a significant part of what is observed.  To reach a final conclusion on the structure of the underlying state, more accurate calculations in strong coupling QCD, either in the continuum or on the lattice, to predict these couplings are required. Here we have established what their answer should be to agree with experiment.

\newpage
\baselineskip=5.3mm

\section{Conclusion}\label{sec:5}

\noindent This paper presents an Amplitude Analysis  of all data on $\gamma\gamma\to\pi\pi$. To make up for the limited angular range in these experiments and the lack of polarization information, we take advantage of the fundamental concepts of $S$-matrix theory. Unitarity relates the two photon reactions we study, both $\pi\pi$ and ${\overline K}K$ production, to hadronic scattering reactions involving pions and kaons. Unitarity determines the interaction of the final state hadrons and so constrains the form of the two photon partial waves. At low energy, Low's theorem for Compton scattering further constrains the amplitudes. So much so that consistency between Compton scattering and the two photon scattering region imposed using dispersion relations restricts the $\gamma\gamma$ partial waves to narrow bands up to 5 or 600~MeV. Data then does the rest. So almost up to 1.5~GeV, we have a rather small band of solutions. The \lq\lq unique'' solution (we label I) has been presented in Sects.~\ref{sec:2},~\ref{sec:3}. That there is such precision comes from the high statistics data on the $\pi^+\pi^-$, $\pi^0\pi^0$ and $K_sK_s$ data from Belle, together with earlier two photon data on $\pi\pi$, as well as some limited information on the $K^+K^-$ and ${\overline K}^0K^0$ channels. This is the first analysis that describes all these data simultaneously, both integrated and differential cross-sections, and imposes coupled channel unitarity to determine the partial waves.

\noindent To achieve this degree of certainty we supplement  the classic meson-meson scattering results from CERN-Munich, Argonne and Brookhaven experiments with the latest hadronic reaction information from the precise dispersive analyses of \cite{Descotes04} and {\it CFDIV}~\cite{KPY},  and the recent BaBar data~\cite{BABAR-pi,BABAR-K}. These constrain our $T$-matrix elements, as shown in Fig.\ref{fig:T}.
With these amplitudes, precise pole locations are given by Table \ref{tab:poles} for the tensor $f_2(1270)$ and the scalars $f_0(500)$/$\sigma$, $f_0(980)$ with a hint of the $f_0$(1370). Since the information on the isovector channel from just $\gamma\gamma\to {\overline K}K$ is highly limited we simply input the $a_2(1320)$ as a Breit-Wigner with parameters fixed from its dominant $3\pi$ decay mode.

\noindent To obtain precise two photon amplitudes, the high statistics Belle datasets on $\gamma\gamma\to\pi^+\pi^-$ and $\pi^0\pi^0$ process~\cite{Belle-pm,Belle-nn} are fitted, together with all earlier two photon results. The solutions are of good quality, as indicated in Table \ref{tab:fit} and Figs~\ref{fig:cspic}-\ref{fig:dcsBelle00}.
The $\gamma\gamma\to \overline{K}K$ datasets~\cite{ARGUS89}-~\cite{TASSO86} have also been fitted. The Belle results~\cite{Belle-KsKs}, in particular, on the $K_sK_s$ channel with their good angular coverage and narrow energy binning are a powerful constraint on our solution space. The results of the fits are shown in Table~\ref{tab:fit} and Fig.~\ref{fig:csK}.

\noindent For this narrow range of amplitudes, the two photon coupling of each state is specified by the residue of its  pole in the complex energy plane. This is the most model-independent result possible. These couplings, listed in Table~\ref{tab:poles}, can be interpreted as radiative widths, but only as an intuitive guide.
We find the width for the  $f_0(500)/\sigma$ width to be $(2.05\pm0.21)$ keV from the pole on the second sheet.
This is roughly half the value predicted for a $({\overline u}u+{\overline d}d)/\sqrt{2}$ bound state. In fact the radiative width of the $\sigma$ is dominated by its coupling to two pions, that underlies the calculation presented here, which updates those in~\cite{MRP06, Oller07, Oller08, MRP08}. The residue for the $f_0(980)$ can be interpreted as a $\gamma\gamma$ width of $(0.32 \pm 0.05)$ keV. As discussed in~\cite{MRP10,morgan92} the $f_0(980)$ may well be a ${\overline K}K$  molecule, and its two photon width is consistent with that, though a combination of  $\overline{n}n$ and $\overline{s}s$ components is still possible.

\noindent  For the tensor
$f_2(1270)$ we interpret its $\gamma\gamma$ residue on the third sheet as a radiative width of $(2.93 \pm 0.40)$~keV for the present Sol.~I. This is a definitive result.
It is between the values given by Sol.~A and Sol.~B of~\cite{MRP08}.

\noindent Limited results for the $f_0(1370)$ are also listed in Table~\ref{tab:poles}. However, this state appears at the very edge of where our analysis with just $\pi\pi$ and ${\overline K}K$ channels can be trusted. Consequently, we regard these results as having only a single star reliability, rather then the four star results we have for other states in ~Table~\ref{tab:poles}.

\noindent The determination here of a very narrow range of partial wave solutions has brought a precision to two photon studies not previously achieved.
The partial wave analysis we have presented gives the individual cross-sections shown in Fig.~\ref{fig:csiso;pi}.
For the $\pi\pi$ $S$-waves we show their Argand plots in Fig.~\ref{fig:argand}. These are our main outputs.
The resonance couplings we have discussed are just consequences of these. In a separate publication we will deduce what our amplitudes imply for
the left hand cut discontinuity and its representation in terms of crossed-channel exchanges  in detail. The precision of the Belle data,
and the fact that their $\pi^0\pi^0$ results go out to $\cos \theta = 0.8$ has allowed a robust separation of helicity zero and two partial waves. Only by a complete partial wave analysis can we fill in the cross-sections for the whole angular range. It is
these that are inputs into the light-by-light sum-rules derived by Pascalutsa and Vanderhaeghen~\cite{PV1012}. Their contribution to this will also be considered in a future work.

\newpage
\noindent With plans advanced for new experiments on the anomalous magnetic moment of the muon at both Fermilab and J-PARC, there has been renewed theoretical effort in limiting the uncertainties from the Standard Model to this fundamental quantity. One of the largest uncertainties comes from hadronic light-by-light scattering. While the need is for theoretical, phenomenological and experimental guidance on the  scattering of virtual photons, results for real photons presented here provide a new precision with which to anchor these studies. Indeed, our amplitudes are a natural input into the newly developed dispersive framework~\cite{Hoferichter} for calculating light-by-light contributions. The inclusion of the full $\gamma\gamma\to MM$ amplitudes, constructed from experiment here, should bring greater certainty than the presently used artificial separation of pion and kaon loops from scalar and tensor resonances, see {\it e.g.}~\cite{Nyffeler}. They are all included here. Indeed, it is this energy domain up to 2~GeV$^2$ that is believed to dominate the light-by-light contributions.

\noindent There has long been speculation about the nature of the scalar mesons,
and whether the lightest ones are examples of multiquark states, molecules or glueballs, and not simply ${\overline q}q$ states~\cite{jaffe4q}-\cite{weinberg}. Two photon couplings serve as a guide to their composition. While the photons do excite their \lq\lq primordial'' seeds, the fact that these  photons at low energy  have long wavelengths means that they couple as much to the hadronic decay modes of these light states. Thus to compare our  {\it experimental} results with models requires more detailed computation than has hitherto been possible. Without such dynamical calculations, any further remarks are mere speculation. Nevertheless, as discussed in \cite{MRP10}, the $\sigma$ and $f_0(980)$ may indeed be seeded by {\it bare} ${\overline n}n$ and ${\overline s}s$ states of higher mass. The inclusion in their dynamics of the hadronic channels, to which they couple, may generate a very short-lived state close to $\pi\pi$ threshold and one close to the opening of the ${\overline K}K$ channel~\cite{vanbeveren}. It is clear these states have something in common with tetraquark configurations, but particularly for the $f_0(980)$ its pole structure reflects rather that of a molecule. Then it is natural that the two photon couplings of these states are dominated by couplings to $\pi\pi$ and ${\overline K}K$ systems, respectively, rather than their speculated inner core. Of course, only in explicit models, like $1/N_c$~\cite{MRP11,weinberg}, is it meaningful to ask what this inner core is. In the real world with $N_c = 3$, the hadronic modes surely dominate. Our understanding of such dynamics is being challenged by the discovery of a range of charmonium and bottomonium states~\cite{eidelman} that are similarly correlated with nearby hadronic channels, and upon which their very existence depends. The light scalar states studied here are a key window on this dynamics.

\noindent Two photon running to come at KLOE-2~\cite{MRP-KLOE2,KLOE2} should provide confirmation of our results for the isoscalar and isotensor channels. This will demand better $\mu^+\mu^-$ separation than even Belle have below 900~MeV. However, results of comparable precision for isovector states must await a corresponding coupled channel analysis combining data on $\gamma\gamma\to\pi^0\eta$, ${K^+K^-}$ and ${\overline K^0}K^0$ with that on $\pi\pi$. While the two photon production of $\pi\pi$ and $\eta\pi$ channels, of course, access different isospins, the ${\overline K}K$ channels involve both $I=0,1$. Thus a larger global analysis would be required, which would inevitably involve multi-pion hadronic scattering channels too. Nevertheless, such analysis would enable a full flavour description of the resonant states that dominate meson-meson interactions up to 1.5~GeV, and inevitably feed into the contribution of hadronic light-by-light to the anomalous magnetic moment of the muon. That is for the future.

\vspace{1cm}

\noindent{\bf Acknowledgment}
\noindent  MRP thanks Belle colleagues, Yasushi Watanabe and Sadaharu Uehara  for access to the Belle $\pi^0\pi^0$ data and for previous discussions on the systematic uncertainties involved. We would like to thank Han-Qing Zheng and David Wilson for helpful discussions.
This paper has been authored by Jefferson Science Associates, LLC under U.S. DOE Contract No. DE-AC05-06OR23177.

\newpage
\appendix
\setcounter{equation}{0}
\setcounter{table}{0}
\renewcommand{\theequation}{\Alph{section}.\arabic{equation}}
\renewcommand{\thetable}{\Alph{section}.\arabic{table}}

\section{Born terms and amplitudes for exchanges}\label{app:A}
\noindent
The Born term with one pion exchange is calculated from the Chiral Lagrangian $\mathcal{L}_2={F^2}\langle u_\mu u^\mu\rangle/4$.  With $\rho_1(s)\,=\,\sqrt{1\,-\,4m_\pi^2/s}$, its partial waves are
\bea\label{eq:Born}
&B^\pi_{S}&=\frac{1-\rho_1(s)^2}{2\rho_1(s)}\,\ln\left(\frac{1+\rho_1(s)}{1-\rho_1(s)}\right) \,\, , \nonumber\\[3mm]
&B^\pi_{D0}&=\sqrt{\frac{5}{16}}\,\frac{1-\rho_1(s)^2}{\rho_1(s)^2}\left(\frac{3-\rho_1(s)^2}{\rho_1(s)}\ln\frac{1+\rho_1(s)}{1-\rho_1(s)}-6 \right) \,\, , \nonumber\\[3mm]
&B^\pi_{D2}&=\sqrt{\frac{15}{8}}\,\left( \frac{[1-\rho_1(s)^2]^2}{2\rho_1(s)^3}\ln\frac{1+\rho_1(s)}{1-\rho_1(s)}
              +\frac{5}{3}-\frac{1}{\rho_1(s)^2}\right) \,\, . \nonumber
\label{eq:Born}
\eea
The higher spin partial waves are determined in terms from the full Born amplitudes using
\bea\label{eq:Born;high}
\sum_{J}~B_{J0}Y_{J}^{0}(\theta,\phi)&=&\sqrt{\frac{1}{4\pi}}\frac{1-\rho_1(s)^2}{1-\rho_1(s)^2 \cos^2\theta} \,\, , \nonumber\\[3mm]
\sum_{J\geq2}~B_{J2}Y_{J}^{2}(\theta,\phi)&=&\sqrt{\frac{1}{4\pi}}\frac{(\rho_1(s)^2\,\sin^2\theta)e^{i2\phi}}{1-\rho_1(s)^2 \cos^2\theta} \,\,,
\eea
by subtracting the lower partial waves,
where $Y_J^m(\theta,\phi)$ are the spherical harmonics function.

\noindent
The contribution of other exchanges are calculated from the Resonance Chiral Lagrangian~\cite{yumao09,Ko90}:
\bea
\mathcal{L}_{VP\gamma}&=&e\, C_V\, \epsilon_{\mu\nu\alpha\beta}F^{\mu\nu}\langle \Phi\{Q,\partial^\alpha V^\beta  \}\rangle, \nonumber\\
\mathcal{L}_{AP\gamma}&=&e\, C_A\, F^{\mu\nu}\langle \Phi [ Q,\partial_\mu A^\nu  ]\rangle \,\, , \nonumber\\
\mathcal{L}_{BP\gamma}&=&e\, C_B\, F^{\mu\nu}\langle \Phi \{ Q,\partial_\mu B^\nu  \}\rangle \,\, ,\label{eq:LV}
\eea
where $\Phi$ is the Pseudoscalar ($0^{--}$) Octet,  V is the ($1^{--}$) Octet, A the ($1^{++}$) Octet, and B the ($1^{+-}$) Octet.
The partial wave projection of these gives
\bea
L_{S}(C_R^2,M_R,s) &=&C_R^2\left(-\frac{M_R^2}{\rho_1(s)}\,LF(M_R,s)+s\right)\, ,\nonumber\\[3mm]
L_{D0}(C_R^2,M_R,s)&=&\frac{C_R^2~M_R^2}{\rho_1(s)}\left( [1-3 XF^2(M_R,s)]~LF(M_R,s)+6 XF(M_R,s) \right)\,\, ,\nonumber\\[3mm]
L_{D2}(C_R^2,M_R,s)&=&C_R^2~s~\rho_1(s)\left(\; [1-XF^2(M_R,s)]^2 LF(M_R,s)\right. \,\,\nonumber\\[3mm]
                   & &\;\;\;\;\;\;\;\;\;\;\;+\frac{2}{3} XF(M_R,s)\left.[5-3 XF^2(M_R,s)]\;\right)\,\, ,\label{eq:L;partial}
\eea
where we follow the functional forms given by~\cite{Moussallam10}:
\bea
XF(M,s)&=&\frac{2M^2-2m_\pi^2+s}{s~\rho_1(s)},\nonumber\\[3mm]
LF(M,s)&=&\ln\left( \frac{XF(M,s)+1}{XF(M,s)-1} \right). \nonumber
\eea
The contributions of Vector and Axial Vector are given by
\bea
&L_{\omega, S}(s)&=L_{S}(C_\omega^2,M_\omega,s)\,\,,\nonumber\\
&L_{\omega, D_0}(s)&=L_{D_0}(\sqrt{5}/2~C_\omega^2,M_\omega,s)\,\,,\nonumber\\
&L_{\omega, D_2}(s)&=L_{D_2}(\sqrt{30}/16~C_\omega^2,M_\omega,s)\,\, ,\nonumber\\[3mm]
&L_{\rho, S}(s)&=L_{S}(1/9~C_\rho^2,M_\rho,s)\,\,,\nonumber\\
&L_{\rho, D_0}(s)&=L_{D_0}(\sqrt{5}/18~C_\rho^2,M_\rho,s)\,\,,\nonumber\\
&L_{\rho, D_2}(s)&=L_{D_2}(\sqrt{30}/144~C_\rho^2,M_\rho,s)\,\, ,\nonumber\\[3mm]
&L_{a_1, S}(s)&=L_{S}(1/4~C_A^2,M_a,s)\,\,,\nonumber\\
&L_{a_1, D_0}(s)&=L_{D_0}(\sqrt{5}/8~C_A^2,M_a,s)\,\,,\nonumber\\
&L_{a_1, D_2}(s)&=L_{D_2}(-\sqrt{30}/64~C_A^2,M_a,s)\,\, ,\nonumber\\[3mm]
&L_{b_1, S}(s)&=L_{S}(-1/36~C_B^2,M_b,s)\,\,,\nonumber\\
&L_{b_1, D_0}(s)&=L_{D_0}(-\sqrt{5}/72~C_B^2,M_b,s)\,\,,\nonumber\\
&L_{b_1, D_2}(s)&=L_{D_2}(\sqrt{30}/576~C_B^2,M_b,s)\,\, ,\nonumber\\[3mm]
&L_{h_1, S}(s)&=L_{S}(-1/4~C_B^2,M_{h},s)\,\,,\nonumber\\
&L_{h_1, D_0}(s)&=L_{D_0}(-\sqrt{5}/8~C_B^2,M_h,s)\,\,,\nonumber\\
&L_{h_1, D_2}(s)&=L_{D_2}(\sqrt{30}/64~C_B^2,M_h,s)\,\, .\label{eq:L;R}
\eea
Here for $L_{R, J_{\lambda}}$, the subscripts R, J, and $\lambda$ represent for the crossed-channel exchange, spin and helicity respectively.
The coefficients $C_\omega$, $C_\rho$, $C_A$, $C_B$ shown in Table~\ref{tab:VMDc} are fixed from the decay widths $R(A,B)\rightarrow \pi \gamma$.
\begin{table}
\begin{center}
 \begin{tabular}  {|c|c|c|}
 \hline
                                     &      $\Gamma(keV)$    &  $C_R$($GeV^{-1}$)          \\[0.5mm] \hline
$\omega\to\pi^0\gamma$       &       703             &  1.15$\pm0.02$    \\[0.5mm] \hline
$\rho^0\to\pi^0\gamma$       &       89.5            &  1.25$\pm0.08$    \\[0.5mm] \hline
$a_1^+\to\pi^+\gamma$        &       640             &  1.08$\pm0.21$    \\[0.5mm] \hline
$b_1^+\to\pi^+\gamma$        &       230             &  1.95$\pm0.25$    \\[0.5mm] \hline
 \end{tabular}
 \caption{\label{tab:VMDc}Decay widths and fitting parameter.}
 \end{center}
\end{table}

\noindent For higher mass exchanges such as the tensors, we use an effective pole ($M_T, C_T^2$) approximation. Their contribution follows the form of Eq.~(\ref{eq:L;partial})
in terms of the two parameters, $M_T$ and $C_T$, we now fix.
Since we have considered all other allowed single particle exchanges up to a mass of 1.3~GeV, we can (without loss of generality) simply set $M_T=1.4\pm0.2$~GeV. To ensure the convergence of the  partial wave amplitudes
we demand the cancellation of these exchange contributions cancel  when $s \rightarrow \infty$. This imposes the requirement:
\bea
C_T^{\,2}(IJ\lambda=000)&=&+\frac{\sqrt{6}}{18}C_\rho^2+\frac{\sqrt{6}}{6}C_\omega^2-\frac{\sqrt{6}}{72}C_B^2
                     -\frac{\sqrt{6}}{24}C_B^2+\frac{\sqrt{6}}{12}C_A^2=0.477\,\, ,\nonumber\\
C_T^{\,2}(IJ\lambda=020)&=&\frac{\sqrt{30}}{36}C_\rho^2+\frac{\sqrt{30}}{12}C_\omega^2-\frac{\sqrt{30}}{144}C_B^2
                     +\frac{\sqrt{30}}{48}C_B^2+\frac{\sqrt{30}}{24}C_A^2=1.403\,\, ,\nonumber\\
C_T^{\,2}(IJ\lambda=022)&=&\frac{\sqrt{5}}{48}C_\rho^2+\frac{\sqrt{5}}{16}C_\omega^2+\frac{\sqrt{5}}{192}C_B^2
+\frac{\sqrt{5}}{64}C_B^2-\frac{\sqrt{5}}{32}C_A^2=0.354\,\, ,\nonumber\\
C_T^{\,2}(IJ\lambda=200)&=&-\frac{\sqrt{3}}{3}C_\omega^2+\frac{\sqrt{3}}{12}C_B^2+\frac{\sqrt{3}}{12}C_A^2=-0.048\,\, ,\nonumber\\
C_T^{\,2}(IJ\lambda=220)&=&-\frac{\sqrt{15}}{6}C_\omega^2+\frac{\sqrt{15}}{24}C_B^2+\frac{\sqrt{15}}{24}C_A^2=-0.053\,\, ,\nonumber\\
C_T^{\,2}(IJ\lambda=222)&=&-\frac{\sqrt{10}}{16}C_\omega^2-\frac{\sqrt{10}}{64}C_B^2-\frac{\sqrt{10}}{64}C_A^2=-0.509\,\, .\label{eq:CT}
\eea



\end{document}